\documentclass[10pt,a4paper]{amsart}

\usepackage{amsmath,amsthm,amssymb}
\usepackage{epic,eepic}
\usepackage{epsfig,indentfirst}
\usepackage{float,array,varioref}

\restylefloat{figure}

\begin{document}

\title[Slavnov determinants, Yang-Mills, and KP]
{Slavnov determinants, Yang-Mills structure constants, and discrete KP}

\author[O Foda and M Wheeler]{Omar Foda and Michael Wheeler}

\address{Department of Mathematics and Statistics,
University of Melbourne, Parkville, 
Victoria 3010, Australia. 
}
\email{omar.foda@unimelb.edu.au, m.wheeler@ms.unimelb.edu.au}


\begin{abstract}
Using Slavnov's scalar product of a Bethe eigenstate and a generic 
state in closed XXZ spin-$\frac{1}{2}$ chains, with possibly twisted 
boundary conditions, we obtain determinant expressions for tree-level 
structure constants in 1-loop conformally-invariant sectors in various 
planar (super) Yang-Mills theories.
When certain rapidity variables are allowed to be free rather than 
satisfy Bethe equations, these determinants become discrete KP 
$\tau$-functions.
\end{abstract}

\maketitle

\def\ll{ \left\lgroup}
\def\rr{\right\rgroup}

\newcommand{\0}{{\bf 0.}}
\newcommand{\1}{{\bf 1.}}
\newcommand{\2}{{\bf 2.}}
\newcommand{\3}{{\bf 3.}}
\newcommand{\4}{{\bf 4.}}
\newcommand{\5}{{\bf 5.}}
\newcommand{\6}{{\bf 6.}}
\newcommand{\7}{{\bf 7.}}
\newcommand{\8}{{\bf 8.}}
\newcommand{\9}{{\bf 9.}}

\newcommand{\Y }{\mathcal{Y}}
\newcommand{\Tr}{{\rm Tr}}
\newcommand{\< }{{\langle}}

\newcommand{\cA}{\mathcal A}
\newcommand{\cB}{\mathcal B}
\newcommand{\cC}{\mathcal C}
\newcommand{\cH}{\mathcal H}
\newcommand{\cL}{\mathcal L}
\newcommand{\cM}{\mathcal M}
\newcommand{\cN}{\mathcal N}
\newcommand{\cO}{\mathcal O}
\newcommand{\cF}{\mathcal F}
\newcommand{\cf}{\mathcal f}
\newcommand{\cQ}{\mathcal Q}

\newcommand{\CC}{\field{C}}
\newcommand{\NN}{\field{N}}
\newcommand{\ZZ}{\field{Z}}

\newcommand{\Det}{{\rm Det}}

\renewcommand{\P}{{\mathcal P}}

\hyphenation{boson-ic
             ferm-ion-ic
             para-ferm-ion-ic
             two-dim-ension-al
             two-dim-ension-al
             rep-resent-ative
             par-tition}

\hyphenation{And-rews
             config-ura-tion
             config-ura-tions
             Gor-don
             boson-ic
             ferm-ion-ic
             para-ferm-ion-ic
             two-dim-ension-al
             two-dim-ension-al}
             
\begin{center}
{\it Dedicated to Professor M Jimbo on his 60th birthday.}
\end{center}

\setcounter{section}{-1}

%SECTION00
\section{Overview}
\label{section-overview-FW}

Classical integrable models, in the sense of integrable hierarchies 
of nonlinear partial differential equations that admit soliton 
solutions, and quantum integrable models, in the sense of 
Yang-Baxter integrability, are topics that Prof M Jimbo continues 
to make profound contributions to since more than three decades. 

They are also topics that, since the late 1980's, have made 
increasingly frequent contacts with, and have lead to definite 
advances in modern quantum field theory. Amongst the most 
important of these contacts are discoveries of integrable 
structures on both sides of Maldacena's conjectured AdS/CFT 
correspondence \cite{maldacena-FW}. 
From 2002 onward, classical integrability was discovered 
in free superstrings \footnote{Superstrings with 
tree-level interactions only, and no spacetime loops.} 
on the AdS side of AdS/CFT 
\cite{bena-polchinski-roiban-FW, tseytlin-review-FW}, and 
quantum integrability 
in the planar limit \footnote{The limit in which 
the number of colours 
$N_c$             $\rightarrow$ $\infty$, the gauge coupling 
$g_{\textit{YM}}$ $\rightarrow$ $0$, 
while the \rq t Hooft coupling 
$\lambda = g^2_{\textit{YM}} N_c$ remains finite.} 
of $\mathcal{N}
$ $= 4$ supersymmetric Yang-Mills on the CFT side 
\cite{MZ-FW, BKS-FW, BS0307-FW}.  
Further, examples of integrability that are restricted 
1-loop level were discovered in planar Yang-Mills theories 
with fewer supersymmetries and in 
QCD \cite{korchemsky-FW, korchemsky-review-FW}.
In the sequel, we use YM for Yang-Mills theories in general, 
and SYM$_{\mathcal{N}
}$ for $\mathcal{N}
$-extended supersymmetric Yang-Mills.

\subsection{Scope of this work}
In this work, we restrict our attention to quantum field theories 
that are 
{\bf 1.}
 planar, so that the methods of integrability have 
a chance to work, 
{\bf 2.}
 weakly-coupled, so that perturbation theory makes sense and
we can focus our attention to 1-loop level, and 
{\bf 3.}
 conformally-invariant at 1-loop level, so they allow an exact 
mapping to Heisenberg spin-chains, that is spin-chains with 
nearest neighbour interactions that can be solved using the 
algebraic Bethe Ansatz. In the sequel, we consider only 
Heisenberg spin-$\frac{1}{2}$ chains. 

Even within the above restrictions, our subject is still very broad 
and we can only review the basics needed to obtain our results. For 
an introduction to the vast subject of integrability in AdS/CFT, we 
refer to \cite{beisert-review-FW} and references 
therein \footnote{Further highlights of integrability in modern quantum 
field theory and in string theory include 
{\bf 1.} Classical integrable hierarchies in matrix models of non-critical 
strings, from the late 1980's \cite{ginsparg-two-reviews-FW}, 
{\bf 2.} Finite gap solutions in Seiberg-Witten theory of low-energy 
SYM$_2$ in 
the mid 1990's \cite{seiberg-witten-FW, gorsky-FW, marshakov-book-FW}, 
{\bf 3.} Integrability in QCD scattering amplitudes in the mid 1990's
\cite{lipatov-faddeev-korchemsky-FW, korchemsky-review-FW},
{\bf 4.} Free fermion methods in works of Nekrasov, Okounkov, Nakatsu, 
Takasaki and others on Seiberg-Witten theory, in the 2000's 
\cite{NO-FW, NT-FW}, 
{\bf 5.} Integrable spin chains in works of Nekrasov, Shatashvili and 
others on SYM$_2$, in the 2000's \cite{NS-FW}, 
{\bf 6.} Integrable structures, particularly the Yangian, that appear 
in recent studies of SYM$_4$ scattering 
amplitudes \cite{drummond-alday-reviews-FW}. There are many more.}. 

\subsection{Conformal invariance and 2-point functions}
Any 1-loop conformally-invar\-iant quantum field theory 
contains (up to 1-loop order) a basis of local scalar 
primary conformal composite operators  \footnote{In this 
work, we restrict our attention to this class of
local composite operators. In particular, we do not 
consider descendants or operators with non-zero spin,
for which the 2-point and 3-point functions are different.} 
$\{\mathcal{O}\}$ such that the 2-point functions can be written as 
\begin{equation}
\langle
\mathcal{O}_i(x) \overline{\mathcal{O}}_j(y)\rangle 
=
\delta_{ij}
\mathcal{N}_i 
|x - y|^{- 2 \Delta_i}
\label{2-point-FW}
\end{equation}
\noindent where 
$\overline{\mathcal{O}}_i$ is the Wick conjugate of 
$\mathcal{O}_i$, 
$\Delta_i$ is the conformal dimension of $\mathcal{O}_i$ 
and $\mathcal{N}_i$ is a normalization factor. Later, 
we choose $\mathcal{N}_i$ to be (the square root of) 
the Gaudin norm of the corresponding spin-chain 
state.

The primary goal of studies of integrability on the CFT 
side of AdS/CFT in the past ten years has arguably been 
the calculation of the spectrum of conformal dimensions 
$\{\Delta_{\mathcal{O}
}\}$ of local composite operators $\{\mathcal{O}
\}$, 
and matching them with corresponding results from 
the strong coupling AdS side of AdS/CFT. This goal has 
by and large been achieved \cite{beisert-review-FW}, and 
the next logical step is 
to study 3-point functions and their structure constants 
\cite{earlier-papers-on-3-pt-funs-FW, E1-FW, F-FW}.

\subsection{3-point functions and structure constants} 
The 3-point function of three basis local operators 
such as those that appear in ({\ref{2-point-FW}}) is 
restricted (up to 1-loop order) by conformal symmetry 
to be of the form
\begin{multline}
\label{3pt-FW}
\langle
 \mathcal{O}
_i (x_i) 
   \mathcal{O}
_j (x_j) 
   \mathcal{O}
_k (x_k) 
\rangle
= 
%\\
\ll
{\mathcal{N}
_i \ \mathcal{N}
_j \ \mathcal{N}
_k}
\rr^{1/2}
\frac{
C_{ijk}
}
{
 |x_{ij}|^{\Delta_i + \Delta_j - \Delta_k} 
 |x_{jk}|^{\Delta_j + \Delta_k - \Delta_i} 
 |x_{ki}|^{\Delta_k + \Delta_i - \Delta_j}
}
\end{multline}
\noindent where $x_{ij} = x_i - x_j$, and $C_{ijk}$ are structure 
constants. The structure constants $C_{ijk}$ are the subject 
of this work.
In \cite{E1-FW}, Escobedo, Gromov, Sever and Vieira (EGSV) obtained 
sum expressions for the structure constants of non-extremal 
single-trace operators in the scalar sector of SYM$_4$. 
In \cite{F-FW}, the sum expressions of EGSV were evaluated, and  
determinant expressions for the same structure constants were 
obtained \footnote{Three operators $\mathcal{O}
_i$, of length $L_i$, 
$i \in \{1, 2, 3\}$, are non-extremal if 
$l_{ij} = L_i + L_j - L_k > 0$.}.  

\subsection{Aims of this work}
We extend the results of \cite{F-FW} to a number of YM 
theories that are conformally invariant at least up to 1-loop 
level. We also show that the determinants that we 
obtain are discrete KP $\tau$-functions.

More precisely, 
{\bf 1.} We recall, and make explicit, 
a generalization of the restricted Slavnov scalar product 
used in \cite{F-FW} to twisted, 
closed and homogeneous XXZ spin-$\frac{1}{2}$ chains. That is, 
we allow for an anisotropy parameter $\Delta \neq 1$, as well 
as a twist parameter $\theta \neq 0$ in the boundary conditions. 
The result is still a determinant. We use this result to obtain 
determinant expressions for the YM theories listed in 
subsection {\bf \ref{theories-FW}} \footnote{The SYM$_4$ 
expression of \cite{F-FW} is a special case of the general 
expression obtained here.}. 
{\bf 2.} Allowing certain rapidity variables in the determinant 
expressions to be free, rather than satisfy Bethe equations, 
we show that these rapidities can be regarded as Miwa variables. 
In terms of these Miwa variables, the determinants satisfy 
Hirota-Miwa equations and become discrete KP $\tau$-functions.
The structure constants are recovered by requiring that 
the free variables are rapidities that label a gauge-invariant 
composite operator and satisfy Bethe equations.

\subsection{Type-A and Type-B YM theories} 
\label{theories-FW}

We consider six planar, weakly-coupled YM theories. 
{\bf 1.} SYM$_4$ \cite{brink-green-schwarz-FW, minahan-review-FW}, 
{\bf 2.} SYM$_4^{M}$, which is an order-$M$ Abelian orbifold 
of SYM$_4$ that is $\mathcal{N} = 2$ supersymmetric
\cite{leigh-strassler-FW, zoubos-review-FW}, and 
{\bf 3.} SYM$_4^{\beta}$, which is a Leigh-Strassler marginal 
real-$\beta$ deformation of SYM$_4$ that is $\mathcal{N} = 1$ 
supersymmetric \cite{orbifold-papers-FW,zoubos-review-FW}. 
{\bf 4.} The complex scalar sector of pure SYM$_2$ 
\cite{DiV-FW, korchemsky-FW}, 
{\bf 5.} The gluino sector of pure SYM$_1$ \cite{korchemsky-FW}, 
and 
{\bf 6.} The gauge sector of QCD 
\cite{korchemsky-FW, korchemsky-review-FW}.

These six theories are naturally divisible into two types.
Type-{\bf A} contains theories {\bf 1}, {\bf 2} and {\bf 3},
which are conformally-invariant to all orders in perturbation 
theory. 
Type-{\bf B} contains theories {\bf 4}, {\bf 5} and {\bf 6},
which are conformally-invariant to 1-loop level 
only \footnote{There are definitely more gauge theories 
that are conformally-invariant at 1-loop or more, with $SU(2)$ 
sectors that map to states in spin-$\frac{1}{2}$ chains. 
Here we consider only samples of theories with different 
supersymmetries and operator content.}. 

Conformal invariance at 1-loop level, which is the case in 
all theories that we consider, is necessary and sufficient 
for our purposes because the mapping to spin-$\frac{1}{2}$ 
chains with nearest neighbour interactions breaks down at 
higher loops. Our results are valid only up to 1-loop level.

\subsection{Non-extremal operators}
\label{condition-FW}
In \cite{E1-FW, F-FW}, structure constants of three operators 
$\mathcal{O}
_i$ of length $L_i$, $i \in \{1, 2, 3\}$ were considered, 
and the condition that the operators are non-extremal, 
that is $l_{ij} = L_i + L_j - L_k > 0$, for all distinct 
$i$, $j$ and $k$, was emphasized. 
The reason is that, in these works, one wished to compute 
the structure constants of three non-BPS operators. Using 
the analysis presented in this work, one can show that  
this requires the condition $l_{ij} > 0$. One can of course 
consider the special case where one of these parameters 
$l_{ij} = 0$, but then at least one of the three operators 
has to be BPS.

In type-{\bf A} theories, which include SYM$_4$, we can compute 
non-trivial structure constants of three non-BPS operators, so 
we do that, and the condition $l_{ij} > 0$ is satisfied. The case 
where one of these parameters vanishes, for example 
$l_{23} = L_2 + L_3 - L_1 = 0$, is allowed, but then either 
$\mathcal{O}_2$ or $\mathcal{O}_3$ has to be BPS.
In type-{\bf B} theories, we find that one of the three 
operators, which we choose to be $\mathcal{O}_3$, 
has to be BPS, hence the condition $l_{ij} > 0$ is no 
longer significant and we consider operators such 
that $l_{23} = L_2 + L_3 - L_1 = 0$. 

\subsection{$SU(2)$ sectors that map to spin-$\frac{1}{2}$ chains}
\label{doublets-FW}
We will not list the full set of fundamental fields in the gauge 
theories that we consider, but only those fundamental fields that 
form $SU(2)$ doublets that map to states in spin-$\frac{1}{2}$ 
chains. All fields are in the adjoint of $SU(N_c)$ and can be 
represented in terms of $N_c \! \times \! N_c$ matrices. 

\smallskip

\noindent {\bf 1.} SYM$_4$ contains six real scalars that form 
three complex scalars $\{X, Y, Z\}$, and their charge conjugates 
$\{ \bar{X}$, $\bar{Y}$, $\bar{Z}\}$. Any pair of non-charge-conjugate 
scalars, {\it e.g.} $\{Z, X\}$, or $\{Z, \bar{X}\}$, 
forms a doublet that maps to a state in a closed periodic 
XXX spin-$\frac{1}{2}$ chain \footnote{XXX spin-$\frac{1}{2}$ 
chains are XXZ spin-$\frac{1}{2}$ chains with an anisotropy 
parameter $\Delta$ = 1.} 
\cite{MZ-FW, minahan-review-FW}.

\noindent {\bf 2.} SYM$_4^M$ has the same fundamental charged 
scalar fields $\{X, Y, Z\}$ and their charge conjugates, as 
SYM$_4$, so the same scalars form $SU(2)$ doublets. 
Due to the orbifolding of the $SU(2)$ sectors by the action of 
the discrete group $\Gamma_M$, these doublets map to states in 
a closed twisted XXX spin-$\frac{1}{2}$ chain. The twist parameter 
is a (real) phase $\theta = \frac{2 \pi}{M}$ \cite{zoubos-review-FW}. 

\noindent {\bf 3.} SYM$_{4}^{\beta}$ has the same fundamental charged 
scalar fields $\{X, Y, Z\}$ and their charge conjugates, as SYM$_4$, 
so the same scalars form $SU(2)$ doublets. 
Due to the real-$\beta$ deformation, these doublets 
map to states in a closed twisted XXX spin-$\frac{1}{2}$ chain. 
The twist parameter is a (real) phase $\theta = \beta$,
where $\beta$ is the deformation parameter.
\cite{berenstein-chekris-FW, zoubos-review-FW}.

\noindent {\bf 4.} SYM$_2$ has a gluino field $\lambda$ and  
its conjugate $\bar{\lambda}$ that form a doublet that 
maps to a state in a closed untwisted XXZ spin-$\frac{1}{2}$ 
chain with $\Delta = 3$ \cite{DiV-FW, korchemsky-FW}.

\noindent {\bf 5.} SYM$_1$ has a complex scalar $\phi$ and 
its conjugate $\bar{\phi}$ that form a doublet that 
maps to a state in a closed untwisted XXZ 
spin-$\frac{1}{2}$ chain with $\Delta = \frac{1}{2}$ 
\cite{korchemsky-FW}.

\noindent {\bf 6.} Pure QCD has light-cone 
derivatives $\{\partial_{+} A, \partial_{+} \bar{A} \}$, 
where $A$ and $\bar{A}$ are the transverse components 
of the gauge field $A_{\mu}$, that form a doublet that 
maps to a state 
in a closed untwisted XXZ spin--$\frac{1}{2}$ chain with 
$\Delta = - \frac{11}{3}$ \cite{korchemsky-FW}.

\subsection{Remark} 
Theories {\bf 1}, {\bf 2} and {\bf 3}, that are conformally 
invariant to all orders, contain three charged scalars and 
their conjugates. These combine into various $SU(2)$ doublets.
Theories {\bf 4}, {\bf 5} and {\bf 6}, on the other hand, 
contain only one doublet. This fact affects the type 
of structure constants that we can compute in determinant 
form in Section {\bf \ref{section-A-structure-constants-FW}} 
and {\bf \ref{section-B-structure-constants-FW}} \footnote{The 
fact that the structure constants in these two types of theories 
should be handled differently was pointed out to us by C Ahn 
and R Nepomechie.}.

\subsection{Outline of contents} 
%OUTLINE1
In Section {\bf \ref{section-background-FW}}, we recall 
basic background information related to integrability in 
weakly coupled YM.
%
%OUTLINE2
In Section {\bf \ref{section-spin-chain-FW}}, we review standard 
facts on closed XXZ spin-$\frac{1}{2}$ chains with twisted 
boundary conditions. In particular, following \cite{KMT-FW}, we 
introduce restricted versions $S[L, N_1, N_2]$ of Slavnov's scalar 
product, that can be evaluated in determinant 
form \footnote{In \cite{F-FW}, $S[L, N_1, N_2]$ was denoted by 
$S[L, \{N\}]$.}. 

%OUTLINE3
In Section {\bf \ref{section-six-vertex-FW}}, we review standard 
facts on the trigonometric six-vertex model, which is regarded 
as another way to view XXZ spin-$\frac{1}{2}$ chains in terms
of diagrams that are convenient for our purposes. Following 
\cite{MW-FW}, we introduce the $[L, N_1, N_2]$-configurations 
that are central to our result.
The determinant $S[L, N_1, N_2]$, obtained in Section 
{\bf \ref{section-spin-chain-FW}}, turns out to be the partition 
function of these $[L, N_1, N_2]$-configurations. 

%OUTLINE4
In Section {\bf \ref{section-A-structure-constants-FW}}, 
we recall the EGSV formulation of the structure constants 
of three non-extremal composite operators in the scalar 
sector of SYM$_4$. Since all Type-{\bf A} theories, 
which include SYM$_4$ and two other theories that 
are closely related to it, share the same set of 
fundamental charged scalar fields, namely $\{X, Y, Z\}$ 
and their charge conjugates $\{\bar{X}, \bar{Y}, \bar{Z}\}$, 
our discussion applies to all of them in one go. 
Since the composite operators that we are interested in
map to states in (generally twisted) XXX spin-$\frac{1}{2}$ 
chains, we express these structure functions in  
terms of rational six-vertex model configurations, and 
obtain determinant expressions for them. 

%OUTLINE5
In Section {\bf \ref{section-B-structure-constants-FW}}, 
we extend the above discussion to Type-{\bf B} theories, 
which contain theories with only one $SU(2)$ doublet 
that we can work with. 
Since the composite operators that we are interested in
map to states in periodic XXZ spin-$\frac{1}{2}$ 
chains, we express these structure functions in 
terms of trigonometric six-vertex model configurations. 
We find that our method applies only when one of the 
operators is BPS-like (a single-trace of a power of 
one type of fundamental fields).
We obtain determinant expressions for these objects, 
and find that the result is identical to that in 
type-{\bf A}, apart from the fact that one of the 
operators in BPS-like.

%OUTLINE6
In Section {\bf \ref{section-kp-FW}}, we show that the 
determinant expressions are solutions of Hirota-Miwa 
equations, and thereby $\tau$-functions of the discrete 
KP hierarchy. 
%
%OUTLINE7
In Section {\bf \ref{section-summary-FW}}, we summarize 
our results.

%SECTION01
\section{Background}
\label{section-background-FW}
Let us recall basic facts on integrability on 
the CFT side of AdS/CFT.

\subsection{Integrability in AdS/CFT} 
In its strongest sense, the anti-de Sitter/conformal field 
theory (AdS/CFT) correspondence is the postulate that all 
physics, including gravity, in an anti-de Sitter space can 
be reproduced in terms of a conformal field theory that lives 
on the boundary of that space \cite{maldacena-review-FW}.
The first and most thoroughly studied example of the 
correspondence is Maldacena's original proposal that 
type-IIB superstring theory in an AdS$^5$$\times$$S^5$ 
geometry is equivalent to planar SYM$_4$ on the 
4-dimensional boundary of AdS$^5$ \cite{maldacena-FW}. 

Since its proposal in 1997, the AdS/CFT correspondence has 
passed every single check that it was subject to, and there 
was a large number of these. However, because the 
correspondence typically identifies one theory in a regime 
that is easy to study (for example, a weakly-coupled planar 
quantum field theory) to another theory in a regime that is 
hard to study (for example, a quantum free superstring 
theory in a strongly curved geometry), it has so far not 
been possible to prove it \cite{beisert-review-FW}.

\subsection{The dilatation operator} 
The generators of the conformal group in 4-dimensional space\-time, 
$SO(4, 2)$, 
contain a dilatation operator $D$ \cite{beisert-short-review-FW}. 
Every gauge-invariant operator $\mathcal{O}$ in a YM theory, that 
is 1-loop conformally-invariant, is an eigenstate of $D$ to that 
order in perturbation theory. The corresponding eigenvalue 
$\Delta_{\mathcal{O}}$, which is the conformal dimension of 
$\mathcal{O}$, is the analogue of mass in massive, non-conformal 
theories. 

\subsection{SYM$_4$ and spin chains. 1-loop results}
An $SU(2)$ doublet of fundamental fields $\{u, d\}$, which could 
be any of those discussed in Subsection {\bf \ref{doublets-FW}} 
above, is analogous to the $\{\uparrow, \downarrow\}$ states of 
a spin variable on a single site in a spin-$\frac{1}{2}$ chain.
Furthermore, the local gauge-invariant operators formed by taking 
single traces of a product of an arbitrary combination of $u$ and 
$d$ fields, such as $\mathrm{Tr} [uududduu \cdots uu]$, is analogous 
to a state in a closed spin-$\frac{1}{2}$ chain.

In \cite{MZ-FW}, Minahan and Zarembo made the above intuitive
analogies exact correspondences by showing that the action 
of the 1-loop dilatation operator on single-trace operators 
in the $SU(2)$ scalar subsector of SYM$_4$ is identical to 
the action of the nearest-neighbour Hamiltonian on the states 
in a closed periodic XXX spin-$\frac{1}{2}$ 
chain \footnote{Minahan 
and Zarembo obtained their remarkable result in the context 
of the complete scalar sector of SYM$_4$. The relevant spin 
chain in that case is $SO(6)$ symmetric. Here we focus our 
attention on the restriction of their result to the $SU(2)$ 
scalar subsector.}. In this mapping, valid up to 1-loop 
level \footnote{We are interested in local single-trace 
composite operators that consist of many fundamental fields. 
These fields are interacting. In a weakly-interacting quantum 
field theory, one can consistently choose to ignore all 
interactions beyond a chosen order in perturbation theory. In 
the planar theory under consideration, perturbation theory can 
be arranged according to the number of loops in Feynman 
diagrams computed. In a 1-loop approximation, one keeps only 
1-loop diagrams.} single-trace operators with well-defined 
conformal dimensions map to eigenstates of the XXX Hamiltonian.
The corresponding eigenvalues are the conformal dimensions 
$\Delta_{\mathcal{O}}$.

The above brief outline is all we need for the purposes
of this work. For an in-depth overview, we refer the reader 
to \cite{beisert-review-FW}.

%SECTION02
\section{The XXZ spin-$\frac{1}{2}$ chain}
\label{section-spin-chain-FW}

In this section, we recall basic facts related to the XXZ 
spin-$\frac{1}{2}$ chain that are needed in later sections. 
The presentation closely follows that in \cite{MW-FW, F-FW}, 
but adapted to closed XXZ spin chains with twisted 
boundary conditions.

\subsection{1-dimensional lattice segments and spin variables}
Consider a length-$L$ 1-dimension\-al lattice, and label the 
sites with $i \in \{1, 2, \dots, L\}$. Assign site $i$ 
a 2-dimensional vector space $h_i$ with the basis

\begin{align}
| \wedge \rangle
_i
=
\ll \begin{array}{c} 1 \\ 0 \end{array} \rr_i,
\quad
|\vee \rangle
_i
=
\ll \begin{array}{c} 0 \\ 1 \end{array} \rr_i
\label{up-down-FW}
\end{align}
which we refer to as `up' and `down' states, and a spin variable 
$s_i$ which can be equal to either of these states. The space of 
states ${\mathcal{H}}$ is the tensor product 
$\mathcal{H}  = h_1 \otimes \cdots \otimes \ h_L$.
Every state in ${\mathcal{H}}$ is 
an assignment $\{s_1,s_2,\dots,s_L\}$ of $L$ 
definite-value (either up or down) spin variables to the sites of 
the spin chain. In computing scalar products, as we do shortly, 
we think of states in ${\mathcal{H}}$ as initial states. 

\subsection{Initial spin-up and spin-down reference states}
${\mathcal{H}}$ contains two distinguished states, 
\begin{align}
| L^{\wedge}  \rangle
 = \bigotimes_{i=1}^{L} 
\ll
\begin{array}{c}
1 \\ 0
\end{array}
\rr_i
,
\quad
|L^{\vee} \rangle
 = \bigotimes_{i=1}^{L} 
\ll
\begin{array}{c}
0 \\ 1
\end{array}
\rr_i
\end{align}
\noindent where 
$L^{\wedge}$ indicates $L$ spin states that are all up, and 
$L^{\vee  }$ indicates $L$ spin states that are all down. 
These are the initial spin-up and spin-down reference states, 
respectively. 

\subsection{Final spin-up and spin-down reference states, 
and a variation}
Consider a length-$L$ spin chain, and assign each site $i$ 
a 2-dimensional vector space $h_i^{*}$ with the basis
\begin{align}
_i \langle
 \wedge |
=
\ll 1 \ \ 0 \rr_i,
\quad
_i \langle
 \vee |
=
\ll 0 \ \ 1 \rr_i
\label{up-down*-FW}
\end{align}
We construct a final space of states as the tensor product
${\mathcal{H}}^{*} = h_1^{*} \otimes \cdots \otimes h_L^{*}$.
$\mathcal{H}^{*}$ contains two distinguished states
\begin{align}
\langle
  L^{\wedge}  | = \bigotimes_{i=1}^{L} 
\ll
\begin{array}{cc}
1 & 0
\end{array}
\rr_i
,
\quad 
\langle
  L^{\vee} | = \bigotimes_{i=1}^{L} 
\ll
\begin{array}{cc}
0 & 1
\end{array}
\rr_i
\end{align}
\noindent where all spins are up, and all spins are down. 
These are the final spin-up and spin-down reference states.
respectively.
Finally, we consider the variation 
\begin{equation}
\langle
 {N_3}^{\vee}, (L-N_3)^{\wedge} |
=
\bigotimes_{1 \leq i \leq N_3}
\ll
\begin{array}{cc}
0 & 1
\end{array}
\rr_i
\bigotimes_{(N_3 + 1) \leq i \leq L}
\ll
\begin{array}{cc}
1 & 0
\end{array}
\rr_i
\end{equation}

\noindent where the first $N_3$ spins from the left are down, and all 
remaining spins are up. 

\subsection{Pauli matrices}

We define the Pauli matrices
\begin{align}
\sigma_m^{x}
=
\ll
\begin{array}{rr}
0 & 1 
\\
1 & 0 
\end{array}
\rr_m,
\quad
\sigma_m^{y}
=
\ll
\begin{array}{rr}
0 & -i 
\\
i & 0 
\end{array}
\rr_m,
\quad
\sigma_m^{z}
=
\ll
\begin{array}{rr}
1 & 0 
\\
0 & -1
\end{array}
\rr_m
\label{pauli-FW}
\end{align}
with $i = \sqrt{-1}$, and the spin raising/lowering matrices
\begin{align}
\sigma_m^{+}
=
\frac{1}{2}
(\sigma_m^{x} + i \sigma_m^{y})
=
\ll
\begin{array}{cc}
0 & 1
\\
0 & 0 
\end{array}
\rr_m,
\quad
\sigma_m^{-}
=
\frac{1}{2}
(\sigma_m^{x}-i\sigma_m^{y})
=
\ll
\begin{array}{cc}
0 & 0
\\
1 & 0
\end{array}
\rr_m
\label{spin-chang-FW} 
\end{align}
where in all cases the subscript $m$ is used to indicate that 
the matrices act in the vector space $h_m$.

\subsection{The Hamiltonian $H$}
\label{s-ham-FW}

The Hamiltonian of the finite length XXZ spin-$\frac{1}{2}$ 
chain is given by the equivalent expressions 
\begin{multline}
H
=
\frac{1}{2}
\sum_{m=1}^{L} 
\ll
\sigma_m^x \sigma_{m+1}^x
+
\sigma_m^y \sigma_{m+1}^y
+
\Delta (\sigma_m^z \sigma_{m+1}^z-1)
\rr
\label{hamiltonian-FW}
\\
=
\sum_{m=1}^{L} 
\ll
\sigma_m^{+} \sigma_{m+1}^{-}
+
\sigma_m^{-} \sigma_{m+1}^{+}
+
\frac{\Delta}{2} (\sigma_m^z \sigma_{m+1}^z-1)
\rr
\end{multline}
where $\Delta$ is the anisotropy parameter of the model, and 
where we assume the \lq twisted\rq\ periodicity conditions 
\begin{align}
\sigma^{\pm}_{L+1} = e^{\pm i \theta} \sigma^{\pm}_1,
\quad
\sigma^{z}_{L+1} = \sigma^{z}_1
\end{align}

\subsection{The $R$-matrix}

From an initial reference state, we can generate all other states 
in $\mathcal{H}$ using operators that flip the spin variables, one 
spin at a time. Defining these operators requires defining a sequence
of objects.
{\bf 1.} The $R$-matrix,  
{\bf 2.} The $L$-matrix, and finally,
{\bf 3.} The monodromy or $M$-matrix.

The $R$-matrix is an element of ${\rm End}(h_a\otimes h_b)$, 
where $h_a,h_b$ are two 2-dimensional auxiliary vector spaces. 
The variables $u_a, u_b$ are the corresponding rapidity variables. 
The $R$-matrix intertwines these spaces, and it has the $(4\times 4)$ 
structure
\begin{align}
R_{ab}(u_a, u_b)
=
\ll
\begin{array}{cccc}
1               & 0              & 0            & 0    \\
0               & b[u_a, u_b]    & c[u_a, u_b]  & 0    \\
0               & c[u_a, u_b]    & b[u_a, u_b]  & 0    \\
0               & 0              & 0            & 1 
\end{array}
\rr_{ab}
\label{R-matrix-FW}
\end{align}
where we have defined the functions
\begin{align}
b[u_a,u_b]
=
\frac{[u_a - u_b]}{[u_a - u_b + \eta]},
\quad\quad
c[u_a,u_b]
=
\frac{[\eta]}{[u_a - u_b + \eta]},
\quad\quad
[u] \equiv \sinh(u)
\end{align}
The $R$-matrix satisfies unitarity, crossing symmetry and the crucial 
Yang-Baxter equation that is required for integrability, given by
\begin{align}
R_{ab}(u_a,u_b) R_{ac}(u_a,u_c) R_{bc}(u_b,u_c)
=
R_{bc}(u_b,u_c) R_{ac}(u_a,u_c) R_{ab}(u_a,u_b)
\label{YB-FW}
\end{align}
which holds in ${\rm End}(h_a \otimes h_b \otimes h_c)$ for all 
$u_a, u_b, u_c$. 

As we will see in Section {\bf\ref{section-six-vertex-FW}}, the elements 
of the $R$-matrix (\ref{R-matrix-FW}) are the weights of the vertices of 
the trigonometric six-vertex model. This is the origin of the connection 
of the two models. One can graphically represent the elements of 
(\ref{R-matrix-FW}) to obtain the six vertices of the trigonometric 
six-vertex model in Figure {\bf \ref{six-vertices-FW}}. 
 
\subsection{The $L$-matrix}

The $L$-matrix of the XXZ spin chain is a local operator that depends 
on a single rapidity $u_a$, and acts in the auxiliary space $h_a$. Its 
entries are operators acting at the $m$-th lattice site, and identically 
everywhere else. It has the form
\begin{align}
L_{am}(u_a)
=
\ll
\begin{array}{cc}
[u_a+\frac{\eta}{2}\sigma_m^{z}]
&
[\eta] \sigma_m^{-}
\\
\phantom{.}
[\eta] \sigma_m^{+}
&
[u_a-\frac{\eta}{2} \sigma_m^{z}]
\end{array}
\rr_a
\label{xxz-Lmat-FW}
\end{align}
Using the definition of the $R$-matrix and the $L$-matrix, 
(\ref{R-matrix-FW}) and (\ref{xxz-Lmat-FW}) respectively, 
the local intertwining equation is given by
\begin{align}
R_{ab}(u_a,u_b) L_{am}(u_a) L_{bm}(u_b)
=
L_{bm}(u_b) L_{am}(u_a) R_{ab}(u_a,u_b)
\label{xxz-Lint-FW}
\end{align}
The proof of (\ref{xxz-Lint-FW}) is immediate, if one uses 
the matrix representations of 
$\sigma_m^{z},\sigma_m^{+},\sigma_m^{-}$ to write
\begin{multline}
L_{am}(u_a)
=
\ll
\begin{array}{cccc}
[u_a+\frac{\eta}{2}] & 0 & 0 & 0
\\
0 & [u_a-\frac{\eta}{2}] & [\eta] & 0
\\
0 & [\eta] & [u_a-\frac{\eta}{2}] & 0
\\
0 & 0 & 0 & [u_a+\frac{\eta}{2}]
\end{array}
\rr_{am}
%\\
=
[u_a+\eta/2] R_{am}(u_a,\eta/2)
\end{multline}
This means that the $L$-matrix is equal to the $R$-matrix 
$R_{am}(u_a,z_m)$ with $z_m = \eta/2$, up to an overall 
multiplicative factor. Cancelling 
these common factors from (\ref{xxz-Lint-FW}), it becomes
\begin{equation}
R_{ab}(u_a,u_b) R_{am}(u_a,\eta/2) R_{bm}(u_b,\eta/2)
=
%\\
R_{bm}(u_b,\eta/2) R_{am}(u_a,\eta/2) R_{ab}(u_a,u_b)
\end{equation}
which is simply a corollary of the Yang-Baxter equation 
(\ref{YB-FW}).

\subsection{The monodromy matrix $M$}

The monodromy or $M$-matrix is a global operator that acts 
on all sites in the spin chain. It is constructed as 
an ordered direct product of the $L$-matrices that act on 
single sites, 
\begin{align}
M_a(u_a)
=
L_{a1}
(u_a) \dots L_{aL}(u_a) \Omega_a(\theta)
\label{hom-M-FW}
\end{align}
where $\Omega_a(\theta)$ is a twist matrix given by
\begin{align}
\Omega_a(\theta)
=
\ll
\begin{array}{cc}
e^{i\theta} & 0
\\
0 & e^{-i\theta}
\end{array}
\rr_a
\label{twist-FW}
\end{align}
The monodromy matrix is essential in the algebraic Bethe Ansatz 
approach to the diagonalization of the Hamiltonian $H$. It is 
convenient to define an inhomogeneous version, as an ordered 
direct product of $R$-matrices $R_{am}(u_a,z_m)$,
\begin{align}
M_{a}(u_a,\{z\}_L)
=
R_{a1}
(u_a,z_1) \ldots R_{aL}(u_a,z_L) \Omega_a(\theta)
\label{inhom-M-FW}
\end{align}
The variables $\{z_1,\dots,z_L\}$ are parameters corresponding 
with the sites of the spin chain and the homogeneous monodromy 
matrix, given by (\ref{hom-M-FW}), is recovered by setting 
$z_m = \eta /2$ for all $1\leq m \leq L$. The inclusion of the 
variables $\{z_1,\dots,z_L\}$ simplifies many later calculations, 
even though it is the homogeneous limit which actually interests 
us. We write the inhomogeneous monodromy matrix in 
$(2 \! \times  \! 2)$ block form, by defining 
\begin{align}
M_{a}(u_a, \{z\}_L)
=
\ll
\begin{array}{cc}
e^{i\theta} A(u_a) & e^{-i\theta} B(u_a) \\
e^{i\theta} C(u_a) & e^{-i\theta} D(u_a)
\end{array}
\rr_a
\label{monodromy-entry-FW}
\end{align}
\noindent where the matrix entries are operators that act in 
${\mathcal{H}} = h_1 \otimes \cdots \otimes h_L$. To simplify 
the notation, we have omitted the dependence of the elements 
of the $M$-matrix on the quantum rapidities $\{z_1,\dots,z_L\}$. 
This dependence is implied from now on.

The operator entries of the monodromy matrix satisfy a set of 
commutation relations, which are determined by the equation
\begin{equation}
R_{ab}(u_a,u_b)
M_{a}(u_a,\{z\}_L)
M_{b}(u_b,\{z\}_L)
=
%\\
M_{b}(u_b,\{z\}_L)
M_{a}(u_a,\{z\}_L)
R_{ab}(u_a,u_b)
\end{equation}
which is a direct consequence of the Yang-Baxter equation 
(\ref{YB-FW}) and the property
\begin{align}
[
R_{ab}(u_a,u_b)
,
\Omega_a(\theta)
\Omega_b(\theta)
]
=
0
\end{align}
of the twist matrix. Typical examples of these commutation 
relations, which are particularly important in the algebraic 
Bethe Ansatz, are
\begin{align}
&B(u) B(v) = B(v) B(u) \label{BB-FW} \\ 
&[u-v+\eta] B(u) A(v) =  [\eta] B(v) A(u) + [u-v] A(v) B(u) \label{AB-FW} \\
&[\eta] B(u) D(v) + [u-v] D(u) B(v) = [u-v+\eta] B(v) D(u) \label{DB-FW} 
\end{align}
In Section {\bf\ref{section-six-vertex-FW}}, we identify 
the operator entries of the monodromy matrix (\ref{monodromy-entry-FW}) 
with rows of vertices from the six-vertex model, see 
Figure {\bf\ref{four-lines-FW}}.

\subsection{The transfer matrix $T$}

The transfer matrix $T\ll u_a,\{z\}_L\rr$ is defined as the trace 
of the inhomogeneous monodromy matrix 
\begin{align}
T\ll u_a,\{z\}_L \rr 
=
{\rm Tr}_a M_a(u_a,\{z\}_L)
= 
e^{i\theta} A(u_a)
+
e^{-i\theta} D(u_a)
\end{align}
The Hamiltonian (\ref{hamiltonian-FW}) is recovered via the quantum 
trace identity
\begin{align}
H
=
[\eta]
\frac{d}{du}
\log T(u)
\Big|_{u = \frac{\eta}{2}},
\quad {\rm where}\quad
T(u)
=
T \ll u,\{z\}_L \rr
\Big|_{z_1=\cdots = z_L =\frac{\eta}{2}}
\label{trace-identity-FW}
\end{align}
where the anisotropy parameter in (\ref{hamiltonian-FW}) is defined as 
$\Delta = \cosh(\eta)$. In this equation all quantum parameters 
have been set equal, so for the 
purpose of reconstructing the Hamiltonian $H$ we see that the homogeneous 
monodromy matrix is sufficient. However, in all subsequent calculations 
we preserve the variables $\{z_1,\dots,z_L\}$ and seek eigenvectors of 
$T \ll u,\{z\}_L \rr$. By (\ref{trace-identity-FW}), they are 
clearly also eigenvectors of $H$. 

\subsection{Generic states, eigenstates and Bethe equations}
The initial and final spin-up reference states 
$|  L^{\wedge}  \rangle$ and $\langle L^{\wedge}  |$ are 
eigenstates of the diagonal elements of the monodromy matrix. 
They satisfy the equations
\begin{align}
A(u,\{z\}_L) |  L^{\wedge}  \rangle
&=
a(u) 
|  L^{\wedge}  \rangle
,
\quad
D(u,\{z\}_L) |  L^{\wedge}  \rangle
=
d(u)
|  L^{\wedge}  \rangle
\label{diagonal1-FW}
\\
\langle
  L^{\wedge}  | A(u,\{z\}_L)
&=
a(u)
\langle
  L^{\wedge}  |,
\quad
\langle
  L^{\wedge}  | D(u,\{z\}_L)
=
d(u)
\langle
  L^{\wedge}  | 
\label{diagonal1*-FW}
\end{align}
where we have defined the eigenvalues
\begin{align}
a(u) =1,
\quad
d(u) =
\prod_{i=1}^{L}
\frac{[u-z_i]}
{[u-z_i+\eta]}
\end{align}
This makes $|  L^{\wedge}  \rangle
$ and $\langle
  L^{\wedge}  |$ 
eigenstates of the transfer matrix $T\ll u,\{z\}_L\rr$. 
The rest of the eigenstates $\{ \mathcal{O}
 \}$ of 
$T\ll u,\{z\}_L \rr$, that is, states that satisfy 
\begin{equation}
    T \ll u,\{z\}_L \rr                |\mathcal{O}
 \rangle
_{\beta}
=
\ll e^{i\theta} A(u) + e^{-i\theta} D(u) \rr     |\mathcal{O}
 \rangle
_{\beta} 
= 
E_{\mathcal{O}
}(u)               |\mathcal{O}
 \rangle
_{\beta}
\label{eigenstate-FW}
\end{equation}
where $E_{\mathcal{O}
}(u)$ is the corresponding eigenvalue, are generated using 
the Bethe Ansatz. This is the statement that all eigenstates 
of $T\ll u,\{z\}_L \rr$ are created in two steps. 
{\bf 1.} One acts on the initial reference state with the 
$B$-element of the monodromy matrix
\begin{equation}
|\mathcal{O}
 \rangle
_{\beta} = 
B(u_{\beta_N}) \cdots B(u_{\beta_1}) |  L^{\wedge}  \rangle
\end{equation}
where $N \leq L$, since acting on $| L^{\wedge} \rangle$ with 
more $B$-operators than the number of sites in the spin chain 
annihilates it. This generates a `generic Bethe state'. 
{\bf 2.} We require that the auxiliary space rapidity variables  
$\{u_{\beta_1}, \dots, u_{\beta_N}\}$ satisfy Bethe equations, 
hence the use of the subscript $\beta$ \footnote{We use $\beta$ 
in two different ways. 
{\bf 1.} To indicate the deformation parameter in SYM$_4^{\beta}$ 
theories, and 
{\bf 2.} To indicate that a certain state is a Bethe eigenstate 
of the spin-chain Hamiltonian. There should be no confusion 
with {\bf 1}, in which $\beta$ is a parameter but never 
a subscript, while in {\bf 2} it is always a subscript.}. 
We call the resulting state a `Bethe eigenstate'. That is, 
$|\mathcal{O}\rangle_{\beta}$ is an eigenstate of 
$T\ll u,\{z\}_L \rr$ if and only if 
\begin{equation}
\frac{a(u_{\beta_i})}
{d(u_{\beta_i})}
=
\prod_{j=1}^{L}
\frac{
[u_{\beta_i} - z_j + \eta]
}{
[u_{\beta_i} - z_j]
}
=
e^{- 2 i \theta}
\prod_{j \not= i}^{N}
\frac{
[u_{\beta_j} - u_{\beta_i} - \eta]
}{
[u_{\beta_j} - u_{\beta_i} + \eta]
},
\label{bethe-equations-FW} 
\end{equation}
for all $1\leq i \leq N$. This fact can be proved using 
the commutation relations (\ref{AB-FW}) and (\ref{DB-FW}), 
as well as (\ref{diagonal1-FW}) and (\ref{diagonal1*-FW}). 
As remarked earlier, by virtue of (\ref{trace-identity-FW}), 
eigenstates of the transfer matrix $T\ll u,\{z\}_L\rr$ are 
also eigenstates of the spin-chain Hamiltonian $H$. 
The latter is the spin-chain version of the 1-loop dilatation 
operator in SYM$_4$. We construct eigenstates 
of $T \ll u,\{z\}_L \rr$ in $\mathcal{H}^{*}$ using 
the $C$-element of the $M$-matrix 
\begin{equation}
_{\beta}\langle
 \mathcal{O}
|
=
\langle
  L^{\wedge} |
C(u_{\beta_1})
\ldots
C(u_{\beta_N})
\end{equation}
where $N \leq L$ to obtain a non vanishing result, 
and requiring that the auxiliary space rapidity variables 
satisfy the Bethe equations.

\subsection{Scalar products that are determinants}
Following \cite{KMT-FW, MW-FW} we define the scalar product 
$S[L,N_1,N_2]$, $0 \leq N_2 \leq N_1$, that involves $(N_1 + N_2)$ 
operators, $N_1$ $B$-operators with auxiliary rapidities that 
satisfy Bethe equations, and 
$N_2$ $C$-operators with auxiliary rapidities that are 
free \footnote{To simplify the notation, 
we use $N_1$, $N_2$ and $N_3 = N_1 - N_2$, instead of the 
corresponding notation used in \cite{KMT-FW,MW-FW}. These variables 
match the corresponding ones in 
Section {\bf \ref{section-A-structure-constants-FW}}.}. 
For $N_2 = 0$, we obtain, up to a non-dynamical factor, the domain 
wall partition function. For $N_2 = N_1$, we obtain Slavnov's scalar 
product \cite{Slavnov-FW}. As we will see in Section 
{\bf \ref{section-six-vertex-FW}}, $S[L,N_1,N_2]$ is the partition 
function (weighted sum over all internal configurations) of 
a lattice in an $[L,N_1,N_2]$-configuration, see Figure 
{\bf\ref{restricted-bc-FW}}. 

Let $\{u_{\beta}\}_{N_1}$ $=$ $\{u_{\beta_1}, \dots, u_{\beta_{N_1}}\}$, 
$\{v\}_{N_2}$ $=$ $\{v_1, \dots, v_{N_2}\}$, 
$\{z\}_L$ $=$ $\{z_1, \dots, z_L\}$ 
be three sets of variables the first of which satisfies Bethe equations, 
$0 \leq N_2 \leq N_1$ and $1\leq N_1 \leq L$. 
We define the scalar products 
\begin{equation}
\label{restricted-sp-FW}
S[L,N_1,N_2] 
\ll 
\{u_{\beta}\}_{N_1}, 
\{v\}_{N_2},
\{z\}_{L  }
\rr
%\\
=
%\\
\langle
  N_3^{\vee}, (L-N_3)^{\wedge}  |
\prod_{i=1}^{N_2} \mathbb{C}(v_i) \prod_{j=1}^{N_1} 
                  \mathbb{B}(u_{\beta_j}) | L^{\wedge}  \rangle
\end{equation}
with $N_3 = N_1 - N_2$, and where we have defined the normalized 
$B$- and $C$-operators
\begin{align}
\mathbb{B}(u) = \frac{B(u)}{d(u)},
\quad
\mathbb{C}(v) = \frac{C(v)}{d(v)}
\end{align}
which are introduced only as a matter of convention. It is clear 
that for $N_2=0$, we obtain a domain wall partition function, 
while for $N_2=N_1$, we obtain Slavnov's scalar product. In all 
cases, we assume that the auxiliary rapidities  
$\{u_{\beta}\}_{N_1}$ obey the Bethe equations 
(\ref{bethe-equations-FW}), and use the subscript $\beta$ 
to emphasize that, while the auxiliary rapidities $\{v\}_{N_2}$ 
are either free or also satisfy their own set of Bethe equations. 
When the latter is the case, this fact is not used. The quantum 
rapidities $\{z\}_L$ are taken to be equal to the same constant value 
in the homogeneous limit. 

\subsection{A determinant expression for the Slavnov scalar product 
$S[L,N_1,N_2]$} 
Following \cite{KMT-FW,MW-FW}, we consider the $(N_1 \! \times \! N_1)$ 
matrix 
\begin{equation}
\mathcal{S}
\ll 
\{u_{\beta}\}_{N_1},
\{v\}_{N_2},
\{z\}_L 
\rr
=
%\\
\ll   
\begin{array}{cccccc}
f_{  1}(z_1) & \cdots & f_{  1}(z_{N_3}) & g_{  1}(v_{N_2}) & \cdots & g_{  1}(v_1)
\\
\vdots       &        & \vdots           & \vdots       &        & \vdots
\\
f_{N_1}(z_1) & \cdots & f_{N_1}(z_{N_3}) 
                                         & g_{N_1}(v_{N_2}) 
				                        & \cdots & g_{N_1}(v_1)
\end{array}
\rr
\end{equation}
whose entries are the functions
\begin{equation}
f_i(z_j)
=
\ll \frac{ [\eta] }{[u_{\beta_i} - z_j + \eta] [u_{\beta_i} - z_j]} \rr
\prod_{k=1}^{N_2}
\frac{1}{[v_k - z_j]} 
\end{equation}

\begin{multline}
g_i(v_j)
=
\ll
\frac{[\eta]}{[u_{\beta_i} - v_j]}
\rr 
\times 
\\
\ll   
\ll
\prod_{k=1}^{L} 
\frac{[v_j - z_k + \eta]}{[v_j - z_k]}
\prod_{k \not = i}^{N_1} [u_{\beta_k} - v_j + \eta]  
\rr
-
e^{- 2 i \theta}
\prod_{k \not = i}^{N_1} [u_{\beta_k} - v_j - \eta] 
\rr
\end{multline}
and where $N_3 = N_1 - N_2$. 
Since the auxiliary rapidities $\{u_{\beta}\}_{N_1}$ satisfy 
Bethe equations (\ref{bethe-equations-FW}), following 
\cite{KMT-FW,MW-FW} it is possible to show that
\begin{equation}
\label{restricted-slavnov-FW}
S[ L, N_1, N_2 ] =
%\\
\frac{\displaystyle{
\prod_{i=1}^{N_1} \prod_{j=1}^{N_3} [u_{\beta_i} - z_j + \eta]
\det \mathcal{S} \ll \{u_{\beta}\}_{N_1}, \{v\}_{N_2}, \{z\}_L \rr
}}
{\displaystyle{
\prod_{1 \leq i < j \leq  N_1} [u_{\beta_j}-u_{\beta_i}]
\prod_{1 \leq i < j \leq  N_2} [v_i-v_j] 
\prod_{1 \leq i < j \leq  N_3} [z_i-z_j]
}} 
\end{equation}

\subsection{The Slavnov scalar product $S[L,N_1,N_1]$}

Consider the special case $N_1 \! = \! N_2 \! = \! N$, which corresponds to 
Slavnov's scalar product itself. In this case we obtain 
the $(N \! \times \! N)$ matrix 
\begin{align}
\mathcal{S}
\ll 
\{u_{\beta}\}_{N},
\{v\}_{N},
\{z\}_L 
\rr
=
\ll   
\begin{array}{ccc}
g_{  1}(v_{N}) & \cdots & g_{  1}(v_1)
\\
\vdots       &        & \vdots
\\
g_{N}(v_{N}) & \cdots  & g_{N}(v_1)
\end{array}
\rr
\end{align}
whose entries are the functions
\begin{multline}
g_i(v_j)
=
\ll
\frac{[\eta]}{[u_{\beta_i} - v_j]}
\rr \times 
\\
\ll   
\ll
\prod_{k=1}^{L} 
\frac{[v_j - z_k + \eta]}{[v_j - z_k]}
\prod_{k \not = i}^{N} [u_{\beta_k} - v_j + \eta]  
\rr
-
e^{- 2 i \theta}
\prod_{k \not = i}^{N} [u_{\beta_k} - v_j - \eta] 
\rr
\end{multline}
The Slavnov scalar product $S[L,N,N]$ is then given by 
\begin{equation} 
S[ L, N,N ] =
\frac{\displaystyle{
\det \mathcal{S} \ll \{u_{\beta}\}_{N}, \{v\}_{N}, \{z\}_L \rr
}}
{\displaystyle{
\prod_{1 \leq i < j \leq  N} [u_{\beta_i}-u_{\beta_j}]
\prod_{1 \leq i < j \leq  N} [v_i-v_j] 
}}
\label{slavnov-FW} 
\end{equation}

\subsection{Restrictions}

There is a simple relation between the scalar products $S[L,N_1,N_1]$ 
and $S[L,N_1,N_2]$, which was used in \cite{MW-FW} to provide a recursive 
proof of Slavnov's scalar product formula \cite{Slavnov-FW}. It is easy 
to show that by restricting the free variables $v_{N_1},\dots,v_{N_2+1}$ 
in (\ref{slavnov-FW}) to the values $z_1,\dots,z_{N_3}$, one obtains the 
recursion relation
\begin{equation}
\label{restrict1-FW}
\left.
\ll
\prod_{i=N_2+1}^{N_1}
\prod_{j=1}^{L}
[v_i-z_j]
S[L,N_1,N_1]
\rr
\right|_{\substack{ v_{N_1} = z_1 \\ \vdots \\ v_{(N_2+1)} = z_{N_3} }}
%\\
=
\prod_{i=1}^{N_3}
\prod_{j=1}^{L}
[z_i-z_j+\eta]
S[L,N_1,N_2]
\end{equation}
As we show in Section {\bf\ref{section-six-vertex-FW}}, 
the scalar products $S[L,N_1,N_1]$ and $S[L,N_1,N_2]$ are 
in direct correspondence with the partition function of an 
$[L,N_1,N_1]$- and $[L,N_1,N_2]$-configuration, respectively. 
Accordingly, we expect that the recursion 
relation (\ref{restrict1-FW}) has a natural interpretation 
at the level of six-vertex model lattice configurations, and 
indeed this turns out to be the case. 

\subsection{The homogeneous limit of $S[L,N_1,N_2]$}
For the result in this paper, 
we need the homogeneous limit of $S[L,N_1,N_2]$, 
which we denote by $S^{\textit{hom}}[L, N_1,N_2]$. 
Taking the limit $z_i \rightarrow z$, $i \in \{1, \dots, L\}$, 
the result is
\begin{equation}
S^{\textit{hom}}[L, N_1,N_2] 
=
\frac{\displaystyle{
\prod_{i=1}^{N_1} [u_{\beta_i} - z + \eta]^{N_3}
\det
\mathcal{S}^{\textit{hom}} 
\ll \{u_{\beta}\}_{N_1}, \{v\}_{N_2}, z \rr
}}
{\displaystyle{
\prod_{1 \leq i < j \leq  N_1 } [u_{\beta_j} - u_{\beta_i}]
\prod_{1 \leq i < j \leq  N_2 } [v_i - v_j]
}}
\label{restricted-slavnov-homogeneous-FW}
\end{equation}

\begin{multline}
\mathcal{S}^{\textit{hom}} 
\ll 
\{u_{\beta}\}_{N_1},
\{v\}_{N_2}, 
  z 
\rr =
%
%\\
\ll   
\begin{array}{cccccc}
\Phi^{(0)}_{  1}(z) & \cdots & \Phi^{(N_3 - 1)}_{ 1}(z) & 
g^{\textit{hom}}_{  1}(v_{N_2}) & \cdots & g^{\textit{hom}}_{ 1}(v_1)
\\
\vdots & & \vdots & \vdots & & \vdots
\\
\Phi^{(0)}_{N_1}(z) & \cdots & \Phi^{(N_3 - 1)}_{N_1}(z) & 
g^{\textit{hom}}_{N_1}(v_{N_2}) & \cdots & g^{\textit{hom}}_{N_1}(v_1)
\end{array}
\rr 
%\hspace{1.2 in}
\end{multline}
where $\Phi^{(j)}_{i} = \frac{1}{j!} \partial^{(j)}_z  f_i (z)$, and
\begin{equation}
g^{\textit{hom}}_i(v_j) =
\frac{[\eta]}{[u_{\beta_i} - v_j]} 
%\times 
%\\
\ll
\ll
\frac{
[v_j - z + \eta]
}
{
[v_j - z]
}
\rr^L
\prod_{k \not = i}^{N_1} [u_{\beta_k} - v_j   + \eta]
-
e^{- 2 i \theta}
\prod_{k \not = i}^{N_1} [u_{\beta_k} - v_j   - \eta]
\rr
\end{equation}

\subsection{The Gaudin norm}
Let us consider the original, unrestricted Slavnov scalar 
product in the homogeneous limit $z_i \rightarrow z$, 
$S[L, N_1, N_1] 
\ll \{u_{\beta}\}_{N_1}, \{v\}_{N_1}, z \rr$, 
and set $\{v\}_{N_1}$ $=$ $\{u_{\beta}\}_{ N_1}$ to obtain the 
Gaudin norm 
$\mathcal{N}
 \ll \{u_{\beta}\}_{N_1} \rr$ which is the square of the norm of 
the Bethe eigenstate with auxiliary rapidities $\{u_{\beta}\}_{N_1}$. 
It inherits a determinant expression that can be computed 
starting from that of the Slavnov scalar product that we begin 
with and taking the limit $\{v\}_{N_1} \rightarrow \{u_{\beta}\}_{N_1}$. 
Following \cite{KMT-FW}, one obtains 
\begin{equation}
\label{Gaudin-FW}
\mathcal{N}
\ll
\{u_{\beta}\}_{N_1}
\rr
=
\ll
e^{-2 i \theta} 
[\eta]
\rr^{N_1}
\ll
\prod_{i \neq j}^{N_1}
\frac{[u_i - u_j + \eta]}
     {[u_i - u_j ]}
\rr
\det \Phi^{\prime} \ll \{u_{\beta}\}_{ N_1} \rr
\end{equation}
where
\begin{equation}
\Phi^{\prime}_{ij} \ll \{u_{\beta}\}_{ N_1} \rr
=
- 
\partial_{u_j}
\ln 
\ll 
\ll 
\frac{[u_i - z + \eta]}{[u_i - z]}
\rr^{L}
\prod_{ \substack{ k = 1 \\ k \neq i }}^{N_1}
\frac{[u_k - u_i + \eta]}
     {[u_k - u_i - \eta]}
\rr
\end{equation}

\noindent We need the Gaudin norm to normalize the Bethe 
eigenstates that form the 3-point functions whose structure 
constants we are interested in. 

%SECTION03
\section{The trigonometric six-vertex model}
\label{section-six-vertex-FW}

This section follows almost {\it verbatim} the exposition in 
\cite{F-FW}, up to straightforward adjustments to account for the 
fact that here we are interested in the trigonometric, rather 
than the rational six-vertex model. We recall the 2-dimensional 
trigonometric six-vertex model in the absence of external fields. 
From now on, `six-vertex model' refers to that.
It is equivalent to the XXZ spin-$\frac{1}{2}$ chain that appears 
in \cite{E1-FW}, but affords a diagrammatic representation that suits 
our purposes. We introduce quite a few terms to make this correspondence
clear and the presentation precise, but the reader with basic 
familiarity with exactly solvable lattice models can skip all these.  
\subsection{Lattice lines, orientations, and rapidity variables}
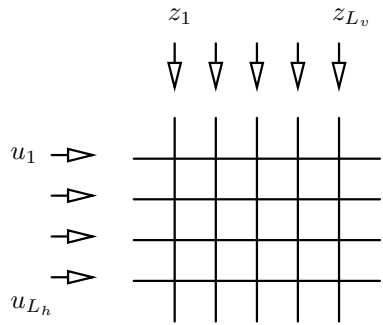
\begin{figure}
\setlength{\unitlength}{0.0009cm}
\begin{picture}(5000,5000)(1500, 1500)
% negative x shifts the figure to the right
% negative y shifts the figure up
% positive y shifts the figure down
\thicklines
%
%The following are vertical lines
\path(2400,4800)(2400,1800)
\path(3000,4800)(3000,1800)
\path(3600,4800)(3600,1800)
\path(4200,4800)(4200,1800)
\path(4800,4800)(4800,1800)
%
%The following are horizontal lines
\path(1800,4200)(5400,4200)
\path(1800,3600)(5400,3600)
\path(1800,3000)(5400,3000)
\path(1800,2400)(5400,2400)
%
%The folowing are the horizontal arrows
%
\path(0600,4254)(1200,4254)
\whiten\path(840,4164)(1200,4254)(840,4344)(840,4164)
\path(0600,3654)(1200,3654)
\whiten\path(840,3564)(1200,3654)(840,3744)(840,3564)
\path(0600,3054)(1200,3054)
\whiten\path(840,2964)(1200,3054)(840,3144)(840,2964)
\path(0600,2454)(1200,2454)
\whiten\path(840,2364)(1200,2454)(840,2544)(840,2364)
%
%The following are the vertical arrows
%
\path(2400,5300)(2400,5900)
\whiten\path(2490,5610)(2400,5250)(2310,5610)(2490,5610)
\path(3000,5300)(3000,5900)
\whiten\path(3090,5610)(3000,5250)(2910,5610)(3090,5610)
\path(3600,5300)(3600,5900)
\whiten\path(3690,5610)(3600,5250)(3510,5610)(3690,5610)
\path(4200,5300)(4200,5900)
\whiten\path(4290,5610)(4200,5250)(4110,5610)(4290,5610)
\path(4800,5300)(4800,5900)
\whiten\path(4890,5610)(4800,5250)(4710,5610)(4890,5610)
\put(0000,4200){$u_1$}
\put(0000,2000){$u_{L_h}$}
\put(2300,6250){$z_1$}
%\put(2900,6250){$z_2$}
%
\put(4700,6250){$z_{L_v}$}
\end{picture}
%
%\begin{ca}
\caption{A square lattice with oriented lines and rapidity variables. 
Lattice lines are assigned the orientations indicated by the 
white arrows.}
%\end{ca}
%
%\end{minipage}
\label{lattice-FW}
\end{figure}

Consider a square lattice with $L_h$ horizontal lines and $L_v$ 
vertical lines that intersect at $L_h \! \times  \! L_v$ points.
There is no restriction, at this stage, on $L_h$ or $L_v$. 
We order the horizontal lines from top to bottom and assign the 
$i$-th line an orientation from left to right and a rapidity 
variable $u_i$. 
We order the vertical lines from left to right and assign
the $j$-th line an orientation from top to bottom and a rapidity 
variable $z_j$. See Figure {\bf \ref{lattice-FW}}. The orientations 
that we assign to the lattice lines are matters of convention and 
are only meant to make the vertices of the six-vertex model, that 
we introduce shortly, unambiguous. We orient the vertical 
lines from top to bottom to agree with the direction of the 
`spin set evolution' that we introduce shortly.

%FIG01
%
%\begin{center}
%\begin{minipage}{4.4in}

\subsection{Line segments, arrows, and vertices} 
Each lattice line is split into segments by all other lines that 
are perpendicular to it. `Bulk segments' are attached to two 
intersection points, and `boundary segments' are attached to 
one intersection point only.
Assign each segment an arrow that can point in either direction, 
and define the vertex $v_{ij}$ as the union of 
{\bf 1.} The intersection point of the $i$-th horizontal line 
and the $j$-th vertical line, 
{\bf 2.} The four line segments attached to this intersection 
point, and 
{\bf 3.} The arrows on these segments (regardless of their 
orientations). Assign $v_{ij}$ a weight that depends on the 
specific orientations of its arrows, and the rapidities $u_i$ 
and $z_j$ that flow through it.

\subsection{Six vertices that conserve `arrow flow'}
Since every arrow can point in either direction, there are 
$2^4 = 16$ possible types of vertices. In this note, we are 
interested in a model such that only those vertices that 
conserves `arrow flow' (that is, the number of arrows that 
point toward the intersection point is equal to the number 
of arrows that point away from it) have non-zero weights. 
There are six such vertices. They are shown in 
Figure {\bf \ref{six-vertices-FW}}. We assign these vertices 
non-vanishing weights. We assign the rest of the 16 possible 
vertices zero weights \cite{baxter-book-FW}. 

In the trigonometric six-vertex model, and in the absence of 
external fields, the six vertices with non-zero weights form 
three equal-weight pairs of vertices, as in Figure 
{\bf \ref{six-vertices-FW}}. 
Two vertices that form a pair are related by reversing all arrows,
thus the vertex weights are invariant under reversing all arrows.
In the notation of Figure {\bf \ref{six-vertices-FW}}, the weights 
of the trigonometric six-vertex model, in the absence of external 
fields, are 
\begin{equation}
\label{weights-FW}
a[u_i, z_j] =  1, 
\quad
b[u_i, z_j] =  \frac{[u_i - z_j       ]}
                    {[u_i - z_j + \eta]},
\quad
c[u_i, z_j] =  \frac{            [ \eta ]}
                    {[u_i - z_j + \eta]}
\end{equation}
where we use the definition $[x] = \sinh (x)$ to simplify 
notation \footnote{The weights of the six-vertex model (\ref{weights-FW}) 
and the entries of the XXZ $R$-matrix (\ref{R-matrix-FW}) are identical. 
This is the origin of the connection between the two models. We have 
chosen to write down these functions twice for clarity and to emphasize 
this fact.}.
The assignment of weights in (\ref{weights-FW}) satisfies unitarity, 
crossing symmetry, and most importantly the Yang-Baxter equations 
\cite{baxter-book-FW}. It is not unique since one can multiply all 
weights by the same factor without changing the final physical results.

\subsection{Correspondence with the XXZ $R$-matrix}
\begin{figure}
\setlength{\unitlength}{0.0009cm}
\begin{picture}(10000,7000)(1000, 1500)
\thicklines
%
%This arrow points to the left
\blacken\path( 1580,3490)( 1220,3400)( 1580,3310)( 1580,3490)
%
%This arrow points to the right
\blacken\path( 1220,5610)( 1580,5700)( 1220,5790)( 1220,5610)
%
%This arrow points to the left
\blacken\path( 2480,3490)( 2120,3400)( 2480,3310)( 2480,3490)
%
%This arrow points to the right
\blacken\path( 2120,5610)( 2480,5700)( 2120,5790)( 2120,5610)
%
%This arrow points to the left
\blacken\path( 5480,5790)( 5120,5700)( 5480,5610)( 5480,5790)
%
%This arrow points to the right
\blacken\path( 5120,3310)( 5480,3400)( 5120,3490)( 5120,3310)
%
%This arrow points to the left
\blacken\path( 6380,5790)( 6020,5700)( 6380,5610)( 6380,5790)
%
%This arrow points to the right
\blacken\path( 6020,3310)( 6380,3400)( 6020,3490)( 6020,3310)
%
%This arrow points to the left
\blacken\path( 9380,3490)( 9020,3400)( 9380,3310)( 9380,3490)
%
%This arrow points to the right
\blacken\path( 9020,5610)( 9380,5700)( 9020,5790)( 9020,5610)
%
%This arrow points to the right
\blacken\path( 9920,3310)(10280,3400)( 9920,3490)( 9920,3310)
%
%This arrow point to the left
\blacken\path(10280,5790)( 9920,5700)(10280,5610)(10280,5790)
%
%This arrow points down
\blacken\path( 1710,3160)( 1800,2800)( 1890,3160)( 1710,3160)
%
%This arrow point down
\blacken\path( 1710,3985)( 1800,3625)( 1890,3985)( 1710,3985)
%
%This arrow point down
\blacken\path( 5610,3160)( 5700,2800)( 5790,3160)( 5610,3160)
%
%This arrow point down
\blacken\path( 5610,4060)( 5700,3700)( 5790,4060)( 5610,4060)
%
%This arrow point down
\blacken\path( 9510,4060)( 9600,3700)( 9690,4060)( 9510,4060)
%
%This arrow point down
\blacken\path( 9510,5460)( 9600,5100)( 9690,5460)( 9510,5460)
%
%This arrow point up
\blacken\path( 1890,5040)( 1800,5400)( 1710,5040)( 1890,5040)
%
%This arrow point up
\blacken\path( 1890,5940)( 1800,6300)( 1720,5940)( 1890,5940)
%
%This arrow point up
\blacken\path( 5790,5040)( 5700,5400)( 5610,5040)( 5790,5040)
%
%This arrow point up
\blacken\path( 5790,5940)( 5700,6300)( 5620,5940)( 5790,5940)
%
%This arrow point up
\blacken\path( 9690,2740)( 9600,3100)( 9520,2740)( 9690,2740)
%
%This arrow point up
\blacken\path( 9690,5940)( 9600,6300)( 9510,5940)( 9690,5940)
%
%vertical segments
\path(01800,7650)(01800,7025)
\path(01800,6600)(01800,4800)
\path(01800,4300)(01800,2500)
%
%vertical segments
\path(05700,7650)(05700,7025)
\path(05700,6600)(05700,4800)
\path(05700,4300)(05700,2500)
%
%vertical segments
\path(09600,7650)(09600,7025)
\path(09600,6600)(09600,4800)
\path(09600,4300)(09600,2500)
%
%the horizontal segments from the right.
%they come in pairs
%
%belongs to white arrows
\path(00000,3400)(00600,3400)
\path(00000,5700)(00600,5700)
%
%these lines hold two arrows
\path(00900,3400)(02700,3400)
\path(00900,5700)(02700,5700)
%
%these belong to white arrows
\path(03900,3400)(04500,3400)
\path(03900,5700)(04500,5700)
%
%these lines hold two black arrows
\path(04800,3400)(06600,3400)
\path(04800,5700)(06600,5700)
%
%belong to white arrows
\path(07800,3400)(08400,3400)
\path(07800,5700)(08400,5700)
\path(08700,3400)(10500,3400)
\path(08700,5700)(10500,5700)
%
%
%The vertical arrows
%
\whiten\path(1890,7360)(1800,7000)(1710,7360)(1890,7360)
\whiten\path(5790,7360)(5700,7000)(5610,7360)(5790,7360)
\whiten\path(9690,7360)(9600,7000)(9510,7360)(9690,7360)
%
%The horizontal arrows
%
\whiten\path(0260,3310)(0620,3400)(0260,3490)(0260,3310)
\whiten\path(0260,5610)(0620,5700)(0260,5790)(0260,5610)
\whiten\path(4160,3310)(4520,3400)(4160,3490)(4160,3310)
\whiten\path(4160,5610)(4520,5700)(4160,5790)(4160,5610)
\whiten\path(8060,3310)(8420,3400)(8060,3490)(8060,3310)
\whiten\path(8060,5610)(8420,5700)(8060,5790)(8060,5610)
\put(1000,1800){$a[u_i, z_j]$}
\put(5000,1800){$b[u_i, z_j]$}
\put(9000,1800){$c[u_i, z_j]$}
\put(0300,5000){$u_i$}
\put(4200,5000){$u_i$}
\put(8100,5000){$u_i$}
\put(0300,2700){$u_i$}
\put(4200,2700){$u_i$}
\put(8100,2700){$u_i$}
\put(1700,8000){$z_j$}
\put(5600,8000){$z_j$}
\put(9500,8000){$z_j$}
\end{picture}
%
%\begin{ca}
\caption{The non-vanishing-weight vertices of the six-vertex model. 
Pairs of vertices in the same column share the weight that 
is shown below that column. 
The white arrows indicate the line orientations needed to 
specify the vertices without ambiguity.}
%\end{ca}
%
%\end{minipage}
\label{six-vertices-FW}
\end{figure}
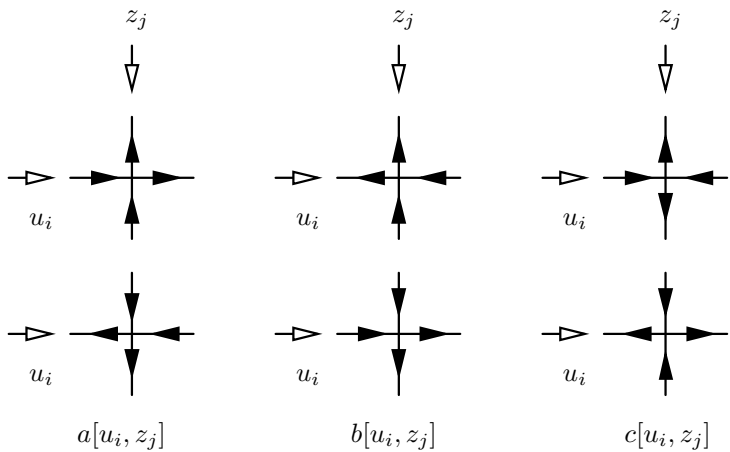

The connection with the $R$-matrix of the XXZ spin-$\frac{1}{2}$ chain 
is straightforward. One can think of the $R$-matrix (\ref{R-matrix-FW}) 
as assigning a weight to the transition from a pair of initial spin 
states (for example, the definite spin states on the right and upper 
segments that meet at a certain vertex) to a pair of final spin states 
(the definite spin states on 
the left and lower segments that meet at the same vertex as 
the initial ones). In the case of the trigonometric XXZ 
spin-$\frac{1}{2}$ chain, this is a transition between four possible 
initial spin states and four final spin states, and accordingly the 
$R$-matrix is $(4 \! \times  \! 4)$. The six non-zero entries of 
(\ref{R-matrix-FW}) correspond with the vertices in Figure 
{\bf \ref{six-vertices-FW}}. 

%FIG02
%
%\begin{center}
%\begin{minipage}{5.0in}

\subsection{Remarks} 
{\bf 1.} The spin chains that are relevant to integrability in 
YM theories are typically homogeneous since all quantum 
rapidities are set equal to the same constant value $z$. 
In our conventions, $z$ $=$ $\frac{1}{2} \sqrt{-1}$. 
{\bf 2.} The trigonometric six-vertex model that corresponds 
to the homogeneous XXZ spin-$\frac{1}{2}$ chain used in 
\cite{E1-FW} has, in our conventions, all vertical 
rapidity variables equal to $\frac{1}{2} \sqrt{-1}$. 
In this note, we start with inhomogeneous vertical rapidities, 
then take the homogeneous limit at the end.
{\bf 3.} In a 2-dimensional vertex model with no external fields, 
the horizontal lines are on equal footing with the vertical 
lines. To make contact with spin chains, we treat these 
two sets of lines differently.
{\bf 4.} In all figures in this note, a line segment with 
an arrow on it obviously indicates a definite arrow assignment. 
A line segment with no arrow on it implies a sum over both arrow 
assignments. 

\subsection{Weighted configurations and partition functions} 
By assigning every vertex $v_{ij}$ a weight $w_{ij}$, a vertex 
model lattice configuration with a definite assignment of arrows 
is assigned a weight equal to the product of the weights of its 
vertices. 
The partition function of a lattice configuration is the sum of 
the weights of all possible configurations that the vertices can 
take and that respect the boundary conditions. Since the vertex 
weights are invariant under reversal of all arrows, the partition 
function is also invariant under reversal of all arrows.

\subsection{Rows of segments, spin systems, spin system states and 
net spin} 
A `row of segments' is a set of {\it vertical} line segments that 
start and/or end on the same horizontal line(s).
An $L_h \! \times  \! L_v$ six-vertex lattice configuration 
has $(L_v + 1)$ rows of segments. On every length-$L_h$ row of 
segments, one can assign a definite spin configuration, whereby 
each segment carries a spin variable (an arrow) that can point 
either up or down. 
A `spin system' on a specific row of segments is a set of all 
possible definite spin configurations that one can assign to 
that row. 
A `spin system state' is one such definite configuration.
Two neighbouring spin systems (or spin system states) are 
separated by a horizontal lattice line. The spin systems 
that live on the top and the bottom rows of segments are 
initial and final spin systems, respectively. Consider a 
specific spin system state. Assign each up-spin the value 
$+1$ and each down-spin the value $-1$. The sum of these 
values is the net spin of this spin system state. In this 
paper, we only consider six-vertex model configurations 
such that all elements in a spin system have the same net 
spin.

\subsection{Initial and final spin-up and spin-down reference 
states, and a variation}

An initial (final) spin-up reference state 
$| L^{\wedge} \rangle
$ ($ \langle
 L^{\wedge} | $)
is a spin system state on a top (bottom) row
of segments with $L$ arrows that are all up. 
An initial (final) spin-down reference state 
$| L^{\vee} \rangle
$ ($ \langle
 L^{\vee} | $)
is a spin system state on a top (bottom) row of segments with 
$L$ arrows that are all down. 
The state $ \langle
  N_{3}^{\vee}, (L-N_3)^{\wedge} |$ 
is a spin system state on a bottom row of segments with 
$L$ arrows such that the first $N_3$ 
arrows from the left are down, while the remaining $(L-N_3)$ 
arrows are up. We do not need the initial version 
of this state.

\subsection{Correspondence with XXZ spin chain states}
The connection to the XXZ spin-$\frac{1}{2}$ chain of Section 
{\bf\ref{section-spin-chain-FW}} is clear. Every state of the 
periodic spin chain is analogous to a spin system state in 
the six-vertex model. Periodicity is not manifest in the 
latter representation for the same reason that it is not 
manifest once we choose a labeling system.  
The initial and final spin-up/down reference states are 
the six-vertex analogues of those discussed in 
Section {\bf {\ref{section-spin-chain-FW}}}.

\subsection{Remarks}
{\bf 1.}
 There is of course no `time variable' in the six-vertex model, 
but one can think of a spin system as a dynamical system that 
evolves in discrete steps as one scans a lattice configuration 
from top to bottom. 
Starting from an initial spin set and scanning the configuration 
from top to bottom, the intermediate spin sets are consecutive 
states in the history of a dynamical system, ending with the final 
spin set. This evolution is caused by the action of the horizontal 
line elements. {\bf 2.}
 In this paper, all elements in a spin system, 
that live on a certain row of segments, have the same net spin. 
The reason is that vertically adjacent spin systems are separated 
by horizontal lines of a fixed type that change the net spin by 
the same amount ($\pm 1$) or keep it unchanged. Since we consider 
only lattice configurations with given horizontal lines (and do not 
sum over different types), the net spin of all elements in a spin 
system changes by the same amount.

\subsection{Four types of horizontal lines} 
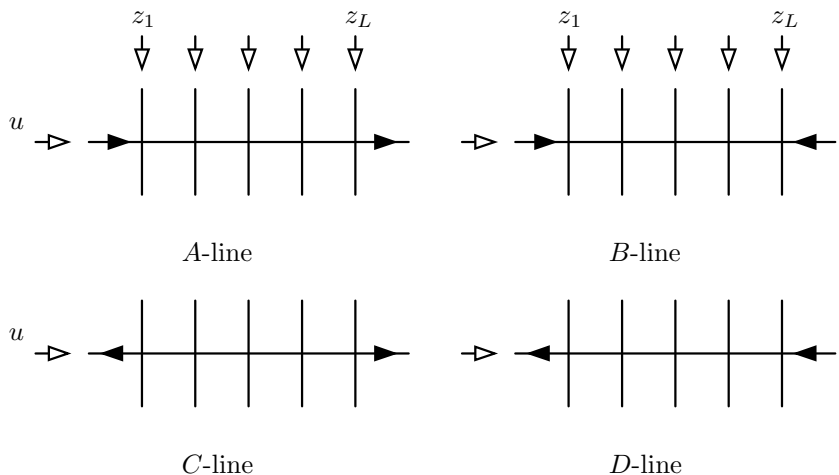
\begin{figure}
\setlength{\unitlength}{0.00035cm}
\begin{picture}(20000,18000)(2500,-13000)
\thicklines
\blacken\path(-1250,00250)(-1250,-0250)(-0530,00000)(-1250,00250)
\blacken\path(-0750,-7750)(-0750,-8250)(-1470,-8000)(-0750,-7750)
\blacken\path(08750,00250)(08750,-0250)(09470,00000)(08750,00250)
\blacken\path(08750,-7750)(08750,-8250)(09470,-8000)(08750,-7750)
\blacken\path(14750,00250)(14750,-0250)(15470,00000)(14750,00250)
\blacken\path(15250,-7750)(15250,-8250)(14530,-8000)(15250,-7750)
\blacken\path(25250,00250)(25250,-0250)(24530,00000)(25250,00250)
\blacken\path(25250,-7750)(25250,-8250)(24530,-8000)(25250,-7750)
\path(00000,-02000)(00000, 2000)
\path(00000, 03000)(00000, 4000)
\path(00000,-10000)(00000,-6000)
\path(02000,-02000)(02000, 2000)
\path(02000, 03000)(02000, 4000)
\path(02000,-10000)(02000,-6000)
\path(04000,-02000)(04000, 2000)
\path(04000, 03000)(04000, 4000)
\path(04000,-10000)(04000,-6000)
\path(06000,-02000)(06000, 2000)
\path(06000, 03000)(06000, 4000)
\path(06000,-10000)(06000,-6000)
\path(08000,-02000)(08000, 2000)
\path(08000, 03000)(08000, 4000)
\path(08000,-10000)(08000,-6000)
\path(12000, 00000)(13000, 0000)
\path(12000,-08000)(13000,-8000)
\path(14000, 00000)(26000, 0000)
\path(14000,-08000)(26000,-8000)
\path(16000,-02000)(16000, 2000)
\path(16000, 03000)(16000, 4000)
\path(16000,-10000)(16000,-6000)
\path(18000,-02000)(18000, 2000)
\path(18000, 03000)(18000, 4000)
\path(18000,-10000)(18000,-6000)
\path(20000,-02000)(20000, 2000)
\path(20000, 03000)(20000, 4000)
\path(20000,-10000)(20000,-6000)
\path(22000,-02000)(22000, 2000)
\path(22000, 03000)(22000, 4000)
\path(22000,-10000)(22000,-6000)
\path(24000,-02000)(24000, 2000)
\path(24000, 03000)(24000, 4000)
\path(24000,-10000)(24000,-6000)
\path(-2000, 00000)(10000, 0000)
\path(-2000,-08000)(10000,-8000)
\path(-4000, 00000)(-3000, 0000)
\path(-4000,-08000)(-3000,-8000)
\put(01500,-04500){$A$-line}
\put(17500,-04500){$B$-line}
\put(01500,-12500){$C$-line}
\put(17500,-12500){$D$-line}
\put(15600, 04500){$z_1$}
\put(23600, 04500){$z_L$}
\put(-0400, 04500){$z_1$}
\put(07600, 04500){$z_L$}
\put(-5000, 00500){$u$}
\put(-5000,-07500){$u$}
\whiten\path(01750,03500)(02250,03500)(02000,02780)(01750,03500)
\whiten\path(03750,03500)(04250,03500)(04000,02780)(03750,03500)
\whiten\path(05750,03500)(06250,03500)(06000,02780)(05750,03500)
\whiten\path(07750,03500)(08250,03500)(08000,02780)(07750,03500)
\whiten\path(12500,00250)(12500,-0250)(13220,00000)(12500,00250)
\whiten\path(12500,-7750)(12500,-8250)(13220,-8000)(12500,-7750)
\whiten\path(15750,03500)(16250,03500)(16000,02780)(15750,03500)
\whiten\path(17750,03500)(18250,03500)(18000,02780)(17750,03500)
\whiten\path(19750,03500)(20250,03500)(20000,02780)(19750,03500)
\whiten\path(21750,03500)(22250,03500)(22000,02780)(21750,03500)
\whiten\path(23750,03500)(24250,03500)(24000,02780)(23750,03500)
\whiten\path(-0250,03500)(00250,03500)(00000,02780)(-0250,03500)
\whiten\path(-3500,00250)(-3500,-0250)(-2780,00000)(-3500,00250)
\whiten\path(-3500,-7750)(-3500,-8250)(-2780,-8000)(-3500,-7750)
\end{picture}
%
%\begin{ca}
\caption{There are four types of horizontal lines in a six-vertex 
model lattice configuration.}
%\end{ca}
%
%\end{minipage}
\label{four-lines-FW}
\end{figure}

Each horizontal line has two boundary segments. Each boundary 
segment has as an arrow that can point into the configuration 
or away from it. Accordingly, we can distinguish four types 
of horizontal lines, as in Figure {\bf \ref{four-lines-FW}}. 
We refer to them as $A$-, $B$-, $C$- and $D$-lines.

%FIG03
%
%\begin{center}
%\begin{minipage}{4.5in}

An important property of a horizontal line is how the net spin 
changes as one moves across it from top to bottom. Given that 
all vertices conserve `arrow flow', one can easily show that, 
scanning a configuration from top to bottom, 
$B$-lines change the net spin by $-1$, 
$C$-lines change it by $+1$, while 
$A$- and $D$-lines preserve the net spin. This can be easily 
understood by working out a few simple examples.

\subsection{Correspondence with monodromy matrix entries}
The $A$-, $B$-, $C$- and $D$-lines in Figure {\bf \ref{four-lines-FW}} 
are the six-vertex model representation of the corresponding elements 
of the $M$-matrix in Section {\bf\ref{section-spin-chain-FW}}. This 
graphical representation is used frequently throughout the 
rest of the paper.
 
\subsection{Four types of configurations} 

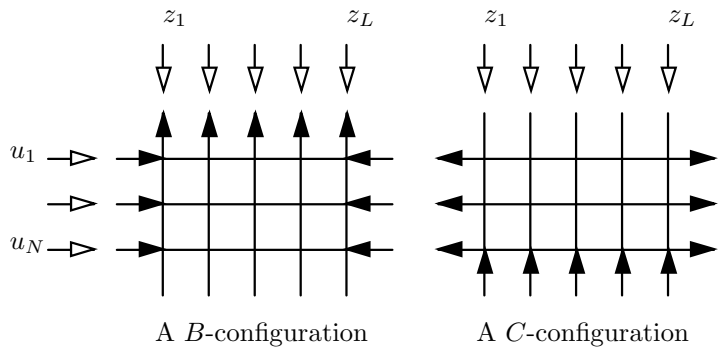
\begin{figure}
\setlength{\unitlength}{0.0010cm}
\begin{picture}(10000,4500)(1000,-0750)
\thicklines
\put(2100,3600){$z_1$}
\put(4500,3600){$z_L$}
\put(6300,3600){$z_1$}
\put(8700,3600){$z_L$}
\put(2000,-0600){A $B$-configuration}
\put(6200,-0600){A $C$-configuration}
\put(100,0610){$u_N$}
\put(100,1810){$u_1$}
\path(0600,0600)(1200,0600)
\path(0600,1200)(1200,1200)
\path(0600,1800)(1200,1800)
\path(1500,0600)(5100,0600)
\path(1500,1200)(5100,1200)
\path(1500,1800)(5100,1800)
%
%Here comes the first 5 vertical lines 
%including those of the arrows
%
\path(2100,2400)(2100,0000)
\path(2100,3300)(2100,3000)
\path(2700,2400)(2700,0000)
\path(2700,3300)(2700,3000)
\path(3300,2400)(3300,0000)
\path(3300,3300)(3300,3000)
\path(3900,2400)(3900,0000)
\path(3900,3300)(3900,3000)
\path(4500,2400)(4500,0000)
\path(4500,3300)(4500,3000)
\path(5700,0600)(9300,0600)
\path(5700,1200)(9300,1200)
\path(5700,1800)(9300,1800)
\path(6300,2400)(6300,0000)
\path(6300,3300)(6300,3000)
\path(6900,2400)(6900,0000)
\path(6900,3300)(6900,3000)
\path(7500,2400)(7500,0000)
\path(7500,3300)(7500,3000)
\path(8100,2400)(8100,0000)
\path(8100,3300)(8100,3000)
\path(8700,2400)(8700,0000)
\path(8700,3300)(8700,3000)
\blacken\path(1800,0700)(2100,0610)(1800,0520)(1800,0700)
\blacken\path(1800,1300)(2100,1210)(1800,1120)(1800,1300)
\blacken\path(1800,1900)(2100,1810)(1800,1720)(1800,1900)
\blacken\path(2020,2100)(2110,2400)(2200,2100)(2020,2100)
\blacken\path(2620,2100)(2710,2400)(2800,2100)(2620,2100)
\blacken\path(3220,2100)(3310,2400)(3400,2100)(3220,2100)
\blacken\path(3820,2100)(3910,2400)(4000,2100)(3820,2100)
\blacken\path(4420,2100)(4510,2400)(4600,2100)(4420,2100)
\blacken\path(4800,0520)(4500,0610)(4800,0700)(4800,0520)
\blacken\path(4800,1120)(4500,1210)(4800,1300)(4800,1120)
\blacken\path(4800,1720)(4500,1810)(4800,1900)(4800,1720)
\blacken\path(6000,0700)(5700,0610)(6000,0520)(6000,0700)
\blacken\path(6000,1300)(5700,1210)(6000,1120)(6000,1300)
\blacken\path(6000,1900)(5700,1810)(6000,1720)(6000,1900)
\blacken\path(6220,0300)(6310,0600)(6400,0300)(6220,0300)
\blacken\path(6820,0300)(6910,0600)(7000,0300)(6820,0300)
\blacken\path(7420,0300)(7510,0600)(7600,0300)(7420,0300)
\blacken\path(8020,0300)(8110,0600)(8200,0300)(8020,0300)
\blacken\path(8620,0300)(8710,0600)(8800,0300)(8620,0300)
\blacken\path(9000,0520)(9300,0610)(9000,0700)(9000,0520)
\blacken\path(9000,1120)(9300,1210)(9000,1300)(9000,1120)
\blacken\path(9000,1720)(9300,1810)(9000,1900)(9000,1720)
\whiten\path(0900,0520)(1200,0610)(0900,0700)(0900,0520)
\whiten\path(0900,1120)(1200,1210)(0900,1300)(0900,1120)
\whiten\path(0900,1720)(1200,1810)(0900,1900)(0900,1720)
\whiten\path(2020,3000)(2110,2700)(2200,3000)(2020,3000)
\whiten\path(2620,3000)(2710,2700)(2800,3000)(2620,3000)
\whiten\path(3220,3000)(3310,2700)(3400,3000)(3220,3000)
\whiten\path(3820,3000)(3910,2700)(4000,3000)(3820,3000)
\whiten\path(4420,3000)(4510,2700)(4600,3000)(4420,3000)
\whiten\path(6220,3000)(6310,2700)(6400,3000)(6220,3000)
\whiten\path(6820,3000)(6910,2700)(7000,3000)(6820,3000)
\whiten\path(7420,3000)(7510,2700)(7600,3000)(7420,3000)
\whiten\path(8020,3000)(8110,2700)(8200,3000)(8020,3000)
\whiten\path(8620,3000)(8710,2700)(8800,3000)(8620,3000)
\end{picture}
%
%\begin{ca}
\caption{On the left, a $B$-configuration, generated by the action 
of $N$ $B$-lines on an initial length-$L$ reference state, 
$N \leq L$. 
A weighted sum over all possible configurations of segments 
with no arrows is implied. On the right, the corresponding 
$C$-configuration.}
%\end{ca}
%
%\end{minipage}
\label{initial-final-state-FW}
\end{figure}

{\bf 1.}
 A $B$-configuration is a lattice configuration with 
$L$ vertical lines and $N$ horizontal lines, 
$N \leq L$, such that
{\bf A.} The initial spin system is an initial reference state
$| L^{\wedge} \rangle
$, and
{\bf B.} All horizontal lines are $B$-lines. 
An example is on the left hand side of Figure 
{\bf \ref{initial-final-state-FW}}. 

{\bf 2.}
 A $C$-configuration is a lattice configuration with 
$L$ vertical lines and $N$ horizontal lines, 
$N \leq L$, such that
{\bf A.} All horizontal lines are $C$-lines, and 
{\bf B.} The final spin system is a final reference state
$\langle
 L^{\wedge} |$.
An example is on the right hand side of Figure 
{\bf \ref{initial-final-state-FW}}.  

%FIG04
%
%\begin{center}
%\begin{minipage}{5.0in}

{\bf 3.} A $BC$-configuration is a lattice configuration with 
$L$ vertical lines and $2N_1$ horizontal lines, $0 \leq N_1 \leq L$,
such that
{\bf A.} The initial spin system is an initial reference state
$|  L^{\wedge}  \rangle
$, 
{\bf B.} The first $N_1$ horizontal lines from top to bottom are $B$-lines,
{\bf C.} The following $N_1$ horizontal lines are $C$-lines,
{\bf D.} The final spin system is a final reference state
$ \langle
  L^{\wedge}  |$. See Figure {\bf \ref{bc-0-FW}} \footnote{For 
visual clarity, we have allowed for a gap between 
the $B$-lines and the $C$-lines in Figure {\bf \ref{bc-0-FW}}. 
There is also a gap between the $N_3$-th and $(N_3 +1)$-th vertical 
lines, where $N_3 = 3$ in the example shown, that indicates 
separate portions of the lattice that will be relevant shortly. 
The reader should ignore this at this stage.}.

{\bf 4.} An $[L,N_1,N_2]$-configuration, $0 \leq N_2 \leq N_1$, 
is identical to a $BC$-configuration except that it has $N_1$ 
$B$-lines, and $N_2$ $C$-lines. When $N_3 = N_1 - N_2 = 0$, we 
evidently recover a $BC$-configuration. The case $N_2 = 0$ is 
discussed below. For intermediate values of $N_2$, we obtain 
restricted $BC$-configurations whose partition functions
turn out to be essentially the structure constants.

\subsection{Correspondence with generic Bethe states} 
An initial (final) generic Bethe state is represented in 
six-vertex model terms as a $B$-configuration ($C$-configuration), 
as illustrated on the left (right) hand side of 
Figure {\bf \ref{initial-final-state-FW}}. Note that the outcome 
of the action of the $N$ $B$-lines ($C$-lines) on the initial 
(final) length-$L$ spin-up reference state is an initial (final) 
spin system that can assume all possible spin states of net spin 
$(L-N)$. Each of these definite spin states is weighted by 
the weight of the corresponding lattice configuration.

\subsection{Correspondence with $S[L,N_1,N_1]$ scalar products 
and $S[L,N_1,N_2]$ restricted scalar products} In the language 
of the six-vertex model, the scalar product $S[L,N_1,N_1]$ 
corresponds with a $BC$-configuration with $N_1$ $B$-lines and 
$N_1$ $C$-lines, as illustrated in Figure {\ref{bc-0-FW}}. 
The restricted scalar product $S[L,N_1,N_2]$ corresponds with 
an $[L,N_1,N_2]$-configuration, as illustrated in Figure 
{\ref{restricted-bc-FW}}. Compared with the definition of 
$S[L,N_1,N_2]$ in (\ref{restricted-sp-FW}), the partition 
function of an $[L,N_1,N_2]$-configuration differs only up 
to an overall normalization. To translate between the two, one 
should divide the latter by $d(u)$ for every $B$-line with rapidity 
$u$ and by $d(v)$ for every $C$-line with rapidity $v$. 

\subsection{$[L, N_1,N_2]$-configurations as restrictions 
of $BC$-\-configurations}
\begin{figure}
\setlength{\unitlength}{0.0008cm}
\begin{picture}(10000,09000)(0000,0750)
\thicklines
\path(0600,1500)(9000,1500)
\path(0600,2100)(9000,2100)
\path(0600,2700)(9000,2700)
\path(0600,3300)(9000,3300)
\path(0600,3900)(9000,3900)
\path(0600,5100)(9000,5100)
\path(0600,5700)(9000,5700)
\path(0600,6300)(9000,6300)
\path(0600,6900)(9000,6900)
\path(0600,7500)(9000,7500)
\path(1210,8100)(1210,0900)
\path(1800,8100)(1800,0900)
\path(2400,8100)(2400,0900)
\path(3600,8100)(3600,0900)
\path(4200,8100)(4200,0900)
\path(4800,8100)(4800,0900)
\path(5400,8100)(5400,0900)
\path(6000,8100)(6000,0900)
\path(6600,8100)(6600,0900)
\path(7200,8100)(7200,0900)
\path(7800,8100)(7800,0900)
\path(8400,8100)(8400,0900)
\blacken\path(0900,1590)(0600,1500)(0900,1410)(0900,1590)
\blacken\path(0900,2190)(0600,2100)(0900,2010)(0900,2190)
\blacken\path(0900,2790)(0600,2700)(0900,2610)(0900,2790)
\blacken\path(0900,3390)(0600,3300)(0900,3210)(0900,3390)
\blacken\path(0900,3990)(0600,3900)(0900,3810)(0900,3990)
\blacken\path(0900,5010)(1200,5110)(0900,5190)(0900,5010)
\blacken\path(0900,5610)(1200,5700)(0900,5790)(0900,5610)
\blacken\path(0900,6210)(1210,6300)(0900,6390)(0900,6210)
\blacken\path(0900,6810)(1200,6900)(0900,6990)(0900,6810)
\blacken\path(0900,7410)(1200,7500)(0900,7590)(0900,7410)
\blacken\path(1300,1200)(1210,1500)(1120,1200)(1300,1200)
\blacken\path(1300,7800)(1210,8100)(1120,7800)(1300,7800)
\blacken\path(1900,1200)(1810,1500)(1720,1200)(1900,1200)
\blacken\path(1900,7800)(1810,8100)(1720,7800)(1900,7800)
\blacken\path(2500,1200)(2410,1500)(2320,1200)(2500,1200)
\blacken\path(2500,7800)(2410,8100)(2320,7800)(2500,7800)
\blacken\path(3700,1200)(3610,1500)(3520,1200)(3700,1200)
\blacken\path(3700,7800)(3610,8100)(3520,7800)(3700,7800)
\blacken\path(4300,1200)(4210,1500)(4120,1200)(4300,1200)
\blacken\path(4300,7800)(4210,8100)(4120,7800)(4300,7800)
\blacken\path(4900,1200)(4810,1506)(4720,1200)(4900,1200)
\blacken\path(4900,7800)(4810,8100)(4720,7800)(4900,7800)
\blacken\path(5500,1200)(5410,1500)(5320,1200)(5500,1200)
\blacken\path(5500,7800)(5410,8100)(5320,7800)(5500,7800)
\blacken\path(6100,1200)(6010,1500)(5920,1200)(6100,1200)
\blacken\path(6100,7800)(6010,8100)(5920,7800)(6100,7800)
\blacken\path(6700,1200)(6610,1500)(6520,1200)(6700,1200)
\blacken\path(6700,7800)(6610,8100)(6520,7800)(6700,7800)
\blacken\path(7300,1200)(7210,1500)(7120,1200)(7300,1200)
\blacken\path(7300,7800)(7210,8100)(7120,7800)(7300,7800)
\blacken\path(7900,1200)(7810,1500)(7720,1200)(7900,1200)
\blacken\path(7900,7800)(7810,8100)(7720,7800)(7900,7800)
\blacken\path(8500,1200)(8410,1500)(8320,1200)(8500,1200)
\blacken\path(8500,7800)(8410,8100)(8320,7800)(8500,7800)
\blacken\path(8700,1400)(9000,1490)(8700,1580)(8700,1400)
\blacken\path(8700,2000)(9000,2090)(8700,2180)(8700,2000)
\blacken\path(8700,2600)(9000,2690)(8700,2780)(8700,2600)
\blacken\path(8700,3200)(9000,3290)(8700,3380)(8700,3200)
\blacken\path(8700,3800)(9000,3890)(8700,3980)(8700,3800)
\blacken\path(8700,5200)(8400,5110)(8700,5020)(8700,5200)
\blacken\path(8700,5800)(8400,5710)(8700,5620)(8700,5800)
\blacken\path(8700,6400)(8400,6310)(8700,6220)(8700,6400)
\blacken\path(8700,7000)(8400,6910)(8700,6820)(8700,7000)
\blacken\path(8700,7600)(8400,7510)(8700,7420)(8700,7600)
%
%The folowing are the horizontal arrows
%
\path(-0360,1500)(0000,1500)
\path(-0360,2100)(0000,2100)
\path(-0360,2700)(0000,2700)
\path(-0360,3300)(0000,3300)
\path(-0360,3900)(0000,3900)
\put(-1200,1500){$v_{N_1}$}
\put(-1200,3900){$v_{1}$}
\whiten\path(0000,1410)(0360,1500)(0000,1590)(0000,1410)
\whiten\path(0000,2010)(0360,2100)(0000,2190)(0000,2010)
\whiten\path(0000,2610)(0360,2700)(0000,2790)(0000,2610)
\whiten\path(0000,3210)(0360,3300)(0000,3390)(0000,3210)
\whiten\path(0000,3810)(0360,3900)(0000,3990)(0000,3810)
\path(-0360,5100)(0000,5100)
\path(-0360,5700)(0000,5700)
\path(-0360,6300)(0000,6300)
\path(-0360,6900)(0000,6900)
\path(-0360,7500)(0000,7500)
\put(-1200,5100){$u_{N_1}$}
\put(-1200,7500){$u_1$}
\whiten\path(0000,5010)(0360,5100)(0000,5190)(0000,5010)
\whiten\path(0000,5610)(0360,5700)(0000,5790)(0000,5610)
\whiten\path(0000,6210)(0360,6300)(0000,6390)(0000,6210)
\whiten\path(0000,6810)(0360,6900)(0000,6990)(0000,6810)
\whiten\path(0000,7410)(0360,7500)(0000,7590)(0000,7410)
%
%The following are the vertical arrows
%
\path(1200,9220)(1200,8500)
\path(1800,9220)(1800,8500)
\path(2400,9220)(2400,8500)
\path(3600,9220)(3600,8500)
\path(4200,9220)(4200,8500)
\path(4800,9220)(4800,8500)
\path(5400,9220)(5400,8500)
\path(6000,9220)(6000,8500)
\path(6600,9220)(6600,8500)
\path(7200,9220)(7200,8500)
\path(7800,9220)(7800,8500)
\path(8400,9220)(8400,8500)
\put(1100,9500){$z_{1  } $}
\put(2300,9500){$z_{N_3  } $}
\put(3500,9500){$z_{N_3 +1} $}
\put(8300,9500){$z_{L  } $}
\whiten\path(1290,8860)(1200,8500)(1110,8860)(1290,8860)
\whiten\path(1890,8860)(1800,8500)(1710,8860)(1890,8860)
\whiten\path(2490,8860)(2400,8500)(2310,8860)(2490,8860)
\whiten\path(3690,8860)(3600,8500)(3510,8860)(3690,8860)
\whiten\path(4290,8860)(4200,8500)(4110,8860)(4290,8860)
\whiten\path(4890,8860)(4800,8500)(4710,8860)(4890,8860)
\whiten\path(5490,8860)(5400,8500)(5310,8860)(5490,8860)
\whiten\path(6090,8860)(6000,8500)(5910,8860)(6090,8860)
\whiten\path(6690,8860)(6600,8500)(6510,8860)(6690,8860)
\whiten\path(7290,8860)(7200,8500)(7110,8860)(7290,8860)
\whiten\path(7890,8860)(7800,8500)(7710,8860)(7890,8860)
\whiten\path(8490,8860)(8400,8500)(8310,8860)(8490,8860)
\end{picture}
%
%\begin{ca} 
\caption{A six-vertex model $BC$-configuration. 
$L   \!  = \! 12$, and 
$N_1   \!  = \!  5$, or equivalently 
$L_h \!  = \!  2  \! \times  \! 5 =\! 10$ and 
$L_v \! =  \! 12$.
The top    $N_1$ horizontal lines represent $B$-operators. 
The bottom $N_1$ horizontal lines represent $C$-operators.
The initial (top) as well as the final (bottom) boundary 
spin systems are reference states.} 
%
%\end{ca}
%
%\end{minipage}
\label{bc-0-FW}
\end{figure}
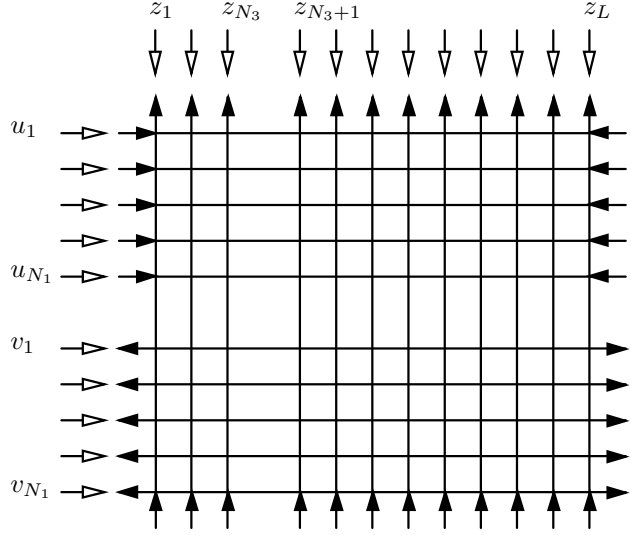

Consider a $BC$-configuration with no restrictions.  
To be specific, let us consider the configuration in 
Figure {\bf \ref{bc-0-FW}}, where $N_1 = 5$ and $L = 12$. 
Both sets of rapidities $\{u\}$ and $\{v\}$ are labeled 
from top to bottom, as usual. 

Consider the vertex at the bottom-left corner of Figure 
{\bf \ref{bc-0-FW}}. 
From Figure {\bf \ref{six-vertices-FW}}, it is easy to see that
this can be either a $b$- or a $c$-vertex. Since the $\{v\}$ 
variables are free, set $v_5 = z_1$, thereby setting the 
weight of all configurations with a $b$-vertex at this 
corner to zero, and forcing the vertex at this corner to be 
a $c$-vertex.

%FIG05
%
%\begin{center}
%\begin{minipage}{4.8in}

\begin{figure}
\thicklines
\setlength{\unitlength}{0.0008cm}
\begin{picture}(10000,10000)(000,0000)
\path(0600,1500)(9000,1500)
\path(0600,2100)(9000,2100)
\path(0600,2700)(9000,2700)
\path(0600,3300)(9000,3300)
\path(0600,3900)(9000,3900)
\path(0600,5100)(9000,5100)
\path(0600,5700)(9000,5700)
\path(0600,6300)(9000,6300)
\path(0600,6900)(9000,6900)
\path(0600,7500)(9000,7500)
\path(1200,8100)(1200,0900)
\path(1800,8100)(1800,0900)
\path(2400,8100)(2400,0900)
\path(3600,8100)(3600,0900)
\path(4200,8100)(4200,0900)
\path(4800,8100)(4800,0900)
\path(5400,8100)(5400,0900)
\path(6000,8100)(6000,0900)
\path(6600,8100)(6600,0900)
\path(7200,8100)(7200,0900)
\path(7800,8100)(7800,0900)
\path(8400,8100)(8400,0900)
\blacken\path(0900,1590)(0600,1500)(0900,1410)(0900,1590)
\blacken\path(0900,2190)(0600,2100)(0900,2010)(0900,2190)
\blacken\path(0900,2790)(0600,2700)(0900,2610)(0900,2790)
\blacken\path(0900,3390)(0600,3300)(0900,3210)(0900,3390)
\blacken\path(0900,3990)(0600,3900)(0900,3810)(0900,3990)
\blacken\path(0900,5010)(1200,5110)(0900,5190)(0900,5010)
\blacken\path(0900,5610)(1200,5700)(0900,5790)(0900,5610)
\blacken\path(0900,6210)(1210,6300)(0900,6390)(0900,6210)
\blacken\path(0900,6810)(1200,6900)(0900,6990)(0900,6810)
\blacken\path(0900,7410)(1200,7500)(0900,7590)(0900,7410)
\blacken\path(1300,1200)(1210,1500)(1120,1200)(1300,1200)
\blacken\path(1300,7800)(1210,8100)(1120,7800)(1300,7800)
\blacken\path(1900,1200)(1810,1500)(1720,1200)(1900,1200)
\blacken\path(1900,7800)(1810,8100)(1720,7800)(1900,7800)
\blacken\path(2500,1200)(2410,1500)(2320,1200)(2500,1200)
\blacken\path(2500,7800)(2410,8100)(2320,7800)(2500,7800)
\blacken\path(3700,1200)(3610,1500)(3520,1200)(3700,1200)
\blacken\path(3700,7800)(3610,8100)(3520,7800)(3700,7800)
\blacken\path(4300,1200)(4210,1500)(4120,1200)(4300,1200)
\blacken\path(4300,7800)(4210,8100)(4120,7800)(4300,7800)
\blacken\path(4900,1200)(4810,1506)(4720,1200)(4900,1200)
\blacken\path(4900,7800)(4810,8100)(4720,7800)(4900,7800)
\blacken\path(5500,1200)(5410,1500)(5320,1200)(5500,1200)
\blacken\path(5500,7800)(5410,8100)(5320,7800)(5500,7800)
\blacken\path(6100,1200)(6010,1500)(5920,1200)(6100,1200)
\blacken\path(6100,7800)(6010,8100)(5920,7800)(6100,7800)
\blacken\path(6700,1200)(6610,1500)(6520,1200)(6700,1200)
\blacken\path(6700,7800)(6610,8100)(6520,7800)(6700,7800)
\blacken\path(7300,1200)(7210,1500)(7120,1200)(7300,1200)
\blacken\path(7300,7800)(7210,8100)(7120,7800)(7300,7800)
\blacken\path(7900,1200)(7810,1500)(7720,1200)(7900,1200)
\blacken\path(7900,7800)(7810,8100)(7720,7800)(7900,7800)
\blacken\path(8500,1200)(8410,1500)(8320,1200)(8500,1200)
\blacken\path(8500,7800)(8410,8100)(8320,7800)(8500,7800)
\blacken\path(8700,1400)(9000,1490)(8700,1580)(8700,1400)
\blacken\path(8700,2000)(9000,2090)(8700,2180)(8700,2000)
\blacken\path(8700,2600)(9000,2690)(8700,2780)(8700,2600)
\blacken\path(8700,3200)(9000,3290)(8700,3380)(8700,3200)
\blacken\path(8700,3800)(9000,3890)(8700,3980)(8700,3800)
\blacken\path(8700,5200)(8400,5110)(8700,5020)(8700,5200)
\blacken\path(8700,5800)(8400,5710)(8700,5620)(8700,5800)
\blacken\path(8700,6400)(8400,6310)(8700,6220)(8700,6400)
\blacken\path(8700,7000)(8400,6910)(8700,6820)(8700,7000)
\blacken\path(8700,7600)(8400,7510)(8700,7420)(8700,7600)
%
%The folowing are the horizontal arrows
%
\path(-0360,1500)(0000,1500)
\path(-0360,2100)(0000,2100)
\path(-0360,2700)(0000,2700)
\path(-0360,3300)(0000,3300)
\path(-0360,3900)(0000,3900)
\put(-1200,1500){$v_{N_1}$}
\put(-1200,3900){$v_1    $}
\whiten\path(0000,1410)(0360,1500)(0000,1590)(0000,1410)
\whiten\path(0000,2010)(0360,2100)(0000,2190)(0000,2010)
\whiten\path(0000,2610)(0360,2700)(0000,2790)(0000,2610)
\whiten\path(0000,3210)(0360,3300)(0000,3390)(0000,3210)
\whiten\path(0000,3810)(0360,3900)(0000,3990)(0000,3810)
\path(-0360,5100)(0000,5100)
\path(-0360,5700)(0000,5700)
\path(-0360,6300)(0000,6300)
\path(-0360,6900)(0000,6900)
\path(-0360,7500)(0000,7500)
\put(-1200,5100){$u_{N_1}$}
\put(-1200,7500){$u_1    $}
\whiten\path(0000,5010)(0360,5100)(0000,5190)(0000,5010)
\whiten\path(0000,5610)(0360,5700)(0000,5790)(0000,5610)
\whiten\path(0000,6210)(0360,6300)(0000,6390)(0000,6210)
\whiten\path(0000,6810)(0360,6900)(0000,6990)(0000,6810)
\whiten\path(0000,7410)(0360,7500)(0000,7590)(0000,7410)
%
%The following are the vertical arrows
%
\path(1200,9220)(1200,8500)
\path(1800,9220)(1800,8500)
\path(2400,9220)(2400,8500)
\path(3600,9220)(3600,8500)
\path(4200,9220)(4200,8500)
\path(4800,9220)(4800,8500)
\path(5400,9220)(5400,8500)
\path(6000,9220)(6000,8500)
\path(6600,9220)(6600,8500)
\path(7200,9220)(7200,8500)
\path(7800,9220)(7800,8500)
\path(8400,9220)(8400,8500)
\put(1100,9500){$z_1   $}
%\put(1700,9500){$z_2   $}
\put(2300,9500){$z_{N_3    }  $}
\put(3500,9500){$z_{N_3 + 1}  $}
%\put(4100,9500){$z_5   $}
%\put(4700,9500){$z_6   $}
%\put(5300,9500){$z_7   $}
%\put(5900,9500){$z_8   $}
%\put(6500,9500){$z_9   $}
%\put(7100,9500){$z_{10}$}
%\put(7700,9500){$z_{11}$}
\put(8300,9500){$z_{L}$}
\whiten\path(1290,8860)(1200,8500)(1110,8860)(1290,8860)
\whiten\path(1890,8860)(1800,8500)(1710,8860)(1890,8860)
\whiten\path(2490,8860)(2400,8500)(2310,8860)(2490,8860)
\whiten\path(3690,8860)(3600,8500)(3510,8860)(3690,8860)
\whiten\path(4290,8860)(4200,8500)(4110,8860)(4290,8860)
\whiten\path(4890,8860)(4800,8500)(4710,8860)(4890,8860)
\whiten\path(5490,8860)(5400,8500)(5310,8860)(5490,8860)
\whiten\path(6090,8860)(6000,8500)(5910,8860)(6090,8860)
\whiten\path(6690,8860)(6600,8500)(6510,8860)(6690,8860)
\whiten\path(7290,8860)(7200,8500)(7110,8860)(7290,8860)
\whiten\path(7890,8860)(7800,8500)(7710,8860)(7890,8860)
\whiten\path(8490,8860)(8400,8500)(8310,8860)(8490,8860)
%
%Arrows on lower boundary
\blacken\path(1320,1410)(1680,1500)(1320,1590)(1320,1410)
\blacken\path(1920,1410)(2280,1500)(1920,1590)(1920,1410)
\blacken\path(2820,1410)(3180,1500)(2820,1590)(2820,1410)
\blacken\path(3720,1410)(4080,1500)(3720,1590)(3720,1410)
\blacken\path(4320,1410)(4680,1500)(4320,1590)(4320,1410)
\blacken\path(4920,1410)(5280,1500)(4920,1590)(4920,1410)
\blacken\path(5520,1410)(5880,1500)(5520,1590)(5520,1410)
\blacken\path(6120,1410)(6480,1500)(6120,1590)(6120,1410)
\blacken\path(6720,1410)(7080,1500)(6720,1590)(6720,1410)
\blacken\path(7320,1410)(7680,1500)(7320,1590)(7320,1410)
\blacken\path(7920,1410)(8280,1500)(7920,1590)(7920,1410)
%Arrows on second lower boundary
\blacken\path(1680,2010)(1320,2100)(1680,2190)(1680,2010)
\blacken\path(1680,2610)(1320,2700)(1680,2790)(1680,2610)
\blacken\path(1680,3210)(1320,3300)(1680,3390)(1680,3210)
\blacken\path(1680,3810)(1320,3900)(1680,3990)(1680,3810)
\blacken\path(1290,1860)(1200,1500)(1110,1860)(1290,1860)
\blacken\path(1890,1740)(1800,2100)(1710,1740)(1890,1740)
\blacken\path(2490,1740)(2400,2100)(2310,1740)(2490,1740)
\blacken\path(3690,1740)(3600,2100)(3510,1740)(3690,1740)
\blacken\path(4290,1740)(4200,2100)(4110,1740)(4290,1740)
\blacken\path(4890,1740)(4800,2100)(4710,1740)(4890,1740)
\blacken\path(5490,1740)(5400,2100)(5310,1740)(5490,1740)
\blacken\path(6090,1740)(6000,2100)(5910,1740)(6090,1740)
\blacken\path(6690,1740)(6600,2100)(6510,1740)(6690,1740)
\blacken\path(7290,1740)(7200,2100)(7110,1740)(7290,1740)
\blacken\path(7890,1740)(7800,2100)(7710,1740)(7890,1740)
\blacken\path(8490,1740)(8400,2100)(8310,1740)(8490,1740)
\blacken\path(1290,1860)(1200,1500)(1110,1860)(1290,1860)
\blacken\path(1290,2460)(1200,2100)(1110,2460)(1290,2460)
\blacken\path(1290,3060)(1200,2700)(1110,3060)(1290,3060)
\blacken\path(1290,3660)(1200,3300)(1110,3660)(1290,3660)
\blacken\path(1290,4560)(1200,4200)(1110,4560)(1290,4560)
\end{picture}
%
%\begin{ca} 
\caption{
Setting $v_{N_1}$ to $z_1$ 
in Figure {\ref{bc-0-FW}}, we freeze 
\1 the vertex at the lower left corner to be type-$c$, 
\2 all vertices to the right of the frozen corner to be 
type-$a$, and 
\3 all vertices above the frozen corner, but on the lower half 
of the diagram, to be type-$b$.
}
%\end{ca}
%
%\end{minipage}
\label{bc-1-FW}
\end{figure}
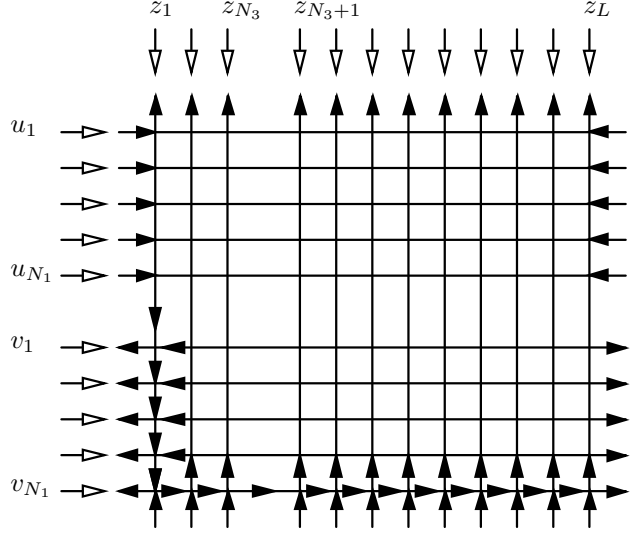

Referring to Figure {\bf \ref{six-vertices-FW}} again, one 
can see that not only is the corner vertex forced to be 
a $c$-vertex, but the orientations of all arrows on 
the horizontal lattice line with rapidity $v_5$, as well 
as all arrows on the vertical line with rapidity $z_1$ 
but below the horizontal line with rapidity $u_{N_1}$ are 
also frozen to fixed values as in Figure {\bf \ref{bc-1-FW}}.

%FIG06
%
%\begin{center}
%\begin{minipage}{4.3in}

\begin{figure}
\thicklines
\setlength{\unitlength}{0.0008cm}
\begin{picture}(10000,10000)(000,0000)
\path(0600,1500)(9000,1500)
\path(0600,2100)(9000,2100)
\path(0600,2700)(9000,2700)
\path(0600,3300)(9000,3300)
\path(0600,3900)(9000,3900)
\path(0600,5100)(9000,5100)
\path(0600,5700)(9000,5700)
\path(0600,6300)(9000,6300)
\path(0600,6900)(9000,6900)
\path(0600,7500)(9000,7500)
\path(1200,8100)(1200,0900)
\path(1800,8100)(1800,0900)
\path(2400,8100)(2400,0900)
\path(3600,8100)(3600,0900)
\path(4200,8100)(4200,0900)
\path(4800,8100)(4800,0900)
\path(5400,8100)(5400,0900)
\path(6000,8100)(6000,0900)
\path(6600,8100)(6600,0900)
\path(7200,8100)(7200,0900)
\path(7800,8100)(7800,0900)
\path(8400,8100)(8400,0900)
\blacken\path(0900,1590)(0600,1500)(0900,1410)(0900,1590)
\blacken\path(0900,2190)(0600,2100)(0900,2010)(0900,2190)
\blacken\path(0900,2790)(0600,2700)(0900,2610)(0900,2790)
\blacken\path(0900,3390)(0600,3300)(0900,3210)(0900,3390)
\blacken\path(0900,3990)(0600,3900)(0900,3810)(0900,3990)
\blacken\path(0900,5010)(1200,5110)(0900,5190)(0900,5010)
\blacken\path(0900,5610)(1200,5700)(0900,5790)(0900,5610)
\blacken\path(0900,6210)(1210,6300)(0900,6390)(0900,6210)
\blacken\path(0900,6810)(1200,6900)(0900,6990)(0900,6810)
\blacken\path(0900,7410)(1200,7500)(0900,7590)(0900,7410)
\blacken\path(1300,1200)(1210,1500)(1120,1200)(1300,1200)
\blacken\path(1300,7800)(1210,8100)(1120,7800)(1300,7800)
\blacken\path(1900,1200)(1810,1500)(1720,1200)(1900,1200)
\blacken\path(1900,7800)(1810,8100)(1720,7800)(1900,7800)
\blacken\path(2500,1200)(2410,1500)(2320,1200)(2500,1200)
\blacken\path(2500,7800)(2410,8100)(2320,7800)(2500,7800)
\blacken\path(3700,1200)(3610,1500)(3520,1200)(3700,1200)
\blacken\path(3700,7800)(3610,8100)(3520,7800)(3700,7800)
\blacken\path(4300,1200)(4210,1500)(4120,1200)(4300,1200)
\blacken\path(4300,7800)(4210,8100)(4120,7800)(4300,7800)
\blacken\path(4900,1200)(4810,1506)(4720,1200)(4900,1200)
\blacken\path(4900,7800)(4810,8100)(4720,7800)(4900,7800)
\blacken\path(5500,1200)(5410,1500)(5320,1200)(5500,1200)
\blacken\path(5500,7800)(5410,8100)(5320,7800)(5500,7800)
\blacken\path(6100,1200)(6010,1500)(5920,1200)(6100,1200)
\blacken\path(6100,7800)(6010,8100)(5920,7800)(6100,7800)
\blacken\path(6700,1200)(6610,1500)(6520,1200)(6700,1200)
\blacken\path(6700,7800)(6610,8100)(6520,7800)(6700,7800)
\blacken\path(7300,1200)(7210,1500)(7120,1200)(7300,1200)
\blacken\path(7300,7800)(7210,8100)(7120,7800)(7300,7800)
\blacken\path(7900,1200)(7810,1500)(7720,1200)(7900,1200)
\blacken\path(7900,7800)(7810,8100)(7720,7800)(7900,7800)
\blacken\path(8500,1200)(8410,1500)(8320,1200)(8500,1200)
\blacken\path(8500,7800)(8410,8100)(8320,7800)(8500,7800)
\blacken\path(8700,1400)(9000,1490)(8700,1580)(8700,1400)
\blacken\path(8700,2000)(9000,2090)(8700,2180)(8700,2000)
\blacken\path(8700,2600)(9000,2690)(8700,2780)(8700,2600)
\blacken\path(8700,3200)(9000,3290)(8700,3380)(8700,3200)
\blacken\path(8700,3800)(9000,3890)(8700,3980)(8700,3800)
\blacken\path(8700,5200)(8400,5110)(8700,5020)(8700,5200)
\blacken\path(8700,5800)(8400,5710)(8700,5620)(8700,5800)
\blacken\path(8700,6400)(8400,6310)(8700,6220)(8700,6400)
\blacken\path(8700,7000)(8400,6910)(8700,6820)(8700,7000)
\blacken\path(8700,7600)(8400,7510)(8700,7420)(8700,7600)
%
%The folowing are the horizontal white arrows
%
\path(-0360,1500)(0000,1500)
\path(-0360,2100)(0000,2100)
\path(-0360,2700)(0000,2700)
\path(-0360,3300)(0000,3300)
\path(-0360,3900)(0000,3900)
\put(-1200,3900){$v_1    $}
%\put(-1200,2100){$v_{N_1 - 1}$}
\put(-1200,1500){$v_{N_1}$}
\whiten\path(0000,1410)(0360,1500)(0000,1590)(0000,1410)
\whiten\path(0000,2010)(0360,2100)(0000,2190)(0000,2010)
\whiten\path(0000,2610)(0360,2700)(0000,2790)(0000,2610)
\whiten\path(0000,3210)(0360,3300)(0000,3390)(0000,3210)
\whiten\path(0000,3810)(0360,3900)(0000,3990)(0000,3810)
\path(-0360,5100)(0000,5100)
\path(-0360,5700)(0000,5700)
\path(-0360,6300)(0000,6300)
\path(-0360,6900)(0000,6900)
\path(-0360,7500)(0000,7500)
\put(-1200,5100){$u_{N_1}$}
\put(-1200,7500){$u_1$    }
\whiten\path(0000,5010)(0360,5100)(0000,5190)(0000,5010)
\whiten\path(0000,5610)(0360,5700)(0000,5790)(0000,5610)
\whiten\path(0000,6210)(0360,6300)(0000,6390)(0000,6210)
\whiten\path(0000,6810)(0360,6900)(0000,6990)(0000,6810)
\whiten\path(0000,7410)(0360,7500)(0000,7590)(0000,7410)
%
%The following are the vertical white arrows
%
\path(1200,9220)(1200,8500)
\path(1800,9220)(1800,8500)
\path(2400,9220)(2400,8500)
\path(3600,9220)(3600,8500)
\path(4200,9220)(4200,8500)
\path(4800,9220)(4800,8500)
\path(5400,9220)(5400,8500)
\path(6000,9220)(6000,8500)
\path(6600,9220)(6600,8500)
\path(7200,9220)(7200,8500)
\path(7800,9220)(7800,8500)
\path(8400,9220)(8400,8500)
\put(1100,9500){$z_1   $}
%\put(1700,9500){$z_2   $}
\put(2300,9500){$z_{N_3    }   $}
\put(3500,9500){$z_{N_3 + 1}   $}
\put(8300,9500){$z_{L}$}
\whiten\path(1290,8860)(1200,8500)(1110,8860)(1290,8860)
\whiten\path(1890,8860)(1800,8500)(1710,8860)(1890,8860)
\whiten\path(2490,8860)(2400,8500)(2310,8860)(2490,8860)
\whiten\path(3690,8860)(3600,8500)(3510,8860)(3690,8860)
\whiten\path(4290,8860)(4200,8500)(4110,8860)(4290,8860)
\whiten\path(4890,8860)(4800,8500)(4710,8860)(4890,8860)
\whiten\path(5490,8860)(5400,8500)(5310,8860)(5490,8860)
\whiten\path(6090,8860)(6000,8500)(5910,8860)(6090,8860)
\whiten\path(6690,8860)(6600,8500)(6510,8860)(6690,8860)
\whiten\path(7290,8860)(7200,8500)(7110,8860)(7290,8860)
\whiten\path(7890,8860)(7800,8500)(7710,8860)(7890,8860)
\whiten\path(8490,8860)(8400,8500)(8310,8860)(8490,8860)
%
%horizontal arrows on y_1 line 
%
\blacken\path(1320,1410)(1680,1500)(1320,1590)(1320,1410)
\blacken\path(1920,1410)(2280,1500)(1920,1590)(1920,1410)
%this is the one on the longer segment
\blacken\path(2820,1410)(3180,1500)(2820,1590)(2820,1410)
\blacken\path(3720,1410)(4080,1500)(3720,1590)(3720,1410)
\blacken\path(4320,1410)(4680,1500)(4320,1590)(4320,1410)
\blacken\path(4920,1410)(5280,1500)(4920,1590)(4920,1410)
\blacken\path(5520,1410)(5880,1500)(5520,1590)(5520,1410)
\blacken\path(6120,1410)(6480,1500)(6120,1590)(6120,1410)
\blacken\path(6720,1410)(7080,1500)(6720,1590)(6720,1410)
\blacken\path(7320,1410)(7680,1500)(7320,1590)(7320,1410)
\blacken\path(7920,1410)(8280,1500)(7920,1590)(7920,1410)
%
%Horizontal arrows on second vertical band from left
%
\blacken\path(1680,2010)(1320,2100)(1680,2190)(1680,2010)
\blacken\path(1680,2610)(1320,2700)(1680,2790)(1680,2610)
\blacken\path(1680,3210)(1320,3300)(1680,3390)(1680,3210)
\blacken\path(1680,3810)(1320,3900)(1680,3990)(1680,3810)
%
%Horizontal arrows on third vertical band from left
%
\blacken\path(1920,2010)(2280,2100)(1920,2190)(1920,2010)
\blacken\path(2280,2610)(1920,2700)(2280,2790)(2280,2610)
\blacken\path(2280,3210)(1920,3300)(2280,3390)(2280,3210)
\blacken\path(2280,3810)(1920,3900)(2280,3990)(2280,3810)
%
%Horizontal arrows on fourth vertical band from left
%
\blacken\path(2820,2010)(3180,2100)(2820,2190)(2820,2010)
\blacken\path(1290,1860)(1200,1500)(1110,1860)(1290,1860)
\blacken\path(1890,1740)(1800,2100)(1710,1740)(1890,1740)
\blacken\path(2490,1740)(2400,2100)(2310,1740)(2490,1740)
\blacken\path(3690,1740)(3600,2100)(3510,1740)(3690,1740)
\blacken\path(4290,1740)(4200,2100)(4110,1740)(4290,1740)
\blacken\path(4890,1740)(4800,2100)(4710,1740)(4890,1740)
\blacken\path(5490,1740)(5400,2100)(5310,1740)(5490,1740)
\blacken\path(6090,1740)(6000,2100)(5910,1740)(6090,1740)
\blacken\path(6690,1740)(6600,2100)(6510,1740)(6690,1740)
\blacken\path(7290,1740)(7200,2100)(7110,1740)(7290,1740)
\blacken\path(7890,1740)(7800,2100)(7710,1740)(7890,1740)
\blacken\path(8490,1740)(8400,2100)(8310,1740)(8490,1740)
\blacken\path(1290,1860)(1200,1500)(1110,1860)(1290,1860)
\blacken\path(1290,2460)(1200,2100)(1110,2460)(1290,2460)
\blacken\path(1290,3060)(1200,2700)(1110,3060)(1290,3060)
\blacken\path(1290,3660)(1200,3300)(1110,3660)(1290,3660)
\blacken\path(1290,4560)(1200,4200)(1110,4560)(1290,4560)
%
%vertical arrows on z_2 vertical line
%
\blacken\path(1890,2460)(1800,2100)(1710,2460)(1890,2460)
\blacken\path(1890,3060)(1800,2700)(1710,3060)(1890,3060)
\blacken\path(1890,3660)(1800,3300)(1710,3660)(1890,3660)
\blacken\path(1890,4560)(1800,4200)(1710,4560)(1890,4560)
%
%vertical arrows on z_3 vertical line
%
\blacken\path(2490,2340)(2400,2700)(2310,2340)(2490,2340)
%
%vertical arrows on third band from below
%
\blacken\path(3700,2340)(3612,2700)(3520,2340)(3700,2340)
\blacken\path(4300,2340)(4212,2700)(4120,2340)(4300,2340)
\blacken\path(4900,2340)(4812,2700)(4720,2340)(4900,2340)
\blacken\path(5500,2340)(5412,2700)(5320,2340)(5500,2340)
\blacken\path(6100,2340)(6012,2700)(5920,2340)(6100,2340)
\blacken\path(6700,2340)(6612,2700)(6520,2340)(6700,2340)
\blacken\path(7300,2340)(7212,2700)(7120,2340)(7300,2340)
\blacken\path(7900,2340)(7812,2700)(7720,2340)(7900,2340)
\blacken\path(8500,2340)(8412,2700)(8320,2340)(8500,2340)
%
%horizontal arrows on y_2 line
\blacken\path(3720,2010)(4080,2100)(3720,2190)(3720,2010)
\blacken\path(4320,2010)(4680,2100)(4320,2190)(4320,2010)
\blacken\path(4920,2010)(5280,2100)(4920,2190)(4920,2010)
\blacken\path(5520,2010)(5880,2100)(5520,2190)(5520,2010)
\blacken\path(6120,2010)(6480,2100)(6120,2190)(6120,2010)
\blacken\path(6720,2010)(7080,2100)(6720,2190)(6720,2010)
\blacken\path(7320,2010)(7680,2100)(7320,2190)(7320,2010)
\blacken\path(7920,2010)(8280,2100)(7920,2190)(7920,2010)
\end{picture}
%
%\begin{ca} 
\caption{
Setting $v_{N_1 - 1}$ (on second horizontal line from below) to 
        $z_2$         (on second vertical   line from left) 
in Figure {\ref{bc-1-FW}}, we freeze 
\1 the vertex at the intersection of the lines that 
carry rapidities $v_{N_1 - 1}$ and $z_2$ to be type-$c$, 
\2 all vertices to the right of the most recently-frozen corner 
to be type-$a$, and
\3 all vertices above the same vertex, but on the lower half
of the diagram, to be type-$b$.
}
%\end{ca}
%
%\end{minipage}
\label{bc-2-FW}
\end{figure}
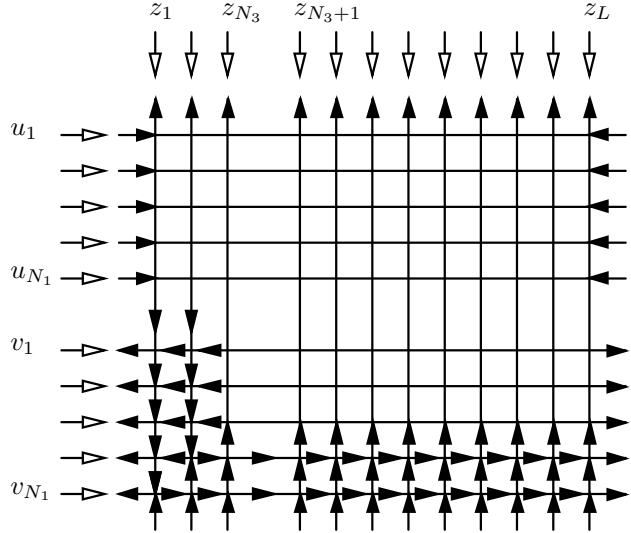

The above exercise in {\it `freezing'} vertices and arrows 
can be repeated and to produce a non-trivial example, we do 
it two more times. 
Setting $v_4 = z_2$ forces the vertex at the intersection 
of the lines carrying the rapidities $v_4$ and $z_2$ to be 
a $c$-vertex and freezes all arrows to the right as well as 
all arrows above that vertex and along $C$-lines, as in 
Figure {\bf \ref{bc-2-FW}}.

%FIG07
%
%\begin{center}
%\begin{minipage}{4.3in}

%FIG08
%
%\begin{center}
%\begin{minipage}{4.8in}
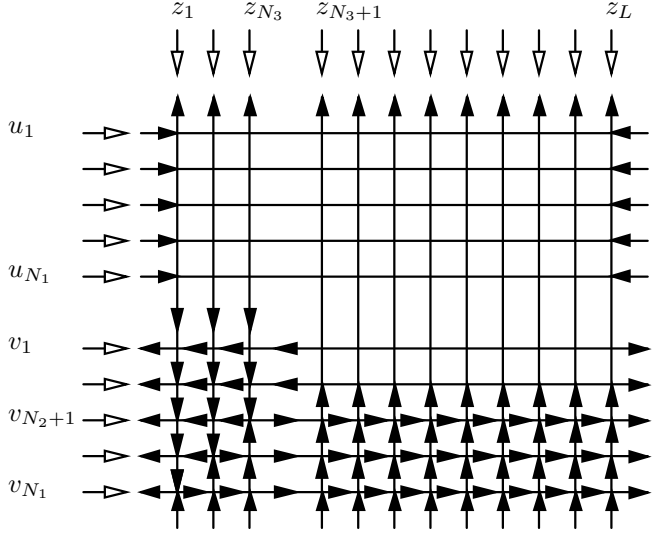
\begin{figure}
\thicklines
\setlength{\unitlength}{0.0008cm}
\begin{picture}(10000,09000)(0000,0750)
\path(0600,1500)(9000,1500)
\path(0600,2100)(9000,2100)
\path(0600,2700)(9000,2700)
\path(0600,3300)(9000,3300)
\path(0600,3900)(9000,3900)
\path(0600,5100)(9000,5100)
\path(0600,5700)(9000,5700)
\path(0600,6300)(9000,6300)
\path(0600,6900)(9000,6900)
\path(0600,7500)(9000,7500)
\path(1200,8100)(1200,0900)
\path(1800,8100)(1800,0900)
\path(2400,8100)(2400,0900)
\path(3600,8100)(3600,0900)
\path(4200,8100)(4200,0900)
\path(4800,8100)(4800,0900)
\path(5400,8100)(5400,0900)
\path(6000,8100)(6000,0900)
\path(6600,8100)(6600,0900)
\path(7200,8100)(7200,0900)
\path(7800,8100)(7800,0900)
\path(8400,8100)(8400,0900)
\blacken\path(0900,1590)(0600,1500)(0900,1410)(0900,1590)
\blacken\path(0900,2190)(0600,2100)(0900,2010)(0900,2190)
\blacken\path(0900,2790)(0600,2700)(0900,2610)(0900,2790)
\blacken\path(0900,3390)(0600,3300)(0900,3210)(0900,3390)
\blacken\path(0900,3990)(0600,3900)(0900,3810)(0900,3990)
\blacken\path(0900,5010)(1200,5110)(0900,5190)(0900,5010)
\blacken\path(0900,5610)(1200,5700)(0900,5790)(0900,5610)
\blacken\path(0900,6210)(1210,6300)(0900,6390)(0900,6210)
\blacken\path(0900,6810)(1200,6900)(0900,6990)(0900,6810)
\blacken\path(0900,7410)(1200,7500)(0900,7590)(0900,7410)
\blacken\path(1300,1200)(1210,1500)(1120,1200)(1300,1200)
\blacken\path(1300,7800)(1210,8100)(1120,7800)(1300,7800)
\blacken\path(1900,1200)(1810,1500)(1720,1200)(1900,1200)
\blacken\path(1900,7800)(1810,8100)(1720,7800)(1900,7800)
\blacken\path(2500,1200)(2410,1500)(2320,1200)(2500,1200)
\blacken\path(2500,7800)(2410,8100)(2320,7800)(2500,7800)
\blacken\path(3700,1200)(3610,1500)(3520,1200)(3700,1200)
\blacken\path(3700,7800)(3610,8100)(3520,7800)(3700,7800)
\blacken\path(4300,1200)(4210,1500)(4120,1200)(4300,1200)
\blacken\path(4300,7800)(4210,8100)(4120,7800)(4300,7800)
\blacken\path(4900,1200)(4810,1506)(4720,1200)(4900,1200)
\blacken\path(4900,7800)(4810,8100)(4720,7800)(4900,7800)
\blacken\path(5500,1200)(5410,1500)(5320,1200)(5500,1200)
\blacken\path(5500,7800)(5410,8100)(5320,7800)(5500,7800)
\blacken\path(6100,1200)(6010,1500)(5920,1200)(6100,1200)
\blacken\path(6100,7800)(6010,8100)(5920,7800)(6100,7800)
\blacken\path(6700,1200)(6610,1500)(6520,1200)(6700,1200)
\blacken\path(6700,7800)(6610,8100)(6520,7800)(6700,7800)
\blacken\path(7300,1200)(7210,1500)(7120,1200)(7300,1200)
\blacken\path(7300,7800)(7210,8100)(7120,7800)(7300,7800)
\blacken\path(7900,1200)(7810,1500)(7720,1200)(7900,1200)
\blacken\path(7900,7800)(7810,8100)(7720,7800)(7900,7800)
\blacken\path(8500,1200)(8410,1500)(8320,1200)(8500,1200)
\blacken\path(8500,7800)(8410,8100)(8320,7800)(8500,7800)
\blacken\path(8700,1400)(9000,1490)(8700,1580)(8700,1400)
\blacken\path(8700,2000)(9000,2090)(8700,2180)(8700,2000)
\blacken\path(8700,2600)(9000,2690)(8700,2780)(8700,2600)
\blacken\path(8700,3200)(9000,3290)(8700,3380)(8700,3200)
\blacken\path(8700,3800)(9000,3890)(8700,3980)(8700,3800)
\blacken\path(8700,5200)(8400,5110)(8700,5020)(8700,5200)
\blacken\path(8700,5800)(8400,5710)(8700,5620)(8700,5800)
\blacken\path(8700,6400)(8400,6310)(8700,6220)(8700,6400)
\blacken\path(8700,7000)(8400,6910)(8700,6820)(8700,7000)
\blacken\path(8700,7600)(8400,7510)(8700,7420)(8700,7600)
%
%The folowing are the horizontal white arrows
%
\path(-0360,1500)(0000,1500)
\path(-0360,2100)(0000,2100)
\path(-0360,2700)(0000,2700)
\path(-0360,3300)(0000,3300)
\path(-0360,3900)(0000,3900)
\put(-1600,3900){$v_{ 1           }$}
%\put(-1000,3100){$v_{(N_1 - N_2 + 1)}$}
\put(-1600,2700){$v_{ N_2+1         }$}
\put(-1600,1500){$v_{ N_1             }$}
\whiten\path(0000,1410)(0360,1500)(0000,1590)(0000,1410)
\whiten\path(0000,2010)(0360,2100)(0000,2190)(0000,2010)
\whiten\path(0000,2610)(0360,2700)(0000,2790)(0000,2610)
\whiten\path(0000,3210)(0360,3300)(0000,3390)(0000,3210)
\whiten\path(0000,3810)(0360,3900)(0000,3990)(0000,3810)
\path(-0360,5100)(0000,5100)
\path(-0360,5700)(0000,5700)
\path(-0360,6300)(0000,6300)
\path(-0360,6900)(0000,6900)
\path(-0360,7500)(0000,7500)
\put(-1600,5100){$u_{N_1}$}
\put(-1600,7500){$u_{  1}$}
\whiten\path(0000,5010)(0360,5100)(0000,5190)(0000,5010)
\whiten\path(0000,5610)(0360,5700)(0000,5790)(0000,5610)
\whiten\path(0000,6210)(0360,6300)(0000,6390)(0000,6210)
\whiten\path(0000,6810)(0360,6900)(0000,6990)(0000,6810)
\whiten\path(0000,7410)(0360,7500)(0000,7590)(0000,7410)
%
%The following are the vertical white arrows
%
\path(1200,9220)(1200,8500)
\path(1800,9220)(1800,8500)
\path(2400,9220)(2400,8500)
\path(3600,9220)(3600,8500)
\path(4200,9220)(4200,8500)
\path(4800,9220)(4800,8500)
\path(5400,9220)(5400,8500)
\path(6000,9220)(6000,8500)
\path(6600,9220)(6600,8500)
\path(7200,9220)(7200,8500)
\path(7800,9220)(7800,8500)
\path(8400,9220)(8400,8500)
\put(1100,9500){$z_1   $}
\put(2300,9500){$z_{ N_3    }$}
\put(3500,9500){$z_{ N_3 + 1 }$}
%\put(4100,9500){$z_5   $}
%\put(4700,9500){$z_6   $}
%\put(5300,9500){$z_7   $}
%\put(5900,9500){$z_8   $}
%\put(6500,9500){$z_9   $}
%\put(7100,9500){$z_{10}$}
%\put(7700,9500){$z_{11}$}
\put(8300,9500){$z_{L}$}
\whiten\path(1290,8860)(1200,8500)(1110,8860)(1290,8860)
\whiten\path(1890,8860)(1800,8500)(1710,8860)(1890,8860)
\whiten\path(2490,8860)(2400,8500)(2310,8860)(2490,8860)
\whiten\path(3690,8860)(3600,8500)(3510,8860)(3690,8860)
\whiten\path(4290,8860)(4200,8500)(4110,8860)(4290,8860)
\whiten\path(4890,8860)(4800,8500)(4710,8860)(4890,8860)
\whiten\path(5490,8860)(5400,8500)(5310,8860)(5490,8860)
\whiten\path(6090,8860)(6000,8500)(5910,8860)(6090,8860)
\whiten\path(6690,8860)(6600,8500)(6510,8860)(6690,8860)
\whiten\path(7290,8860)(7200,8500)(7110,8860)(7290,8860)
\whiten\path(7890,8860)(7800,8500)(7710,8860)(7890,8860)
\whiten\path(8490,8860)(8400,8500)(8310,8860)(8490,8860)
%
%horizontal arrows on v_1 line 
%
\blacken\path(1320,1410)(1680,1500)(1320,1590)(1320,1410)
\blacken\path(1920,1410)(2280,1500)(1920,1590)(1920,1410)
%this is the one on the longer segment
\blacken\path(2820,1410)(3180,1500)(2820,1590)(2820,1410)
\blacken\path(3720,1410)(4080,1500)(3720,1590)(3720,1410)
\blacken\path(4320,1410)(4680,1500)(4320,1590)(4320,1410)
\blacken\path(4920,1410)(5280,1500)(4920,1590)(4920,1410)
\blacken\path(5520,1410)(5880,1500)(5520,1590)(5520,1410)
\blacken\path(6120,1410)(6480,1500)(6120,1590)(6120,1410)
\blacken\path(6720,1410)(7080,1500)(6720,1590)(6720,1410)
\blacken\path(7320,1410)(7680,1500)(7320,1590)(7320,1410)
\blacken\path(7920,1410)(8280,1500)(7920,1590)(7920,1410)
%
%Horizontal arrows on second vertical band from left
%
\blacken\path(1680,2010)(1320,2100)(1680,2190)(1680,2010)
\blacken\path(1680,2610)(1320,2700)(1680,2790)(1680,2610)
\blacken\path(1680,3210)(1320,3300)(1680,3390)(1680,3210)
\blacken\path(1680,3810)(1320,3900)(1680,3990)(1680,3810)
%
%Horizontal arrows on third vertical band from left
%
\blacken\path(1920,2010)(2280,2100)(1920,2190)(1920,2010)
\blacken\path(2280,2610)(1920,2700)(2280,2790)(2280,2610)
\blacken\path(2280,3210)(1920,3300)(2280,3390)(2280,3210)
\blacken\path(2280,3810)(1920,3900)(2280,3990)(2280,3810)
%
%Horizontal arrows on fourth vertical band from left
%
\blacken\path(2820,2010)(3180,2100)(2820,2190)(2820,2010)
\blacken\path(1290,1860)(1200,1500)(1110,1860)(1290,1860)
\blacken\path(1890,1740)(1800,2100)(1710,1740)(1890,1740)
\blacken\path(2490,1740)(2400,2100)(2310,1740)(2490,1740)
\blacken\path(3690,1740)(3600,2100)(3510,1740)(3690,1740)
\blacken\path(4290,1740)(4200,2100)(4110,1740)(4290,1740)
\blacken\path(4890,1740)(4800,2100)(4710,1740)(4890,1740)
\blacken\path(5490,1740)(5400,2100)(5310,1740)(5490,1740)
\blacken\path(6090,1740)(6000,2100)(5910,1740)(6090,1740)
\blacken\path(6690,1740)(6600,2100)(6510,1740)(6690,1740)
\blacken\path(7290,1740)(7200,2100)(7110,1740)(7290,1740)
\blacken\path(7890,1740)(7800,2100)(7710,1740)(7890,1740)
\blacken\path(8490,1740)(8400,2100)(8310,1740)(8490,1740)
\blacken\path(1290,1860)(1200,1500)(1110,1860)(1290,1860)
\blacken\path(1290,2460)(1200,2100)(1110,2460)(1290,2460)
\blacken\path(1290,3060)(1200,2700)(1110,3060)(1290,3060)
\blacken\path(1290,3660)(1200,3300)(1110,3660)(1290,3660)
\blacken\path(1290,4560)(1200,4200)(1110,4560)(1290,4560)
%
%vertical arrows on z_2 vertical line
%
\blacken\path(1890,2460)(1800,2100)(1710,2460)(1890,2460)
\blacken\path(1890,3060)(1800,2700)(1710,3060)(1890,3060)
\blacken\path(1890,3660)(1800,3300)(1710,3660)(1890,3660)
\blacken\path(1890,4560)(1800,4200)(1710,4560)(1890,4560)
%
%vertical arrows on z_3 vertical line
%
\blacken\path(2490,2340)(2400,2700)(2310,2340)(2490,2340)
%
%vertical arrows on third band from below
%
\blacken\path(3700,2340)(3612,2700)(3520,2340)(3700,2340)
\blacken\path(4300,2340)(4212,2700)(4120,2340)(4300,2340)
\blacken\path(4900,2340)(4812,2700)(4720,2340)(4900,2340)
\blacken\path(5500,2340)(5412,2700)(5320,2340)(5500,2340)
\blacken\path(6100,2340)(6012,2700)(5920,2340)(6100,2340)
\blacken\path(6700,2340)(6612,2700)(6520,2340)(6700,2340)
\blacken\path(7300,2340)(7212,2700)(7120,2340)(7300,2340)
\blacken\path(7900,2340)(7812,2700)(7720,2340)(7900,2340)
\blacken\path(8500,2340)(8412,2700)(8320,2340)(8500,2340)
%
%horizontal arrows on v_2 line
\blacken\path(3720,2010)(4080,2100)(3720,2190)(3720,2010)
\blacken\path(4320,2010)(4680,2100)(4320,2190)(4320,2010)
\blacken\path(4920,2010)(5280,2100)(4920,2190)(4920,2010)
\blacken\path(5520,2010)(5880,2100)(5520,2190)(5520,2010)
\blacken\path(6120,2010)(6480,2100)(6120,2190)(6120,2010)
\blacken\path(6720,2010)(7080,2100)(6720,2190)(6720,2010)
\blacken\path(7320,2010)(7680,2100)(7320,2190)(7320,2010)
\blacken\path(7920,2010)(8280,2100)(7920,2190)(7920,2010)
%
%horizontal arrows on v_3 line
\blacken\path(2820,2610)(3180,2700)(2820,2790)(2820,2610)
\blacken\path(3720,2610)(4080,2700)(3720,2790)(3720,2610)
\blacken\path(4320,2610)(4680,2700)(4320,2790)(4320,2610)
\blacken\path(4920,2610)(5280,2700)(4920,2790)(4920,2610)
\blacken\path(5520,2610)(5880,2700)(5520,2790)(5520,2610)
\blacken\path(6120,2610)(6480,2700)(6120,2790)(6120,2610)
\blacken\path(6720,2610)(7080,2700)(6720,2790)(6720,2610)
\blacken\path(7320,2610)(7680,2700)(7320,2790)(7320,2610)
\blacken\path(7920,2610)(8280,2700)(7920,2790)(7920,2610)
%
%vertical arrows on fourth band from below
%
\blacken\path(2500,3060)(2410,2700)(2320,3060)(2500,3060)
%Here are the arrows on top of the above
\blacken\path(2500,3660)(2410,3300)(2320,3660)(2500,3660)
%and here is the one on top of that on a long v segment
\blacken\path(2500,4560)(2410,4200)(2320,4560)(2500,4560)
%
%Horizontal arrows on fourth vertical band from left
%
\blacken\path(3180,3210)(2820,3300)(3180,3390)(3180,3210)
\blacken\path(3180,3810)(2820,3900)(3180,3990)(3180,3810)
\blacken\path(3700,2940)(3610,3300)(3520,2940)(3700,2940)
\blacken\path(4300,2940)(4210,3300)(4120,2940)(4300,2940)
\blacken\path(4900,2940)(4810,3300)(4720,2940)(4900,2940)
\blacken\path(5500,2940)(5410,3300)(5320,2940)(5500,2940)
\blacken\path(6100,2940)(6010,3300)(5920,2940)(6100,2940)
\blacken\path(6700,2940)(6610,3300)(6520,2940)(6700,2940)
\blacken\path(7300,2940)(7210,3300)(7120,2940)(7300,2940)
\blacken\path(7900,2940)(7810,3300)(7720,2940)(7900,2940)
\blacken\path(8500,2940)(8410,3300)(8320,2940)(8500,2940)
\end{picture}
%
%\begin{ca} 
\caption{The effect of forcing the three vertices at the 
intersection of the lines that carry the pairs of rapidities 
$\{v_{N_1    }, z_1\}$, 
$\{v_{N_1 - 1}, z_2\}$ and
$\{v_{N_1 - 2}, z_3\}$
to be $c$-vertices. We used the notation $N_3 = N_1 - N_2$.
}
%\end{ca}
%
%\end{minipage}
\label{bc-3-FW}
\end{figure}

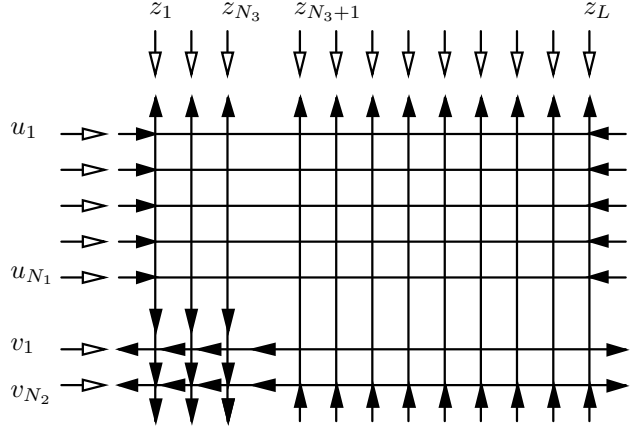
\begin{figure}
\setlength{\unitlength}{0.0008cm}
\begin{picture}(10000,07500)(0000,2500)
\thicklines
%
%The following lines are horizontal
%
\path(0600,3300)(9000,3300)
\path(0600,3900)(9000,3900)
\path(0600,5100)(9000,5100)
\path(0600,5700)(9000,5700)
\path(0600,6300)(9000,6300)
\path(0600,6900)(9000,6900)
\path(0600,7500)(9000,7500)
%
%The following lines are vertical
\path(1200,8100)(1200,2700)
\path(1800,8100)(1800,2700)
\path(2400,8100)(2400,2700)
\path(3600,8100)(3600,2700)
\path(4200,8100)(4200,2700)
\path(4800,8100)(4800,2700)
\path(5400,8100)(5400,2700)
\path(6000,8100)(6000,2700)
\path(6600,8100)(6600,2700)
\path(7200,8100)(7200,2700)
\path(7800,8100)(7800,2700)
\path(8400,8100)(8400,2700)
%
%two horizontaL arrows pointing left on v_4 and v_5 lines
%
\blacken\path(0900,3390)(0600,3300)(0900,3210)(0900,3390)
\blacken\path(0900,3990)(0600,3900)(0900,3810)(0900,3990)
\blacken\path(0900,5010)(1200,5110)(0900,5190)(0900,5010)
\blacken\path(0900,5610)(1200,5700)(0900,5790)(0900,5610)
\blacken\path(0900,6210)(1210,6300)(0900,6390)(0900,6210)
\blacken\path(0900,6810)(1200,6900)(0900,6990)(0900,6810)
\blacken\path(0900,7410)(1200,7500)(0900,7590)(0900,7410)
\blacken\path(1300,7800)(1210,8100)(1120,7800)(1300,7800)
\blacken\path(1900,7800)(1810,8100)(1720,7800)(1900,7800)
\blacken\path(2500,7800)(2410,8100)(2320,7800)(2500,7800)
\blacken\path(3700,7800)(3610,8100)(3520,7800)(3700,7800)
\blacken\path(4300,7800)(4210,8100)(4120,7800)(4300,7800)
\blacken\path(4900,7800)(4810,8100)(4720,7800)(4900,7800)
\blacken\path(5500,7800)(5410,8100)(5320,7800)(5500,7800)
\blacken\path(6100,7800)(6010,8100)(5920,7800)(6100,7800)
\blacken\path(6700,7800)(6610,8100)(6520,7800)(6700,7800)
\blacken\path(7300,7800)(7210,8100)(7120,7800)(7300,7800)
\blacken\path(7900,7800)(7810,8100)(7720,7800)(7900,7800)
\blacken\path(8500,7800)(8410,8100)(8320,7800)(8500,7800)
\blacken\path(8700,3200)(9000,3290)(8700,3380)(8700,3200)
\blacken\path(8700,3800)(9000,3890)(8700,3980)(8700,3800)
\blacken\path(8700,5200)(8400,5110)(8700,5020)(8700,5200)
\blacken\path(8700,5800)(8400,5710)(8700,5620)(8700,5800)
\blacken\path(8700,6400)(8400,6310)(8700,6220)(8700,6400)
\blacken\path(8700,7000)(8400,6910)(8700,6820)(8700,7000)
\blacken\path(8700,7600)(8400,7510)(8700,7420)(8700,7600)
%
%The folowing are the horizontal white arrows
%
\path(-0360,3300)(0000,3300)
\path(-0360,3900)(0000,3900)
\put(-1200,3900){$v_{ 1          }$}
\put(-1200,3100){$v_{N_2      }$}
\whiten\path(0000,3210)(0360,3300)(0000,3390)(0000,3210)
\whiten\path(0000,3810)(0360,3900)(0000,3990)(0000,3810)
\path(-0360,5100)(0000,5100)
\path(-0360,5700)(0000,5700)
\path(-0360,6300)(0000,6300)
\path(-0360,6900)(0000,6900)
\path(-0360,7500)(0000,7500)
\put(-1200,5100){$u_{N_1}$}
\put(-1200,7500){$u_1$}
\whiten\path(0000,5010)(0360,5100)(0000,5190)(0000,5010)
\whiten\path(0000,5610)(0360,5700)(0000,5790)(0000,5610)
\whiten\path(0000,6210)(0360,6300)(0000,6390)(0000,6210)
\whiten\path(0000,6810)(0360,6900)(0000,6990)(0000,6810)
\whiten\path(0000,7410)(0360,7500)(0000,7590)(0000,7410)
%
%The following are the vertical white arrows
%
\path(1200,9220)(1200,8500)
\path(1800,9220)(1800,8500)
\path(2400,9220)(2400,8500)
\path(3600,9220)(3600,8500)
\path(4200,9220)(4200,8500)
\path(4800,9220)(4800,8500)
\path(5400,9220)(5400,8500)
\path(6000,9220)(6000,8500)
\path(6600,9220)(6600,8500)
\path(7200,9220)(7200,8500)
\path(7800,9220)(7800,8500)
\path(8400,9220)(8400,8500)
\put(1100,9500){$z_{      1}$}
\put(2300,9500){$z_{N_3    }$}
\put(3500,9500){$z_{N_3 + 1}$}
\put(8300,9500){$z_{L      }$}
\whiten\path(1290,8860)(1200,8500)(1110,8860)(1290,8860)
\whiten\path(1890,8860)(1800,8500)(1710,8860)(1890,8860)
\whiten\path(2490,8860)(2400,8500)(2310,8860)(2490,8860)
\whiten\path(3690,8860)(3600,8500)(3510,8860)(3690,8860)
\whiten\path(4290,8860)(4200,8500)(4110,8860)(4290,8860)
\whiten\path(4890,8860)(4800,8500)(4710,8860)(4890,8860)
\whiten\path(5490,8860)(5400,8500)(5310,8860)(5490,8860)
\whiten\path(6090,8860)(6000,8500)(5910,8860)(6090,8860)
\whiten\path(6690,8860)(6600,8500)(6510,8860)(6690,8860)
\whiten\path(7290,8860)(7200,8500)(7110,8860)(7290,8860)
\whiten\path(7890,8860)(7800,8500)(7710,8860)(7890,8860)
\whiten\path(8490,8860)(8400,8500)(8310,8860)(8490,8860)
%
%horizontal arrows on v_1 line 
%
%Horizontal arrows on second vertical band from left
%
\blacken\path(1680,3210)(1320,3300)(1680,3390)(1680,3210)
\blacken\path(1680,3810)(1320,3900)(1680,3990)(1680,3810)
%
%Horizontal arrows on third vertical band from left
%
\blacken\path(2280,3210)(1920,3300)(2280,3390)(2280,3210)
\blacken\path(2280,3810)(1920,3900)(2280,3990)(2280,3810)
%
%Horizontal arrows on fourth vertical band from left
%
%\blacken\path(2820,2010)(3180,2100)(2820,2190)(2820,2010)
%
%vertical arrows on z_2 vertical line
%
\blacken\path(1890,3660)(1800,3300)(1710,3660)(1890,3660)
\blacken\path(1890,4560)(1800,4200)(1710,4560)(1890,4560)
%
%vertical arrows on fourth band from below
%
\blacken\path(2500,3060)(2410,2700)(2320,3060)(2500,3060)
%
%These are the current final vertical arrows pointing
%upwards
%
\blacken\path(2500,3660)(2410,3300)(2320,3660)(2500,3660)
%%and here is the one on top of that on a long v segment
\blacken\path(2500,4560)(2410,4200)(2320,4560)(2500,4560)
%
%Horizontal arrows on fourth vertical band from left
%
\blacken\path(3180,3210)(2820,3300)(3180,3390)(3180,3210)
\blacken\path(3180,3810)(2820,3900)(3180,3990)(3180,3810)
%
%Here are the 3 left most vertical arrows in the new initial 
%state. They point downwards
%
\blacken\path(1300,3060)(1210,2700)(1120,3060)(1300,3060)
\blacken\path(1900,3060)(1810,2700)(1720,3060)(1900,3060)
\blacken\path(2500,3060)(2410,2700)(2320,3060)(2500,3060)
%The two arrows on top of the left most initial state arrow
%
\blacken\path(1300,3660)(1210,3300)(1120,3660)(1300,3660)
\blacken\path(1300,4560)(1210,4200)(1120,4560)(1300,4560)
%
%Here is the rest of the arrows in the new initial state.
%They point upwards.
\blacken\path(3700,2940)(3610,3300)(3520,2940)(3700,2940)
\blacken\path(4300,2940)(4210,3300)(4120,2940)(4300,2940)
\blacken\path(4900,2940)(4810,3300)(4720,2940)(4900,2940)
\blacken\path(5500,2940)(5410,3300)(5320,2940)(5500,2940)
\blacken\path(6100,2940)(6010,3300)(5920,2940)(6100,2940)
\blacken\path(6700,2940)(6610,3300)(6520,2940)(6700,2940)
\blacken\path(7300,2940)(7210,3300)(7120,2940)(7300,2940)
\blacken\path(7900,2940)(7810,3300)(7720,2940)(7900,2940)
\blacken\path(8500,2940)(8410,3300)(8320,2940)(8500,2940)
\end{picture}
%
%\begin{ca} 
\caption{An $[L, N_1,N_2]$-configuration. In 
this example, 
$N_1=5$, $N_2 = 2$, and as always $N_3 = N_1 - N_2$.}
%\end{ca}
%
%\end{minipage}
\label{restricted-bc-FW}
\end{figure}

Setting $v_3 = z_3$, we end up with the lattice configuration 
in Figure {\bf \ref{bc-3-FW}}, from which we can see that 
{\bf 1.}
 All arrows on the lower $N_3$ horizontal lines, 
where $N_3=3$ in the specific example shown, are 
frozen, and 
{\bf 2.}
 All lines on the $N_3$ left most vertical lines 
in the lower half of the diagram, where they 
intersect with $C$-lines. Removing the lower 
$N_3$ $C$-lines we obtain the 
configuration in Figure {\bf \ref{restricted-bc-FW}}.
This configuration has a subset (rectangular shape 
on lower left corner) that is also completely frozen. 
All vertices in this part are $a$-vertices, hence 
from (\ref{weights-FW}), their contribution 
to the partition function of this configuration is 
trivial.

An $[L,N_1,N_2]$-configuration, as in 
Figure {\bf \ref{restricted-bc-FW}}, interpolates between 
an initial reference state 
$| L^{\wedge} \rangle
$
and a final 
$ \langle
 {N_3}^{\vee}, (L-N_3)^{\wedge} |$ 
state, using $N_1$ $B$-lines followed by $N_2$ $C$-lines.

%FIG09
%
%\begin{center}
%\begin{minipage}{4.8in}

%FIG10
%
%\begin{center}
%\begin{minipage}{4.8in}
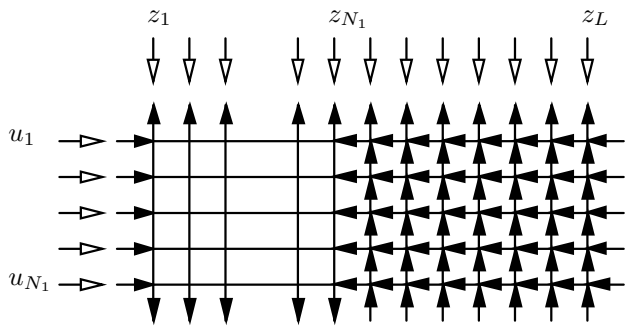
\begin{figure}
\setlength{\unitlength}{0.0008cm}
\begin{picture}(10000,06000)(0000,4000)
\thicklines
%
%The following lines are horizontal
%
\path(0600,5100)(9000,5100)
\path(0600,5700)(9000,5700)
\path(0600,6300)(9000,6300)
\path(0600,6900)(9000,6900)
\path(0600,7500)(9000,7500)
%
%The following lines are vertical
\path(1200,8100)(1200,4500)
\path(1800,8100)(1800,4500)
\path(2400,8100)(2400,4500)
\path(3600,8100)(3600,4500)
\path(4200,8100)(4200,4500)
\path(4800,8100)(4800,4500)
\path(5400,8100)(5400,4500)
\path(6000,8100)(6000,4500)
\path(6600,8100)(6600,4500)
\path(7200,8100)(7200,4500)
\path(7800,8100)(7800,4500)
\path(8400,8100)(8400,4500)
%
%These are the left most z_1 arrows are horizontal, pointing 
%to the right
\blacken\path(0900,5010)(1200,5100)(0900,5190)(0900,5010)
\blacken\path(0900,5610)(1200,5700)(0900,5790)(0900,5610)
\blacken\path(0900,6210)(1210,6300)(0900,6390)(0900,6210)
\blacken\path(0900,6810)(1200,6900)(0900,6990)(0900,6810)
\blacken\path(0900,7410)(1200,7500)(0900,7590)(0900,7410)
\blacken\path(1300,7800)(1210,8100)(1120,7800)(1300,7800)
\blacken\path(1900,7800)(1810,8100)(1720,7800)(1900,7800)
\blacken\path(2500,7800)(2410,8100)(2320,7800)(2500,7800)
\blacken\path(3700,7800)(3610,8100)(3520,7800)(3700,7800)
\blacken\path(4300,7800)(4210,8100)(4120,7800)(4300,7800)
\blacken\path(4900,7800)(4810,8100)(4720,7800)(4900,7800)
\blacken\path(5500,7800)(5410,8100)(5320,7800)(5500,7800)
\blacken\path(6100,7800)(6010,8100)(5920,7800)(6100,7800)
\blacken\path(6700,7800)(6610,8100)(6520,7800)(6700,7800)
\blacken\path(7300,7800)(7210,8100)(7120,7800)(7300,7800)
\blacken\path(7900,7800)(7810,8100)(7720,7800)(7900,7800)
\blacken\path(8500,7800)(8410,8100)(8320,7800)(8500,7800)
%
%The frozen horizontal arrows on the u_5 line
%
\blacken\path(8760,5200)(8400,5110)(8760,5020)(8760,5200)
\blacken\path(8160,5200)(7800,5110)(8160,5020)(8160,5200)
\blacken\path(7560,5200)(7200,5110)(7560,5020)(7560,5200)
\blacken\path(6960,5200)(6600,5110)(6960,5020)(6960,5200)
\blacken\path(6360,5200)(6000,5110)(6360,5020)(6360,5200)
\blacken\path(5760,5200)(5400,5110)(5760,5020)(5760,5200)
\blacken\path(5160,5200)(4800,5110)(5160,5020)(5160,5200)
\blacken\path(4560,5200)(4200,5110)(4560,5020)(4560,5200)
%
%These are the horizontal arrows that form the right 
%boundary of the domain wall configuration
%
%\blacken\path(4560,5200)(4200,5110)(4560,5020)(4560,5200)
%
%this is the one on the u_4 line and all those to its right
%
\blacken\path(4560,5800)(4200,5710)(4560,5620)(4560,5800)
\blacken\path(5160,5800)(4800,5710)(5160,5620)(5160,5800)
\blacken\path(5760,5800)(5400,5710)(5760,5620)(5760,5800)
\blacken\path(6360,5800)(6000,5710)(6360,5620)(6360,5800)
\blacken\path(6960,5800)(6600,5710)(6960,5620)(6960,5800)
\blacken\path(7560,5800)(7200,5710)(7560,5620)(7560,5800)
\blacken\path(8160,5800)(7800,5710)(8160,5620)(8160,5800)
%
%this is the one on the u_3 line and all those to its right
%
\blacken\path(4560,6400)(4200,6310)(4560,6220)(4560,6400)
\blacken\path(5160,6400)(4800,6310)(5160,6220)(5160,6400)
\blacken\path(5760,6400)(5400,6310)(5760,6220)(5760,6400)
\blacken\path(6360,6400)(6000,6310)(6360,6220)(6360,6400)
\blacken\path(6960,6400)(6600,6310)(6960,6220)(6960,6400)
\blacken\path(7560,6400)(7200,6310)(7560,6220)(7560,6400)
\blacken\path(8160,6400)(7800,6310)(8160,6220)(8160,6400)
%
%this is the one on the u_2 line and all those to its right
%
\blacken\path(4560,7000)(4200,6910)(4560,6820)(4560,7000)
\blacken\path(5160,7000)(4800,6910)(5160,6820)(5160,7000)
\blacken\path(5760,7000)(5400,6910)(5760,6820)(5760,7000)
\blacken\path(6360,7000)(6000,6910)(6360,6820)(6360,7000)
\blacken\path(6960,7000)(6600,6910)(6960,6820)(6960,7000)
\blacken\path(7560,7000)(7200,6910)(7560,6820)(7560,7000)
\blacken\path(8160,7000)(7800,6910)(8160,6820)(8160,7000)
%
%this is the one on the u_1 line and all those to its right
%
\blacken\path(4560,7600)(4200,7510)(4560,7420)(4560,7600)
\blacken\path(5160,7600)(4800,7510)(5160,7420)(5160,7600)
\blacken\path(5760,7600)(5400,7510)(5760,7420)(5760,7600)
\blacken\path(6360,7600)(6000,7510)(6360,7420)(6360,7600)
\blacken\path(6960,7600)(6600,7510)(6960,7420)(6960,7600)
\blacken\path(7560,7600)(7200,7510)(7560,7420)(7560,7600)
\blacken\path(8160,7600)(7800,7510)(8160,7420)(8160,7600)
\blacken\path(8700,5800)(8400,5710)(8700,5620)(8700,5800)
\blacken\path(8700,6400)(8400,6310)(8700,6220)(8700,6400)
\blacken\path(8700,7000)(8400,6910)(8700,6820)(8700,7000)
\blacken\path(8700,7600)(8400,7510)(8700,7420)(8700,7600)
%
%The folowing are the horizontal white arrows
%
\path(-0360,5100)(0000,5100)
\path(-0360,5700)(0000,5700)
\path(-0360,6300)(0000,6300)
\path(-0360,6900)(0000,6900)
\path(-0360,7500)(0000,7500)
\put(-1200,5100){$u_{N_1}$}
\put(-1200,7500){$u_{  1}$}
\whiten\path(0000,5010)(0360,5100)(0000,5190)(0000,5010)
\whiten\path(0000,5610)(0360,5700)(0000,5790)(0000,5610)
\whiten\path(0000,6210)(0360,6300)(0000,6390)(0000,6210)
\whiten\path(0000,6810)(0360,6900)(0000,6990)(0000,6810)
\whiten\path(0000,7410)(0360,7500)(0000,7590)(0000,7410)
%
%The following are the vertical white arrows
%
\path(1200,9220)(1200,8500)
\path(1800,9220)(1800,8500)
\path(2400,9220)(2400,8500)
\path(3600,9220)(3600,8500)
\path(4200,9220)(4200,8500)
\path(4800,9220)(4800,8500)
\path(5400,9220)(5400,8500)
\path(6000,9220)(6000,8500)
\path(6600,9220)(6600,8500)
\path(7200,9220)(7200,8500)
\path(7800,9220)(7800,8500)
\path(8400,9220)(8400,8500)
\put(1100,9500){$z_{     1}$}
\put(4100,9500){$z_{N_1   }$}
%\put(4700,9500){$z_{N_1 +1}$}
\put(8300,9500){$z_{L}$}
\whiten\path(1290,8860)(1200,8500)(1110,8860)(1290,8860)
\whiten\path(1890,8860)(1800,8500)(1710,8860)(1890,8860)
\whiten\path(2490,8860)(2400,8500)(2310,8860)(2490,8860)
\whiten\path(3690,8860)(3600,8500)(3510,8860)(3690,8860)
\whiten\path(4290,8860)(4200,8500)(4110,8860)(4290,8860)
\whiten\path(4890,8860)(4800,8500)(4710,8860)(4890,8860)
\whiten\path(5490,8860)(5400,8500)(5310,8860)(5490,8860)
\whiten\path(6090,8860)(6000,8500)(5910,8860)(6090,8860)
\whiten\path(6690,8860)(6600,8500)(6510,8860)(6690,8860)
\whiten\path(7290,8860)(7200,8500)(7110,8860)(7290,8860)
\whiten\path(7890,8860)(7800,8500)(7710,8860)(7890,8860)
\whiten\path(8490,8860)(8400,8500)(8310,8860)(8490,8860)
%
%Here are the 5 left most vertical arrows in the new initial 
%state. They point downwards
%
\blacken\path(1300,4860)(1210,4500)(1120,4860)(1300,4860)
\blacken\path(1900,4860)(1810,4500)(1720,4860)(1900,4860)
\blacken\path(2500,4860)(2410,4500)(2320,4860)(2500,4860)
\blacken\path(3700,4860)(3610,4500)(3520,4860)(3700,4860)
\blacken\path(4300,4860)(4210,4500)(4120,4860)(4300,4860)
%
%Here is the rest of the arrows in the new initial state.
%They point upwards. each followed by those above it.
%
%This is the family on line z_6
%
\blacken\path(4900,4740)(4810,5100)(4720,4740)(4900,4740)
\blacken\path(4900,5340)(4810,5700)(4720,5340)(4900,5340)
\blacken\path(4900,5940)(4810,6300)(4720,5940)(4900,5940)
\blacken\path(4900,6540)(4810,6900)(4720,6540)(4900,6540)
\blacken\path(4900,7140)(4810,7500)(4720,7140)(4900,7140)
%
%This is the family on line z_7
%
\blacken\path(5500,4740)(5410,5100)(5320,4740)(5500,4740)
\blacken\path(5500,5340)(5410,5700)(5320,5340)(5500,5340)
\blacken\path(5500,5940)(5410,6300)(5320,5940)(5500,5940)
\blacken\path(5500,6540)(5410,6900)(5320,6540)(5500,6540)
\blacken\path(5500,7140)(5410,7500)(5320,7140)(5500,7140)
%
%This is the family on line z_8
%
\blacken\path(6100,4740)(6010,5100)(5920,4740)(6100,4740)
\blacken\path(6100,5340)(6010,5700)(5920,5340)(6100,5340)
\blacken\path(6100,5940)(6010,6300)(5920,5940)(6100,5940)
\blacken\path(6100,6540)(6010,6900)(5920,6540)(6100,6540)
\blacken\path(6100,7140)(6010,7500)(5920,7140)(6100,7140)
%
%This is the family on line z_9
%
\blacken\path(6700,4740)(6610,5100)(6520,4740)(6700,4740)
\blacken\path(6700,5340)(6610,5700)(6520,5340)(6700,5340)
\blacken\path(6700,5940)(6610,6300)(6520,5940)(6700,5940)
\blacken\path(6700,6540)(6610,6900)(6520,6540)(6700,6540)
\blacken\path(6700,7140)(6610,7500)(6520,7140)(6700,7140)
%
%This is the family on line z_10
%
\blacken\path(7300,4740)(7210,5100)(7120,4740)(7300,4740)
\blacken\path(7300,5340)(7210,5700)(7120,5340)(7300,5340)
\blacken\path(7300,5940)(7210,6300)(7120,5940)(7300,5940)
\blacken\path(7300,6540)(7210,6900)(7120,6540)(7300,6540)
\blacken\path(7300,7140)(7210,7500)(7120,7140)(7300,7140)
%
%This is the family on line z_11
%
\blacken\path(7900,4740)(7810,5100)(7720,4740)(7900,4740)
\blacken\path(7900,5340)(7810,5700)(7720,5340)(7900,5340)
\blacken\path(7900,5940)(7810,6300)(7720,5940)(7900,5940)
\blacken\path(7900,6540)(7810,6900)(7720,6540)(7900,6540)
\blacken\path(7900,7140)(7810,7500)(7720,7140)(7900,7140)
%
%This is the family on line z_12
%
\blacken\path(8500,4740)(8410,5100)(8320,4740)(8500,4740)
\blacken\path(8500,5340)(8410,5700)(8320,5340)(8500,5340)
\blacken\path(8500,5940)(8410,6300)(8320,5940)(8500,5940)
\blacken\path(8500,6540)(8410,6900)(8320,6540)(8500,6540)
\blacken\path(8500,7140)(8410,7500)(8320,7140)(8500,7140)
\end{picture}
%
%\begin{ca} 
\caption{Another $[L,N_1,N_2]$-configuration. In this example,
$N_2=0$ and $N_1=5$. Equivalently, the left half is an 
$(N_1 \! \times \! N_1)$ 
domain wall configuration, where $N_1=5$, with an additional 
totally frozen lattice configuration to its right.}
%
%\end{ca}
%
%\end{minipage}
\label{dw-extended-FW}
\end{figure}

Setting $v_{N_1-i+1} = z_{i}$ for $i = 1, \dots, N_1$, we 
freeze all arrows that are on $C$-lines or on segments that 
end on $C$-lines.  Discarding these we obtain the lattice 
configuration in Figure {\bf \ref{dw-extended-FW}}.

\begin{figure}
\setlength{\unitlength}{0.0008cm}
\begin{picture}(10000,06000)(0000,4000)
\thicklines
%
%The following lines are horizontal
%
\path(0600,5100)(4800,5100)
\path(0600,5700)(4800,5700)
\path(0600,6300)(4800,6300)
\path(0600,6900)(4800,6900)
\path(0600,7500)(4800,7500)
%
%The following lines are vertical
\path(1200,8100)(1200,4500)
\path(1800,8100)(1800,4500)
\path(2400,8100)(2400,4500)
\path(3600,8100)(3600,4500)
\path(4200,8100)(4200,4500)
%
%These are the left most z_1 arrows are horizontal, pointing 
%to the right
\blacken\path(0900,5010)(1200,5100)(0900,5190)(0900,5010)
\blacken\path(0900,5610)(1200,5700)(0900,5790)(0900,5610)
\blacken\path(0900,6210)(1210,6300)(0900,6390)(0900,6210)
\blacken\path(0900,6810)(1200,6900)(0900,6990)(0900,6810)
\blacken\path(0900,7410)(1200,7500)(0900,7590)(0900,7410)
\blacken\path(1300,7800)(1210,8100)(1120,7800)(1300,7800)
\blacken\path(1900,7800)(1810,8100)(1720,7800)(1900,7800)
\blacken\path(2500,7800)(2410,8100)(2320,7800)(2500,7800)
\blacken\path(3700,7800)(3610,8100)(3520,7800)(3700,7800)
\blacken\path(4300,7800)(4210,8100)(4120,7800)(4300,7800)
\blacken\path(4560,5200)(4200,5110)(4560,5020)(4560,5200)
%
%These are the horizontal arrows that form the right 
%boundary of the domain wall configuration
%
\blacken\path(4560,5800)(4200,5710)(4560,5620)(4560,5800)
\blacken\path(4560,6400)(4200,6310)(4560,6220)(4560,6400)
\blacken\path(4560,7000)(4200,6910)(4560,6820)(4560,7000)
\blacken\path(4560,7600)(4200,7510)(4560,7420)(4560,7600)
%
%The folowing are the horizontal white arrows
%
\path(-0360,5100)(0000,5100)
\path(-0360,5700)(0000,5700)
\path(-0360,6300)(0000,6300)
\path(-0360,6900)(0000,6900)
\path(-0360,7500)(0000,7500)
\put(-1000,5100){$u_N$}
\put(-1000,7500){$u_1$}
\whiten\path(0000,5010)(0360,5100)(0000,5190)(0000,5010)
\whiten\path(0000,5610)(0360,5700)(0000,5790)(0000,5610)
\whiten\path(0000,6210)(0360,6300)(0000,6390)(0000,6210)
\whiten\path(0000,6810)(0360,6900)(0000,6990)(0000,6810)
\whiten\path(0000,7410)(0360,7500)(0000,7590)(0000,7410)
%
%The following are the vertical white arrows
%
\path(1200,9220)(1200,8500)
\path(1800,9220)(1800,8500)
\path(2400,9220)(2400,8500)
\path(3600,9220)(3600,8500)
\path(4200,9220)(4200,8500)
\put(1100,9500){$z_1   $}
\put(4100,9500){$z_N   $}
\whiten\path(1290,8860)(1200,8500)(1110,8860)(1290,8860)
\whiten\path(1890,8860)(1800,8500)(1710,8860)(1890,8860)
\whiten\path(2490,8860)(2400,8500)(2310,8860)(2490,8860)
\whiten\path(3690,8860)(3600,8500)(3510,8860)(3690,8860)
\whiten\path(4290,8860)(4200,8500)(4110,8860)(4290,8860)
%
%Here are the 5 left most vertical arrows in the new initial 
%state. They point downwards
%
\blacken\path(1300,4860)(1210,4500)(1120,4860)(1300,4860)
\blacken\path(1900,4860)(1810,4500)(1720,4860)(1900,4860)
\blacken\path(2500,4860)(2410,4500)(2320,4860)(2500,4860)
\blacken\path(3700,4860)(3610,4500)(3520,4860)(3700,4860)
\blacken\path(4300,4860)(4210,4500)(4120,4860)(4300,4860)
%
%Here is the rest of the arrows in the new initial state.
%They point upwards. each followed by those above it.
%
%%%%%%%%%%%%%%%%%%%%%%%%%%%%%%%%%%%%%%%%%%%%%%%%%%%%%%%%%
%
%This is the diagram on the right
%The following lines are horizontal
%
%These are the horizontal lines
%
\path(5400,7500)(9600,7500)
\path(5400,6900)(9600,6900)
\path(5400,6300)(9600,6300)
\path(5400,5700)(9600,5700)
\path(5400,5100)(9600,5100)
%
%The following lines are vertical
%
\path(6000,8100)(6000,4500)
\path(6600,8100)(6600,4500)
\path(7200,8100)(7200,4500)
\path(8400,8100)(8400,4500)
\path(9000,8100)(9000,4500)
%
%These are the black arrows to the left 
%
\blacken\path(5760,5010)(5400,5100)(5760,5190)(5760,5010)
\blacken\path(5760,5610)(5400,5700)(5760,5790)(5760,5610)
\blacken\path(5760,6210)(5400,6300)(5760,6390)(5760,6210)
\blacken\path(5760,6810)(5400,6900)(5760,6990)(5760,6810)
\blacken\path(5760,7410)(5400,7500)(5760,7590)(5760,7410)
%
%These are the top black arrows
%
\blacken\path(6100,7860)(6010,7500)(5920,7860)(6100,7860)
\blacken\path(6700,7860)(6610,7500)(6520,7860)(6700,7860)
\blacken\path(7300,7860)(7210,7500)(7120,7860)(7300,7860)
\blacken\path(8500,7860)(8410,7500)(8320,7860)(8500,7860)
\blacken\path(9100,7860)(9010,7500)(8920,7860)(9100,7860)
%
%These are the horizontal black arrows that form the right 
%boundary of the domain wall configuration
%
\blacken\path(9240,7600)(9600,7510)(9240,7420)(9240,7600)
\blacken\path(9240,7000)(9600,6910)(9240,6820)(9240,7000)
\blacken\path(9240,6400)(9600,6310)(9240,6220)(9240,6400)
\blacken\path(9240,5800)(9600,5710)(9240,5620)(9240,5800)
\blacken\path(9240,5200)(9600,5110)(9240,5020)(9240,5200)
%
%The following are the vertical white arrows
%
\path(6000,9220)(6000,8500)
\path(6600,9220)(6600,8500)
\path(7200,9220)(7200,8500)
\path(8400,9220)(8400,8500)
\path(9000,9220)(9000,8500)
\put(5900,9500){$z_1   $}
\put(8900,9500){$z_N   $}
\whiten\path(6090,8860)(6000,8500)(5910,8860)(6090,8860)
\whiten\path(6690,8860)(6600,8500)(6510,8860)(6690,8860)
\whiten\path(7290,8860)(7200,8500)(7110,8860)(7290,8860)
\whiten\path(8490,8860)(8400,8500)(8310,8860)(8490,8860)
\whiten\path(9090,8860)(9000,8500)(8910,8860)(9090,8860)
%
%The bottom black arrows
%
\blacken\path(6100,4740)(6010,5100)(5920,4740)(6100,4740)
\blacken\path(6700,4740)(6610,5100)(6520,4740)(6700,4740)
\blacken\path(7300,4740)(7210,5100)(7120,4740)(7300,4740)
\blacken\path(8500,4740)(8410,5100)(8320,4740)(8500,4740)
\blacken\path(9100,4740)(9010,5100)(8920,4740)(9100,4740)
\end{picture}
%
%\begin{ca} 
\caption{The left hand side is an $(N \! \times  \! N)$ domain wall 
configuration, where $N=5$. The right hand side is the 
corresponding dual configuration.}
%\end{ca}
%
%\end{minipage}
\label{dw-FW}
\end{figure}
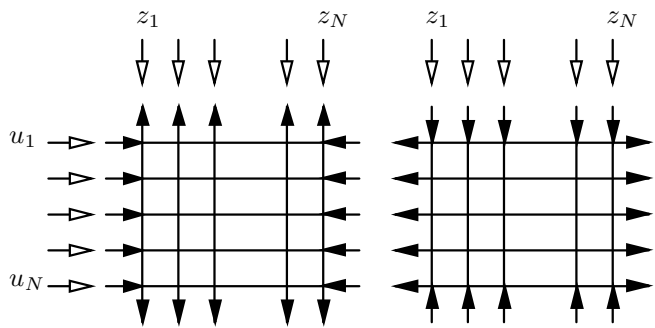

Removing all frozen vertices (as well as the extra space 
between two sets of vertical lines, that is no longer
necessary), one obtains the {\it domain wall configuration} 
in Figure {\bf \ref{dw-FW}}, which is characterized as 
follows. All arrows on the left and right boundaries point 
inwards, and all arrows on the upper and lower boundaries 
point outwards. The internal arrows remain free, and the
configurations that are consistent with the boundary conditions
are summed over. Reversing the orientation of all arrows on all 
boundaries is a dual a domain wall configuration. 

%FIG11
%
%\begin{center}
%\begin{minipage}{4.8in}

\subsection{Remarks on domain wall configurations}
\label{DW-remarks-FW}
{\bf 1.} One can generate a domain wall configuration directly 
starting from a length-$N$ initial reference state followed 
by $N$ $B$-lines, 
{\bf 2.} One can generate a dual domain wall configuration 
directly starting from a length-$N$ dual initial reference 
state followed by $N$ $C$-lines,
{\bf 3.} A $BC$-configuration with length-$L$ initial and final 
reference states, $L$ $B$-lines and $L$ $C$-lines, factorizes 
into a product of a domain wall configuration and a dual domain 
wall configuration,
{\bf 4.} The restriction of $BC$-configurations to 
$[L,N_1,N_2]$-configurations, where $N_2 < N_1$, produces 
a recursion relation that was used in \cite{MW-FW} to provide 
a recursive proof of Slavnov's determinant expression for the 
scalar product of a Bethe eigenstate and a generic state in 
the corresponding spin chain, 
{\bf 5.} The partition function of a domain wall configuration 
has a determinant expression found by Izergin, that can be 
derived in six-vertex model terms (without reference to spin 
chains or the BA) \cite{Izergin-FW}. 

\subsection{Izergin's domain wall partition function}

Let $\{u\}_N =  \{u_1, \dots, u_N\}$ and $\{z\}_N = 
\{z_1, \dots, z_N\}$ be two sets of rapidity 
variables \footnote{The following result does not require that 
any set of rapidities satisfy Bethe equations.} and define 
$Z_N \ll \{u\}_N, \{z\}_N \rr$ to be the partition function 
of the domain wall lattice configuration on the left hand side 
of Figure {\bf \ref{dw-FW}}, after dividing by $d(u)$ for every 
$B$-line with rapidity $u$. Izergin's determinant expression for 
the domain wall partition function is 

\begin{multline}
Z_N \ll \{u\}_N,\{z\}_N \rr
=
%\\
\frac{
\prod_{i, j=1}^{N} 
[u_i - z_j+\eta]
}
{
\prod_{1 \leq i < j \leq N} 
[u_i-u_j] [z_j-z_i]
}
\det
\ll
\frac{
[\eta]
}
{
[u_i - z_j + \eta] [u_i - z_j]
}
\rr_{1\leq i,j \leq N}
\label{dwpf-FW}
\end{multline}

Dual domain wall configurations have the same partition 
functions due to invariance under reversing all arrows.
For the result of this note, we need 
the homogeneous limit of the above expression. Taking 
the limit $z_i \rightarrow z$, $\{i = 1, \cdots, L\}$, 
we obtain 

\begin{equation}
\label{izergin-homogeneous-FW}
Z_{N}^{\textit{hom}} \ll \{u\}_N, z \rr
=
\frac{
\prod_{i=1}^{N}
[u_i - z+\eta]^{N}
}
{
\prod_{1 \leq i < j \leq N}
[u_i - u_j]
}
\det
\ll
\phi^{(j-1)}(u_i, z)
\rr_{1\leq i,j \leq N}
\end{equation}

\noindent where 

\begin{equation}
\phi^{(j)}(u_i, z) = \frac{1}{j!}
\
\partial^{(j)}_z
\ll
\frac{
[\eta]
}{
[u_i - z + \eta] [u_i -  z]
}
\rr
\end{equation}

%SECTION04
\section{Structure constants in Type-{\bf A} theories}
\label{section-A-structure-constants-FW}

In this section, we recall the discussion of SYM$_4$ 
tree-level structure constants of \cite{E1-FW, F-FW} but now 
in the context of the Type-{\bf A} theories in Subsection 
{\bf \ref{theories-FW}}, and construct determinant expressions 
for structure constants of three non-extremal $SU(2)$ 
single-trace operators. 

Since theory {\bf 1} is SYM$_4$, theory {\bf 2} is 
an Abelian orbifolding of SYM$_4$, and theory {\bf 3} 
is a real-${\bf \beta}$-deformation of it, all three 
theories share the same fundamental charged scalar field 
content, that is $\{X, Y, Z\}$ and their charge conjugates 
$\{\bar{X}, \bar{Y}, \bar{Z}\}$,
and all are conformally invariant up to all loops 
\cite{zoubos-review-FW}. This makes our discussion 
a straightforward paraphrasing of that in \cite{E1-FW, F-FW}.

%FIG12
%
%\begin{center}
%\begin{minipage}{10.0cm}
\begin{figure}
\thicklines
\setlength{\unitlength}{0.0008cm}
\begin{picture}(14000,06000)(-1000,0000)
%
%This is the dashed line on the left
\drawline[-30](2400,0300)(4800,2700)
%
%Here are the fill in lines on the left
%
\path(1200,0300)(3600,2700)
\path(1300,0300)(3700,2700)
\path(1400,0300)(3800,2700)
\path(1500,0300)(3900,2700)
\path(1600,0300)(4000,2700)
\path(1700,0300)(4100,2700)
\path(1800,0300)(4200,2700)
\path(1900,0300)(4300,2700)
\path(2000,0300)(4400,2700)
\path(2100,0300)(4500,2700)
\path(2200,0300)(4600,2700)
\path(3600,2700)(3600,3900)
\path(3700,2700)(3700,3900)
\path(3800,2700)(3800,3900)
\path(3900,2700)(3900,3900)
\path(4000,2700)(4000,3900)
\path(4100,2700)(4100,3900)
\path(4200,2700)(4200,3900)
\path(4300,2700)(4300,3900)
\path(4400,2700)(4400,3900)
\path(4500,2700)(4500,3900)
\path(4600,2700)(4600,3900)
%
%Here are the fill in lines on the right
%
\path(5600,2700)(8000,0300)
\path(5800,2700)(8200,0300)
\path(6400,2700)(8800,0300)
\path(6600,2700)(9000,0300)
\path(7200,2700)(9600,0300)
\path(5600,2700)(5600,3900)
\path(5800,2700)(5800,3900)
\path(6400,2700)(6400,3900)
\path(6600,2700)(6600,3900)
\path(7200,2700)(7200,3900)
\drawline[-30](4800,2700)(7200,0300)
\drawline[-40](4800,3900)(4800,2700)
%
% here are the sides
\path(3600,3900)(3600,2700)(1200,300)
\path(7200,3900)(7200,2700)(9600,300)
%
%diagonal pieces
\path(4800,1500)(3600,0300)
\path(4800,1500)(6000,0300)
%
%horizontal ine, bottom right
\path(6000,0300)(9600,0300)
%
%two short vertical pieces
\path(1200,0300)(3600,0300)
\path(3600,3900)(7200,3900)
\put( 2900, 4000){$l_1$}
\put( 2900, 2700){$m_1$}
\put( 0500,-0100){$l_3$}
\put( 7400, 4000){$r_1$}
\put( 7400, 2700){$m_3$}
\put( 9800,-0100){$r_2$}
\put( 4900, 4000){$c_1$}
\put( 4900, 2700){$m_0$}
\put( 2400,-0100){$c_3$}
\put( 7200,-0100){$c_2$}
\put( 3600,-0100){$r_3$}
\put( 4800, 0800){$m_2$}
\put( 6000,-0100){$l_2$}
\end{picture}
%
%\begin{ca} 
\caption{A schematic representation of a 3-point function. 
State $\cO_1$ is on top. 
$\cO_2$ and $\cO_3$ are below, to the right and to the left. 
Type-{\bf A} 3-point functions are (initially) 
in this {\it \lq wide-pants\rq} form.
}
%\end{ca}
%
%\end{minipage}
\label{wide-pants-FW}
\end{figure}
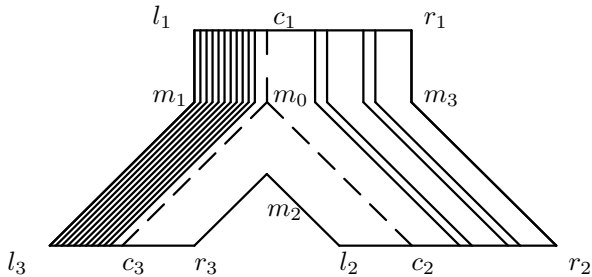

\bigskip

\subsection{Tree-level structure constants}
We consider tree-level 3-point functions 
of $SU(2)$ single-trace operators that 
{\bf 1.}
 have well-defined conformal dimensions at 1-loop level, and 
{\bf 2.}
 can be mapped to Bethe eigenstates in closed spin-$\frac{1}{2}$ 
chains. 

These 3-point functions can be represented schematically 
as in Figure {\bf \ref{wide-pants-FW}}. Identify the pairs of 
corner points
$\{l_1, r_1\}$,
$\{l_2, r_2\}$,
$\{l_3, r_3\}$,
as well as the triple 
$\{m_1, m_2, m_3\}$ to obtain a pants diagram.
The structure constants have a perturbative expansion in 
the \rq t Hooft coupling constant $\lambda$, 

\begin{equation}
C_{ijk} =
c_{ijk}^{(0)} + \lambda c_{ijk}^{(1)} + \dots
\end{equation}

We restrict our attention to the leading coefficient $c_{ijk}^{(0)}$. 
In the limit $\lambda \rightarrow 0$, many single-trace operators 
have the same conformal dimension. This 
degeneracy is lifted at 1-loop level and certain linear combinations 
of single-trace operators have definite 1-loop conformal 
dimension. This is why although we compute 
tree-level structure constants, we insist on 1-loop conformal 
invariance: We identify operators with well defined 
conformal dimensions.

As explained in Section {\bf \ref{section-background-FW}}, 
these linear combinations correspond to eigenstates of 
a closed spin-$\frac{1}{2}$ chain. Their conformal dimensions 
are the corresponding Bethe eigenvalues. These closed spin 
chain states correspond to the circles at the boundaries of 
the pants diagram that can be constructed from 
Figure {\bf \ref{wide-pants-FW}} as discussed above.

\subsection{Remark} In computing 3-point functions, the three 
composite operators {\it may or may not belong to the same} 
$SU(2)$ doublet. In particular, in \cite{E1-FW}, EGSV use 
operators from the doublets 
$\{     Z,       X\}$, 
$\{\bar{Z}, \bar{X}\}$, and 
$\{     Z,  \bar{X}\}$. 
In \cite{F-FW}, this procedure allowed us to construct determinant 
expressions for structure constants of non-extremal 3-point 
functions. This applies to all Type-{\bf A} theories. Type-{\bf B} 
structure constants are constructed differently. In
particular, the non-extremal case $l_{23} = 0$ is considered.

\subsection{Constructing 3-point functions}
To construct three-point functions at the gauge theory operator 
level, the fundamental fields in the operators $\mathcal{O}_i$, 
$i = \{1, 2, 3\}$ are contracted by free propagators. 
Each propagator connects two fields, hence $L_1 + L_2 + L_3$ 
is an even number. The number of propagators between 
$\mathcal{O}_i$ and $\mathcal{O}_j$ is 
\begin{equation}
l_{ij}= \frac{1}{2} (L_i + L_j - L_k)
\end{equation}
\noindent where $(i,j,k)$ take distinct values in $(1,2,3)$. 
We restrict our attention, in this section, to the non-extremal 
case, that is, all $l_{ij}$'s are strictly positive. 
The free propagators reproduce the factor 
$1/|x_i - x_j|^{\Delta_i+\Delta_j-\Delta_k}$ in 
({\ref{3pt-FW}
}), where $\Delta_i$ $=$ $\Delta_i^{(0)}$, 
the tree-level conformal dimension. See Figure {\bf \ref{wide-pants-FW}
} 
for a schematic representation of a three point function of the 
type discussed in this note. The horizontal line segment between 
$l_i$ and $r_i$ represents the operator $\mathcal{O}_i$. 
The lines that start at $\mathcal{O}_1$ and end at either 
$\mathcal{O}_2$ or $\mathcal{O}_3$ 
represent one type of propagator. 

\subsection{From single-trace operators to spin-chain states} 
One represents the single-trace operator $\mathcal{O}_i$ of 
well-defined 1-loop conformal dimension $\Delta_i$ by a closed 
spin-chain Bethe eigenstate 
$| \mathcal{O}_i \rangle_{\beta}$. Its eigenvalue $E_i$ is equal 
to $\Delta_i$. The number of fundamental fields $L_i$ in the trace 
is the length of the spin chain. 

The single-trace operator $\mathcal{O}_i$ is a composite operator 
built from weighted sums over traces of products of two fundamental 
fields $\{u, d\}$. These fundamental fields are mapped to 
definite spin states. 
To perform suitable mappings that lead to non-vanishing results, 
we need to decide on which state(s) are in-state(s) from the 
viewpoint of the lattice representation, and which are 
out-state(s).

\subsection{Type-A. Fundamental field content of the states} 

All three Type-{\bf A} theories have the same fundamental field 
content, namely that of SYM$_4$, and thereby, more than one 
doublet. We focus on the doublets formed from the fields $Z$, 
$X$ and their conjugates. Following \cite{E1-FW}, we identify 
the fundamental field content of $\mathcal{O}_i$, $i \in \{1, 2, 3\}$ 
with spin-chain spin states as shown in Table {\bf 1}, 
where $\bar{Z}$ and $\bar{X}$ are the conjugates of 
$Z$ and $X$. That is, if $Z$ appears on one side of a propagator 
and $\bar{Z}$ appears on the other side, then that propagator 
is not identically vanishing, and $Z$ and $\bar{Z}$ can be 
Wick contracted. Similarly for $X$ and $\bar{X}$. 
\setlength{\extrarowheight}{6.0pt}
\begin{table}[ht]
\begin{center}
\begin{tabular}{| c || c | c | c | c |}
\hline
{Operator} & 
{$\ll \begin{array}{c} 1 \\  0 \end{array} \rr$} & 
{$\ll \begin{array}{c} 0 \\  1 \end{array} \rr$} &
{$\ll                  1 \ \ 0             \rr$} &
{$\ll                  0 \ \ 1             \rr$} 
\\
\hline
\hline
$\mathcal{O}
_1$ & $      Z  $ & $      X  $ & $ \bar{Z} $ & $ \bar{X} $ 
\\ 
\hline
$\mathcal{O}
_2$ & $ \bar{Z} $ & $ \bar{X} $ & $      Z  $ & $      X  $ 
\\
\hline
$\mathcal{O}
_3$ & $      Z  $ & $ \bar{X} $ & $ \bar{Z} $ & $      X  $ 
\\
\hline
\end{tabular}
\bigskip
\caption{Identification of operator content of $\mathcal{O}
_i$,
$i \in \{1, 2, 3\}$ with initial and final spin-chain states.}
\end{center}
\end{table}

In our conventions

\begin{equation}
\label{contractions-FW}
\langle
 \bar{Z} Z \rangle
 = \langle
      Z  | Z \rangle
 = 1, \quad 
\langle
      Z  Z \rangle
 = \langle
 \bar{Z} | Z \rangle
 = 0 
\end{equation}

\noindent and similarly for $X$ and $\bar{X}$. 
In (\ref{contractions-FW}
), $\langle
 \bar{f} \ f\rangle
 $ with no 
vertical bar between the two operators is a propagator, 
while 
$\langle
 f | f \rangle
$ 
with a vertical bar between the two operators is a scalar 
product of an initial state $| f \rangle $ and a final 
state $\langle f |$.

From Table {\bf 1}, one can read the fundamental-scalar 
operator content of each single-trace operator $\mathcal{O}_i$,
$i \in \{1, 2, 3\}$, when it is an initial state and when it is 
a final state.  For example, the fundamental field content of 
the initial state 
$ |\mathcal{O}_1 \rangle
$ is $\{      Z,       X\}$, 
and that of the corresponding final state 
$\langle \mathcal{O}_1  |$ is $\{ \bar{Z}, \bar{X}\}$.
The content of an initial state and the corresponding 
final state are related by the `flipping' operation
of \cite{E1-FW} described below.

\subsection{Structure constants in terms of spin-chains}
Having mapped the single-trace operators $\mathcal{O}_i$, 
$i \in \{1, 2, 3\}$ to spin-chain eigenstates, EGSV 
construct the structure constants in three steps.

\subsection*{Step 1. Split the lattice configurations
that correspond to closed spin-chain eigenstates into two 
parts}
Consider the open 1-dimensional lattice configuration 
that corresponds to the $i$-th closed spin-chain 
eigenstate, $i \in \{1, 2, 3\}$. 
This is schematically represented by a line in Figure 
{\bf \ref{wide-pants-FW}} that starts at $l_i$ and ends 
at $r_i$. Split that, at point $c_i$ into 
{\it left} and {\it right} sub-lattice configurations 
of lengths 
$L_{i, l} = \frac{1}{2} (L_i + L_j - L_k)$ and 
$L_{i, r} = \frac{1}{2} (L_i + L_k -L_j)$ 
respectively. Note that the lengths of the sub-lattices 
is fully determined by $L_1$, $L_2$ and $L_3$ which are 
fixed.

Following \cite{korepin-book-FW}, we express the single 
lattice configuration of the original closed spin chain 
state as a weighted sum of tensor products of states 
that live in two smaller Hilbert spaces. The latter 
correspond to closed spin chains of lengths $L_{i, l}$ 
and $L_{i, r}$ respectively. That is, 
$|\mathcal{O}_i \rangle = 
\sum H_{l, r} |\mathcal{O}_{i} \rangle_l \otimes 
|\mathcal{O} _{i}\rangle _r$.
The factors $H_{l, r}$ were computed in 
\cite{korepin-book-FW} and were needed in \cite{E1-FW}, 
where one of the scalar products is generic and had 
to be expressed as an explicit sum. They are not 
needed in this work as we use Bethe equations to 
evaluate this very sum as a determinant.

\subsection*{Step 2. From initial to final states} 
Map $|\mathcal{O}_i \rangle_l \otimes |\mathcal{O}_i \rangle_r\ 
\to  |\mathcal{O}_i \rangle_l \otimes {_r\langle \mathcal{O}_i|}$, 
using the operator $\mathcal{F}$ that acts as follows.
\begin{equation}
\mathcal{F}
\ll | 
f_1 f_2 \cdots f_{L-1} f_{L}
\rangle 
\rr 
= 
\langle 
\bar{f}_{L} \bar{f}_{L-1} \cdots \bar{f}_2 \bar{f}_1 | 
\label{flipping-FW}
\end{equation}
\noindent In particular,

\begin{equation}
\langle ZZ \cdots Z|     ZZ \cdots Z\rangle = 
\langle \bar{Z} \bar{Z} \cdots \bar{Z}| 
        \bar{Z} \bar{Z} \cdots \bar{Z} 
\rangle = 1, 
%\\
\quad {\rm and} \quad
\langle \bar{Z} \bar{Z} \cdots \bar{Z} | Z Z \cdots Z \rangle 
 = 0 
\end{equation}

\noindent More generally

\begin{equation}
\langle
 f_{i_1} f_{i_2} \cdots f_{i_L} | f_{j_1} f_{j_2} \cdots f_{j_L}  \rangle
\sim 
\delta_{i_1 j_1} \delta_{i_2 j_2} \cdots \delta_{i_L j_L} 
\end{equation}

\noindent The `flipping' operation in ({\ref{flipping-FW}}) 
is the origin of the differences in assignments of fundamental 
fields to initial and final operator states in Table {\bf 1}. 
For example, $| \mathcal{O}_1 \rangle $ has fundamental field 
content $\{Z, X\}$, but $\langle \mathcal{O}_1 |$ has fundamental 
field content $\{ \bar{Z}, \bar{X}\}$.
This agrees with the fact that in computing 
$\langle \mathcal{O}_i | \mathcal{O}_i \rangle $, 
free propagators can only connect conjugate fundamental fields.

\subsection*{Step 3. Compute scalar products}
Wick contract pairs of initial states 
$|\mathcal{O}_{i}\rangle_r$ and final states
$|\mathcal{O}_{i+1}\rangle_l$, where $i \in \{1, 2, 3\}$ and 
$i + 3 \equiv i$. The spin-chain equivalent of that is to compute 
the scalar products 
$\,_r\langle \mathcal{O}_i| \mathcal{O}_{i+1}\rangle_l$,
which in six-vertex model terms are $BC$-configurations.
The most general scalar product that we can consider is the 
generic scalar product between two generic Bethe states

\begin{equation}
S_{\textit{generic}} \ll \{u\},\{v\} \rr 
= 
\langle
0 | 
\prod_{j=1}^N \mathbb{C}(v_j) 
\prod_{j=1}^N \mathbb{B}(u_j) 
| 0 \rangle \, 
\end{equation}

A computationally tractable evaluation of 
$S_{\textit{generic}}(\{u\},\{v\})$
using the commutation relations of BA operators is known 
\cite{Korepin-FW}. Simpler expressions are obtained when 
the auxiliary rapidities of one (or both) states satisfies 
Bethe equations. The result in this case is a determinant. 
When only one set satisfies Bethe equations, one obtains
a Slavnov scalar product. This was discussed in Section
{\bf \ref{section-spin-chain-FW}}.

\subsection{Type-A. An unevaluated expression}
The above three steps lead to the following preliminary, 
unevaluated expression

\begin{equation}
c^{(0)}_{123} = 
\mathcal{N}_{123}
\sum                             
\ _r\langle
 {\mathcal{O}}_{3}| \mathcal{O}_{1} \rangle_l 
\ 
_r\langle {\mathcal{O}}_{1}| \mathcal{O}_{2}\rangle_l \ 
_r\langle {\mathcal{O}}_{2}| \mathcal{O}_{3}\rangle_l \
\label{intermediate-FW}
\end{equation}

\noindent where the normalization factor $\mathcal{N}_{123}$, 
that turns out to be a non-trivial object that depends on 
the norms of the Bethe eigenstates, is 

\begin{equation}
\mathcal{N}_{123} = 
\sqrt{ \frac{L_1 L_2 L_3} {\mathcal{N}_1 
                           \mathcal{N}_2 
                           \mathcal{N}_3} }
\label{normalization-FW}
\end{equation}

In ({\ref{normalization-FW}}), $L_i$ is the number of sites 
in the closed spin chain that represents state $\mathcal{O}_i$. 
$\mathcal{N}_i$ is the Gaudin norm of state $\mathcal{O}_i$ as 
in (\ref{Gaudin-FW}).
The sum in ({\ref{intermediate-FW}}) is to be understood
as follows. 
{\bf 1.} It is a sum over all possible ways to split the sites 
of each closed spin chain (represented as a segment in 
a 1-dimensional lattice) into a left part and a right part. 
We will see shortly that only one term in this sum survives.
{\bf 2.} It is a sum over all possible ways of partitioning 
the $X$ or $\bar{X}$ content of a spin chain state between 
the two parts that that spin chain was split into. We will 
see shortly that only one sum survives.

\subsection{Type-A. Simplifying the unevaluated 
expression}
Wick contracting single-trace operators, we can only contract 
a fundamental field with its conjugate. Given the assignments 
in Table {\bf 1}, one can see that 
{\bf 1.} {\it All} $Z$ fields in $\mathcal{O}_3$ must contract 
with $\bar{Z}$ fields in $\mathcal{O}_2$. 
The reason is that there are $\bar{Z}$ fields only in 
$\mathcal{O}_2$, and none in $\mathcal{O}_1$.
{\bf 2.} All $\bar{X}$ fields in $\mathcal{O}_3$ contract 
with $X$ fields in $\mathcal{O}_1$. 
The reason is that there are $X$ fields only in 
$\mathcal{O}_1$, and none in $\mathcal{O}_2$.
If the total number of scalar fields in $\mathcal{O}_i$ 
is $L_i$, and the number of $\{X, \bar{X}\}$-type scalar 
fields is $N_i$, then 

\begin{equation}
l_{13} = N_3,       \qquad
l_{23} = L_3 - N_3, \qquad 
l_{12} = L_1 - N_3
\label{relations-FW}
\end{equation}

\noindent and we have the constraint 

\begin{equation}
N_1 = N_2 + N_3 
\label{constraint-FW}
\end{equation}

From ({\ref{relations-FW}}) and ({\ref{constraint-FW}}), 
we have the following 4 simplifications.
{\bf 1.} There is only one way to split each lattice 
configuration that represents a spin chain into a left 
part and a right part, 
{\bf 2.} The scalar product 
$_r\langle {\mathcal{O}}_{2}| \mathcal{O}_{3}\rangle_l$ 
involves the fundamental field $Z$ (and only $Z$) 
in the initial state $| \mathcal{O}_{3}\rangle_l$ as well 
as in the final state $_r\langle {\mathcal{O}}_{2} |$.
Using Table {\bf 1}, 
we find that these states translate to an initial and a final 
spin-up reference state, respectively. This is represented in 
Figure {\bf \ref{wide-pants-FW}} by the fact that no connecting 
lines (that stand for propagators of 
$\{X, \bar{X}\}$ states) connect $\mathcal{O}_2$ and 
$\mathcal{O}_3$. The scalar product of the two reference states 
is ${}_r\langle  \mathcal{O}_{2}| 
                 \mathcal{O}_{3}\rangle _l \ = 1$,
{\bf 3.} The scalar product 
$ _r\langle {\mathcal{O}}_{3}| \mathcal{O}_{1}\rangle _l$ 
involves the fundamental fields 
$X$ (and only $X$)
in the initial state $| \mathcal{O}_{1}\rangle_l$ as well as
in the final   state $_r\langle {\mathcal{O} }_{3}|$.
Using Table {\bf 1},
we find that these states translate to an initial spin-up
and a final spin-down reference state, respectively.
This is represented in Figure {\bf \ref{wide-pants-FW}}
by the high density of connecting lines (that stand for
propagators of $\{X, \bar{X}\}$ states) between  
$\mathcal{O}_1$ and $\mathcal{O}_3$.
This scalar product is straightforward to evaluate in 
terms of the domain wall partition function,
{\bf 4.} In the remaining scalar product 
$_r\langle {\mathcal{O} }_{1}  | \mathcal{O}_{2}\rangle_l$, 
both the initial state 
$| \mathcal{O}_{2}\rangle_l$ and the final state 
$_r\langle {\mathcal{O} }_{1} |$
involve $\{\bar{X}, \bar{Z} \}$. These fields translate 
to up and down spin states and the scalar product is generic. 
Using the BA commutation relations, it can be evaluated 
as a weighted sum \cite{korepin-book-FW}.

\subsection{Type-A. Evaluating the expression}
\label{section-determinant-expression-FW}
The idea of \cite{F-FW} is to identify the expression in 
({\ref{intermediate-FW}}), up to simple factors, with the 
partition function of an $[L_1, N_1, N_2]$-configuration.
Since this partition function is a restricted scalar product 
$S[L_1, N_1,N_2]$, it can be evaluated as a determinant. 
This is achieved in two steps.

\subsection*{Step 1. Re-writing one of the scalar products}
\begin{figure}
\thicklines
\setlength{\unitlength}{0.0008cm}
\begin{picture}(14000,06000)(-1000,0000)
\path(2400,0300)(4800,2700)
\path(4800,2700)(7200,0300)
\path(4800,3900)(4800,2700)
%
% here are the sides
\path(3600,3900)(3600,2700)(1200,300)
\path(7200,3900)(7200,2700)(9600,300)
%
%diagonal pieces
%\path(4800,1500)(3600,0300)
%\path(4800,1500)(6000,0300)
%
\path(7200,0300)(9600,0300)
%
%two short horizontal pieces
\path(1200,0300)(2400,0300)
\path(3600,3900)(7200,3900)
\put( 2900, 4000){$l_1$}
\put( 2900, 2700){$m_1$}
\put( 0500,-0100){$r_2$}
\put( 7400, 4000){$r_1$}
\put( 7400, 2700){$m_2$}
\put( 9800,-0100){$r_3$}
\put( 4900, 4000){$c_1$}
\put( 4900, 2700){$m_0$}
\put( 2400,-0100){$c_2$}
\put( 7200,-0100){$c_3$}
%
%Here are the fill in lines on the left
%
\path(1200,0300)(3600,2700)
\path(1300,0300)(3700,2700)
\path(1400,0300)(3800,2700)
\path(1500,0300)(3900,2700)
\path(1600,0300)(4000,2700)
\path(1700,0300)(4100,2700)
\path(1800,0300)(4200,2700)
\path(1900,0300)(4300,2700)
\path(2000,0300)(4400,2700)
\path(2100,0300)(4500,2700)
\path(2200,0300)(4600,2700)
\path(2300,0300)(4700,2700)
\path(2400,0300)(4800,2700)
%\path(2500,0300)(4900,2700)
%
\path(3600,2700)(3600,3900)
\path(3700,2700)(3700,3900)
\path(3800,2700)(3800,3900)
\path(3900,2700)(3900,3900)
\path(4000,2700)(4000,3900)
\path(4100,2700)(4100,3900)
\path(4200,2700)(4200,3900)
\path(4300,2700)(4300,3900)
\path(4400,2700)(4400,3900)
\path(4500,2700)(4500,3900)
\path(4600,2700)(4600,3900)
\path(4700,2700)(4700,3900)
\path(4800,2700)(4800,3900)
%\path(4900,2700)(4900,3900)
%
%Here are the fill in lines on the right
%
%\path(5000,2700)(7400,0300)
\path(5600,2700)(8000,0300)
\path(5800,2700)(8200,0300)
\path(6400,2700)(8800,0300)
\path(6600,2700)(9000,0300)
\path(7200,2700)(9600,0300)
%
%\path(5000,2700)(5000,3900)
\path(5600,2700)(5600,3900)
\path(5800,2700)(5800,3900)
\path(6400,2700)(6400,3900)
\path(6600,2700)(6600,3900)
\path(7200,2700)(7200,3900)
\end{picture}
%
%\begin{ca} 
\caption{
A schematic representation of a 3-point function after 
removal of a contraction between the left part of $\cO_2$ 
and the right part of $\cO_3$, that evaluates to a factor 
of 1. Type-{\bf B} 3-point functions are in this 
{\it \lq narrow pants\rq} form from the outset. 
}
%\end{ca}
%
%\end{minipage}
\label{narrow-pants-FW}
\end{figure}
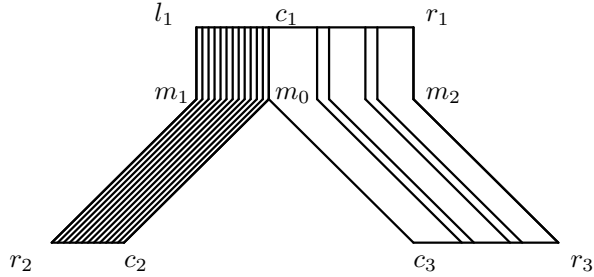

We use the facts that  
{\bf 1.} ${}_r\langle {\mathcal{O}}_{2} | {\mathcal{O}}_{3} \rangle_l$ 
$=$ $1$, and
{\bf 2.} ${}_r\langle {\mathcal{O}}_{2} | {\mathcal{O}}_{1} \rangle_l$ 
$=$ 
${}_l\langle {\mathcal{O}}_{1} | {\mathcal{O}}_{2} \rangle_r$, 
which is true for all scalar products, to re-write 
({\ref{intermediate-FW}}) in the form

\begin{equation}
c^{(0)}_{123} =
\mathcal{N}_{123}
\sum_{\alpha \cup \bar{\alpha} = \{u_{\beta}\}_{N_1}}
{}_r\langle \mathcal{O}_3 | \mathcal{O}_1 \rangle_l \ {}_l\langle 
\mathcal{O}_2 | \mathcal{O}_1 \rangle_r \ 
%\\
=
\mathcal{N}_{123} \ \ll {}_r\langle \mathcal{O}_3 | \otimes {}_l\langle 
\mathcal{O}_2 | \rr | \mathcal{O}_1 \rangle \
\label{unevaluated-more-simplified-FW}
\end{equation}

\noindent where the right hand side of 
({\ref{unevaluated-more-simplified-FW}}) 
is a scalar product of the full initials
tate $| \mathcal{O}_1 \rangle$ (so we no 
longer have a sum over partitions of the 
rapidities $\{u_{\beta}\}_{N_1}$ since we 
no longer split the state $\mathcal{O}_1$) 
and two states that are pieces of original 
states that were split. Deleting the scalar 
product corresponding to contracting the left 
part of state $\mathcal{O}_2$ with the right 
part of state $\mathcal{O}_3$, 
since that contraction leads to a factor of unity,
the object that we are evaluating can be schematically 
drawn as in Figure {\bf \ref{narrow-pants-FW}}.

%FIG13
%
%\begin{center}
%\begin{minipage}{10.0cm}
%

This right hand 
side is identical to an $[L_1, N_1, N_2]$-configuration, apart 
from the fact that it includes an $(N_3 \times N_3)$-domain wall 
configuration, that corresponds to the spin-down reference state 
contribution of $_r\langle {N_3}^{\vee}  |$, that is not included 
in an $[L_1, N_1, N_2]$-configuration. 

\subsection*{Step 2. The domain wall partition functions}
Accounting for the domain wall partition function, and working 
in the homogeneous limit where all quantum rapidities are set 
to $z = \frac{1}{2} \sqrt{-1}$, we obtain our result for the 
structure constants, which up to a factor, is in determinant form.

\begin{equation}
\boxed{
c^{(0)}_{123} =
\cN_{123}
\ \
Z_{N_3}^{\textit{hom}} \ll \{w\}_{N_3}, \frac{1}{2} \sqrt{-1} \rr
\ \
S^{\textit{hom}}[L_1, N_1, N_2]
\ll \{u_{\beta}\}_{N_1}, \{v\}_{N_2}, \frac{1}{2} \sqrt{-1}\rr
}
\label{evaluated-FW}
\end{equation}
\bigskip

\noindent where the normalization
$\mathcal{N}_{123}$ is defined in ({\ref{normalization-FW}}),
the $(N_3 \! \times \! N_3)$ domain wall partition function 
$Z_{N_3}^{\textit{hom}} \ll \{w\}_{N_3}, \frac{1}{2} \sqrt{-1} \rr$ 
is given in ({\ref{izergin-homogeneous-FW}}). The term 
$S^{\textit{hom}}[L_1, N_1, N_2]$ 
$\ll \{u_{\beta}\}_{N_1}, \{v\}_{N_2}, \frac{1}{2} \sqrt{-1}\rr$ 
is an $(N_1 \! \times \! N_1)$ determinant expression of 
the partition function of an $[L_1, N_1, N_2]$-configuration, 
given in ({\ref{restricted-slavnov-homogeneous-FW}}). The auxiliary 
rapidities $\{u\}$, $\{v\}$ and $\{w\}$ are those of the eigenstates 
$\mathcal{O}_1$, $\mathcal{O}_2$ and $\mathcal{O}_3$ in \cite{E1-FW}, 
respectively.  Notice that $\{v\}$ and $\{w\}$ are actually 
$\{v\}_{\beta}$ and $\{w\}_{\beta}$, that is, they 
satisfy Bethe equations, but this fact is not used.

%FIG14
%
%\begin{center}
%\begin{minipage}{4.3in}
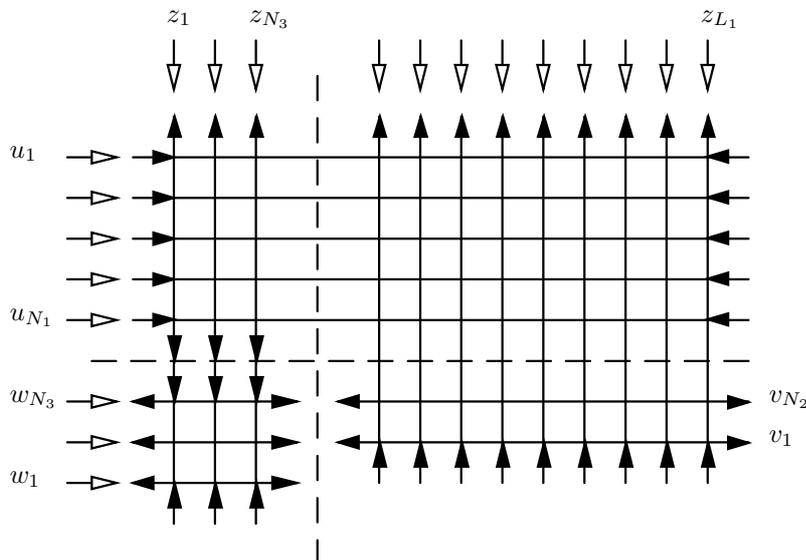
\begin{figure}
\setlength{\unitlength}{0.0009cm}
\begin{picture}(10000,08000)(000,2000)
\thicklines
%
%The following lines are horizontal
%
\path(0600,7500)(9600,7500)
\path(0600,6900)(9600,6900)
\path(0600,6300)(9600,6300)
\path(0600,5700)(9600,5700)
\path(0600,5100)(9600,5100)
\drawline[-30](0000,4500)(9600,4500)
\path(0600,3900)(3000,3900)
\path(0600,3300)(3000,3300)
\path(0600,2700)(3000,2700)
\path(3600,3900)(9600,3900)
\path(3600,3300)(9600,3300)
%
%The following lines are vertical
%
\path(1200,8100)(1200,2100)
\path(1800,8100)(1800,2100)
\path(2400,8100)(2400,2100)
\drawline[-30](3300,8700)(3300,1500)
\path(4200,8100)(4200,2700)
\path(4800,8100)(4800,2700)
\path(5400,8100)(5400,2700)
\path(6000,8100)(6000,2700)
\path(6600,8100)(6600,2700)
\path(7200,8100)(7200,2700)
\path(7800,8100)(7800,2700)
\path(8400,8100)(8400,2700)
\path(9000,8100)(9000,2700)
\blacken\path(0900,3990)(0600,3900)(0900,3810)(0900,3990)
\blacken\path(0900,3390)(0600,3300)(0900,3210)(0900,3390)
\blacken\path(0900,2790)(0600,2700)(0900,2610)(0900,2790)
\blacken\path(3900,3990)(3600,3900)(3900,3810)(3900,3990)
\blacken\path(3900,3390)(3600,3300)(3900,3210)(3900,3390)
\blacken\path(0900,5010)(1200,5110)(0900,5190)(0900,5010)
\blacken\path(0900,5610)(1200,5700)(0900,5790)(0900,5610)
\blacken\path(0900,6210)(1210,6300)(0900,6390)(0900,6210)
\blacken\path(0900,6810)(1200,6900)(0900,6990)(0900,6810)
\blacken\path(0900,7410)(1200,7500)(0900,7590)(0900,7410)
\blacken\path(1300,7800)(1210,8100)(1120,7800)(1300,7800)
\blacken\path(1900,7800)(1810,8100)(1720,7800)(1900,7800)
\blacken\path(2500,7800)(2410,8100)(2320,7800)(2500,7800)
%
%\blacken\path(3700,7800)(3610,8100)(3520,7800)(3700,7800)
\blacken\path(4300,7800)(4210,8100)(4120,7800)(4300,7800)
\blacken\path(4900,7800)(4810,8100)(4720,7800)(4900,7800)
\blacken\path(5500,7800)(5410,8100)(5320,7800)(5500,7800)
\blacken\path(6100,7800)(6010,8100)(5920,7800)(6100,7800)
\blacken\path(6700,7800)(6610,8100)(6520,7800)(6700,7800)
\blacken\path(7300,7800)(7210,8100)(7120,7800)(7300,7800)
\blacken\path(7900,7800)(7810,8100)(7720,7800)(7900,7800)
\blacken\path(8500,7800)(8410,8100)(8320,7800)(8500,7800)
\blacken\path(9100,7800)(9010,8100)(8920,7800)(9100,7800)
%Right most arrows
\blacken\path(9300,7600)(9000,7510)(9300,7420)(9300,7600)
\blacken\path(9300,7000)(9000,6910)(9300,6820)(9300,7000)
\blacken\path(9300,6400)(9000,6310)(9300,6220)(9300,6400)
\blacken\path(9300,5800)(9000,5710)(9300,5620)(9300,5800)
\blacken\path(9300,5200)(9000,5110)(9300,5020)(9300,5200)
%Lower right most arrows
\blacken\path(9300,3800)(9600,3890)(9300,3980)(9300,3800)
\blacken\path(9300,3200)(9600,3290)(9300,3380)(9300,3200)
%
%The folowing are the horizontal white arrows
%
\path(-0360,2700)(0000,2700)
\path(-0360,3300)(0000,3300)
\path(-0360,3900)(0000,3900)
\put(-1200,3900){$w_{N_3}$}
\put(-1200,2700){$w_1$}
\put(9900,3900){$v_{N_2}$}
\put(9900,3300){$v_1$}
\whiten\path(0000,2610)(0360,2700)(0000,2790)(0000,2610)
\whiten\path(0000,3210)(0360,3300)(0000,3390)(0000,3210)
\whiten\path(0000,3810)(0360,3900)(0000,3990)(0000,3810)
\path(-0360,5100)(0000,5100)
\path(-0360,5700)(0000,5700)
\path(-0360,6300)(0000,6300)
\path(-0360,6900)(0000,6900)
\path(-0360,7500)(0000,7500)
\put(-1200,5100){$u_{N_1}$}
\put(-1200,7500){$u_1$}
\whiten\path(0000,5010)(0360,5100)(0000,5190)(0000,5010)
\whiten\path(0000,5610)(0360,5700)(0000,5790)(0000,5610)
\whiten\path(0000,6210)(0360,6300)(0000,6390)(0000,6210)
\whiten\path(0000,6810)(0360,6900)(0000,6990)(0000,6810)
\whiten\path(0000,7410)(0360,7500)(0000,7590)(0000,7410)
%
%The following are the vertical white arrows
%
\path(1200,9220)(1200,8500)
\path(1800,9220)(1800,8500)
\path(2400,9220)(2400,8500)
%
%\path(3600,9220)(3600,8500)
\path(4200,9220)(4200,8500)
\path(4800,9220)(4800,8500)
\path(5400,9220)(5400,8500)
\path(6000,9220)(6000,8500)
\path(6600,9220)(6600,8500)
\path(7200,9220)(7200,8500)
\path(7800,9220)(7800,8500)
\path(8400,9220)(8400,8500)
\path(9000,9220)(9000,8500)
\put(1100,9500){$z_{1}$}
\put(2300,9500){$z_{N_3}$}
\put(8900,9500){$z_{L_1}$}
\whiten\path(1290,8860)(1200,8500)(1110,8860)(1290,8860)
\whiten\path(1890,8860)(1800,8500)(1710,8860)(1890,8860)
\whiten\path(2490,8860)(2400,8500)(2310,8860)(2490,8860)
%
%\whiten\path(3690,8860)(3600,8500)(3510,8860)(3690,8860)
\whiten\path(4290,8860)(4200,8500)(4110,8860)(4290,8860)
\whiten\path(4890,8860)(4800,8500)(4710,8860)(4890,8860)
\whiten\path(5490,8860)(5400,8500)(5310,8860)(5490,8860)
\whiten\path(6090,8860)(6000,8500)(5910,8860)(6090,8860)
\whiten\path(6690,8860)(6600,8500)(6510,8860)(6690,8860)
\whiten\path(7290,8860)(7200,8500)(7110,8860)(7290,8860)
\whiten\path(7890,8860)(7800,8500)(7710,8860)(7890,8860)
\whiten\path(8490,8860)(8400,8500)(8310,8860)(8490,8860)
\whiten\path(9090,8860)(9000,8500)(8910,8860)(9090,8860)
%
%vertical arrow on long vertical segment
%\blacken\path(1890,4560)(1800,4200)(1710,4560)(1890,4560)
%
%vertical arrows on fourth band from below
%
%These are the current final vertical arrows pointing
%upwards
%
%vertical arrow on long vertical segment
%\blacken\path(2500,4560)(2410,4200)(2320,4560)(2500,4560)
%
%Right boundary of domain wall
%
\blacken\path(2640,3810)(3000,3900)(2640,3990)(2640,3810)
\blacken\path(2640,3210)(3000,3300)(2640,3390)(2640,3210)
\blacken\path(2640,2610)(3000,2700)(2640,2790)(2640,2610)
%
%Here are the 3 left most vertical arrows in the new initial 
%state. They point downwards
%
\blacken\path(1300,2340)(1210,2700)(1120,2340)(1300,2340)
\blacken\path(1900,2340)(1810,2700)(1720,2340)(1900,2340)
\blacken\path(2500,2340)(2410,2700)(2320,2340)(2500,2340)
%
%The two arrows on top of the left most initial state arrow
%
% vertical arrow on z_1 long segment
%
%vertical arrow on long vertical segment
%\blacken\path(1300,4560)(1210,4200)(1120,4560)(1300,4560)
%
%the arrows on lower left boundary of upper partition 
%
\blacken\path(1300,4860)(1210,4500)(1120,4860)(1300,4860)
\blacken\path(1900,4860)(1810,4500)(1720,4860)(1900,4860)
\blacken\path(2500,4860)(2410,4500)(2320,4860)(2500,4860)
%
%the arrows on upper boundary of domain wall config
%
\blacken\path(1300,4260)(1210,3900)(1120,4260)(1300,4260)
\blacken\path(1900,4260)(1810,3900)(1720,4260)(1900,4260)
\blacken\path(2500,4260)(2410,3900)(2320,4260)(2500,4260)
%
%Here is the rest of the arrows in the new initial state.
%They point upwards.
%\blacken\path(3700,2940)(3610,3300)(3520,2940)(3700,2940)
\blacken\path(4300,2940)(4210,3300)(4120,2940)(4300,2940)
\blacken\path(4900,2940)(4810,3300)(4720,2940)(4900,2940)
\blacken\path(5500,2940)(5410,3300)(5320,2940)(5500,2940)
\blacken\path(6100,2940)(6010,3300)(5920,2940)(6100,2940)
\blacken\path(6700,2940)(6610,3300)(6520,2940)(6700,2940)
\blacken\path(7300,2940)(7210,3300)(7120,2940)(7300,2940)
\blacken\path(7900,2940)(7810,3300)(7720,2940)(7900,2940)
\blacken\path(8500,2940)(8410,3300)(8320,2940)(8500,2940)
\blacken\path(9100,2940)(9010,3300)(8920,2940)(9100,2940)
\end{picture}
%
%\begin{ca} 
\label{lattice-3-pt-FW}
\caption{The six-vertex lattice configuration that corresponds, 
up to a normalization factor $\cN_{123}$, to the structure constant 
$c^{(0)}_{123}$.} 
%\end{ca}
%
%\end{minipage}
\end{figure}

\bigskip

\subsection{Type-A specializations}
Equation {\bf \ref{evaluated-FW}} is quite general.
To obtain an expression specific to a certain Type-{\bf A} theory, 
we need to use the values of the spin-chain parameters 
appropriate to that theory, as were given in Subsection 
{\bf \ref{doublets-FW}}. 
All Type-{\bf A} theories map to XXX spin-$\frac{1}{2}$ chains, 
hence the anisotropy parameter $\Delta = 1$, but with different 
values for the twist parameter $\theta$. 
Theory {\bf 1} is SYM$_4$ and $\theta = 0$. 
Theory {\bf 2} is SYM$_4^M$ is an Abelian orbifold version 
of SYM$_4$ and $\theta = \frac{2 \pi}{M}$.
Theory {\bf 3} is a real-$\beta$-deformed version  
of SYM$_4$ and $\theta = \beta$.

%SECTION05
\section{Structure constants in Type-B theories}
\label{section-B-structure-constants-FW}

In this section, we consider structure constants in Type-{\bf B} 
theories. Our approach is parallel to that used in Type-{\bf A}. 
The difference is that each Type-{\bf B} theory has 
only one doublet, and therefore requires a slightly modified 
treatment \footnote{The conclusion that, in order to obtain 
a determinant formula, one of the single-trace operators should 
be BPS-like, was obtained in discussions with C Ahn and 
R Nepomechie.}.

In type-{\bf A} theories, the left part of $\mathcal{O}_2$ gets 
trivially contracted with the right part of $\mathcal{O}_3$, and 
the pants diagram is reduced to the `{\it narrow pants} diagram' 
in Figure {\bf \ref{narrow-pants-FW}}. As we will see, the starting 
point in the case of Type-{\bf B} theories is a \lq narrow pants\rq 
diagram. 

This implies that in Type-{\bf B} theories $\mathcal{O}
_3$ must be chosen 
to be a BPS-like state, with {\it one} type of fundamental 
field in the composite operator $\mathcal{O}_3$. On the other hand, 
since the missing contraction (that between the left part of 
$\mathcal{O}_2$ and the right part of $\mathcal{O}_3$) was trivial 
for Type-{\bf A} theories, the final result remains the same.
 
\subsection{Type-B. Fundamental field content of the states} 
As in Type-{\bf A}, we consider single-trace operators in 
an $SU(2)$ sector of a 1-loop conformally-invariant gauge theory, 
that is $\mathrm{Tr} (f_1 f_2 f_3 \cdots)$, where $f_i \in \{u, d\}$ 
is a fundamental field that belongs to an $SU(2)$ doublet. 

The new feature in Type-{\bf B} theories is that we have only 
one doublet to work with. The doublets relevant to Type-{\bf B} 
theories were given in Subsection {\bf \ref{doublets-FW}}. 
Theory {\bf 4} is pure gauge SYM$_2$, and the doublet consists of 
the gluino and its conjugate $\{\lambda, \bar{\lambda}\}$. Theory 
{\bf 5} is pure gauge SYM$_1$, and the doublet consists of the
complex scalar and its conjugate $\{\phi,    \bar{\phi}\}$. Theory 
{\bf 6} is pure QCD and the doublet consists of the light 
cone derivative of the gauge field component $A$ and its 
conjugate $\bar{A}$, that is, 
$\{ \partial_{+} A, \partial_+ \bar{A}\}$. 
In the following, we deal with all three theories in one 
go, using the notation $\{\zeta, \bar{\zeta}\}$ for a generic 
single doublet.

Since we have only one doublet to construct composite operators 
from, we identify the fundamental field content 
of $\mathcal{O}_i$, $i \in \{1, 2, 3\}$ with spin-chain spin 
states as shown in Table {\bf 2}. 

\setlength{\extrarowheight}{6.0pt}

\begin{table}[h]
\begin{center}
\begin{tabular}{| c || c | c | c | c |}
\hline
{Operator} & 
{$\ll \begin{array}{c} 1 \\  0 \end{array} \rr$} & 
{$\ll \begin{array}{c} 0 \\  1 \end{array} \rr$} &
{$\ll                  1 \ \ 0             \rr$} &
{$\ll                  0 \ \ 1             \rr$} 
\\
\hline
\hline
$\mathcal{O}
_1$ & $      \zeta  $ & $  \bar{\zeta} $ & $ \bar{\zeta} $ & $       \zeta  $ 
\\ 
\hline
$\mathcal{O}
_2$ & $ \bar{\zeta} $ & $       \zeta  $ & $      \zeta  $ & $  \bar{\zeta} $ 
\\
\hline
$\mathcal{O}
_3$ & $ \bar{\zeta} $ & $        \zeta $ & $       \zeta $ & $  \bar{\zeta} $ 
\\
\hline
\end{tabular}
\bigskip
\caption{Identification of Type-{\bf B} operator content of $\mathcal{O}
_i$,
$i \in \{1, 2, 3\}$ with initial and final spin-chain states.}
\end{center}
\end{table}

Once again, in our conventions

\begin{equation}
\label{contractions-2-FW}
\langle
 \bar{\zeta} \zeta \rangle
 = \langle
      \zeta  | \zeta \rangle
 = 1, \quad
\langle
      \zeta  \zeta \rangle
 = \langle
 \bar{\zeta} | \zeta \rangle
 = 0 
\end{equation}

From Table {\bf 2}, one can read the fundamental-scalar 
operator content of each single-trace operator $\mathcal{O}
_i$,
$i \in \{1, 2, 3\}$, when it is an initial state and 
when it is a final state.

\subsection{Similarities between Type-A and Type-B theories}
Steps {\bf 1}, {\bf 2} and {\bf 3} from the EGSV construction of 
the structure constants apply unchanged to Type-{\bf B} theories. 
In other words, 
{\bf 1.} The splitting of each lattice, 
{\bf 2.} The flipping procedure, and 
{\bf 3.} The contraction of left and right halves to form scalar 
products, are replicated in the case of Type-{\bf B} theories. 
Therefore we see that equation (\ref{intermediate-FW}) continues 
to hold, and we assume that as our starting point. 

\subsection{Differences between Type-A and Type-B theories} 
{\bf 1.} In the case of Type-{\bf A} theories, 
$\mathcal{O}_3$ contains $Z$ fields that can only contract with 
$\bar{Z}$ fields in $\mathcal{O}_2$. This is because there are no 
fields that they can contract with in $\mathcal{O}_1$. This trivializes 
the $_l\langle \mathcal{O}_2|\mathcal{O}_3 \rangle_r$ scalar product. 

This is not the case in Type-{\bf B}
theories, where we have only a single doublet that 
must be used to populate all three states 
$\mathcal{O}_1$, $\mathcal{O}_2$ and $\mathcal{O}_3$. Because of that, 
one can see that if there is a contraction between 
$\mathcal{O}_2$ and $\mathcal{O}_3$, 
it is in general non-trivial. This is 
sufficient to prevent us from duplicating 
our Type-{\bf A} arguments in the case of  
Type-{\bf B} theories. In fact, there is yet another difference.

{\bf 2.} In the case of Type-{\bf A} theories, 
$\mathcal{O}_3$ contains $\bar{X}$ fields that 
can contract only with $X$ fields in $\mathcal{O}_1$. 
The reason is that there are no $X$ fields in $\mathcal{O}_2$.
This trivializes the scalar product that involves the left 
part of $\mathcal{O}_1$ and the right part of $\mathcal{O}_3$, 
leading to a domain wall partition function.

Once again, in the case of Type-{\bf B} theories, 
the above trivial contraction is no longer the case, 
and contractions between $\mathcal{O}_1$ and 
$\mathcal{O}_3$ are in general non-trivial.

\subsection{One of the operators must be BPS-like}
Because of the above reasons, {\it we cannot map the most 
general} $SU(2)$ structure constants of Type-${\bf B}$ 
operators onto a restricted Slavnov scalar product. 
However, both problems are overcome if we take $\mathcal{O}_3$ 
to be BPS-like, that is, a single-trace operator of 
the form 
${\rm Tr} \ [\bar{\zeta} \ \bar{\zeta} \cdots \bar{\zeta} \ ]$. 
This means that we {\it demand} that $N_3 = L_3$, or equivalently, 
that $l_{23} = L_3 - N_3 = 0$. In other words, the fields 
in $\mathcal{O}_3$ are all of the same type $\bar{\zeta}$ (magnons) 
and they contract with a subset of the fields in $\mathcal{O}_1$, 
while there are no contractions between $\mathcal{O}_3$ and 
$\mathcal{O}_2$. From this, we conclude that the starting point 
of the Type-{\bf B} structure constants that we can compute 
in determinant form is the \lq narrow pants\rq diagram 
in Figure {\bf \ref{narrow-pants-FW}}.

But we know that the partition function of the lattice 
configuration corresponding to Figure {\bf \ref{narrow-pants-FW}} 
is given by a restricted Slavnov scalar product. Therefore for 
Type-{\bf B} structure constants for which $\mathcal{O}_3$ is 
BPS-like, that is $L_3 = N_3$, we obtain 

\begin{equation}
c^{(0)}_{123} =
\mathcal{N}_{123} 
Z_{N_3}^{\textit{hom}} \ll \{w\}_{N_3}, \frac{1}{2} \sqrt{-1} \rr
%\\
%\times
S^{\textit{hom}}[L_1, N_1, N_2] 
\ll \{u_{\beta}\}_{N_1}, \{v\}_{N_2}, \frac{1}{2} \sqrt{-1}\rr
\label{evaluated-2-FW}
\end{equation}

\noindent This is the same result as the Type-{\bf A} case, but 
with the caveat that we are restricting our attention to the situation 
$L_3 = N_3$. As a result the Gaudin norm $\mathcal{N}_3$, which occurs 
in the normalization factor $\mathcal{N}_{123}$, is equal to the partition 
function of a $BC$-configuration with length-$N_3$ initial and final 
reference states, and $N_3$ $B$-lines and $C$-lines. As we commented 
in Subsection {\bf \ref{DW-remarks-FW}}, such a configuration factorizes 
into a product of domain wall partition functions. Hence we are able to 
cancel the factor 
$Z_{N_3}^{\textit{hom}} \ll \{w\}_{N_3}, \frac{1}{2} \sqrt{-1} \rr$ 
in (\ref{evaluated-2-FW}) at the expense of the factor 
$\sqrt{\mathcal{N}_3}$ in the denominator, and obtain the final 
expression 
\begin{equation}
\boxed{
c^{(0)}_{123} =
\sqrt{\frac{
L_1 L_2 L_3
}
{
\mathcal{N}
_1 \mathcal{N}
_2
}}
\ \ 
S^{\textit{hom}}[L_1, N_1, N_2] 
\ll \{u_{\beta}\}_{N_1}, \{v\}_{N_2}, \frac{1}{2} \sqrt{-1}\rr
}
\label{evaluated-3-FW}
\end{equation}

\subsection{Type-B specializations}
As in the previous section, (\ref{evaluated-2-FW}) is quite general.
To obtain an expression specific to a certain Type-{\bf B} theory, 
we need to use the values of the spin-chain parameters 
appropriate to that theory, as were given in Subsection 
{\bf \ref{doublets-FW}}. 
All Type-{\bf B} theories map to periodic XXZ spin-$\frac{1}{2}$ 
chains, hence the twist parameter $\theta = 0$, but with 
different values of the anisotropy parameter $\Delta$.
Theory {\bf 4} is pure SYM$_2$ and $\Delta = 3$ 
\cite{DiV-FW, korchemsky-FW}. 
Theory {\bf 5} is pure SYM$_1$ and $\Delta = \frac{1}{2}$ 
\cite{korchemsky-FW}. 
Theory {\bf 6} is pure gauge QCD and $\Delta =  - \frac{11}{3}$ 
\cite{korchemsky-FW}.

%SECTION06
\section{Discrete KP $\tau$-functions}
\label{section-kp-FW}

In this section we closely follow \cite{FS-FW}, where it was shown 
that Slavnov's scalar product is a $\tau$-function of the discrete 
KP hierarchy. The only differences in this work are {\bf 1.} A more 
compact expression for the $\tau$-function itself, see 
(\ref{final-tau-FW}), {\bf 2.} The inclusion of the twist parameter 
$\theta$ in the $\tau$-function, and {\bf 3.} A discussion of 
restricting the Miwa variables to the values of the quantum 
inhomogeneities.   

\subsection{Notation related to sets of variables} 
We use $\{x\}$ for the set of finitely many variables 
$\{x_1, x_2, \dots, x_N\}$, and 
$\{\widehat{x}_m\}$ for $\{x\}$ with 
the element $x_m$ omitted. In the case of sets with 
a repeated variable $x_i$, we use the superscript 
$(m_i)$ to indicate the multiplicity of $x_i$, as 
in $x_i^{(m_i)}$. For example,
$\{x_1^{(3)}, x_2, x_3^{(2)}, x_4, \dots \}$ 
is the same as
$\{x_1, x_1, x_1, x_2, x_3, x_3, x_4, \dots \}$
and $f\{\dots, x_{i}^{(m_i)}, \dots\}$ 
is equivalent to saying that $f$ depends on $m_i$ distinct 
variables all of which have the same value $x_i$. 
For simplicity, we use $x_i$ to indicate $x_i^{(1)}$. 

\subsection{The complete symmetric function $h_i\{x\}$} 
Let $\{x\}$ denote a set of $N$ variables 
$\{x_1, x_2, $ $\dots, x_N\}$. 
The complete symmetric function $h_i\{x\}$ is the coefficient of 
$k^i$ in the power series expansion 
\begin{align}
\prod_{i=1}^{N}
\frac{1}{1-x_i \, k}
=
\sum_{i=0}^{\infty}
h_i\{x\}
\, 
k^i
\label{complete-symmetric-FW}
\end{align}
For example, $h_0\{x\} = 1$, $h_1(x_1,x_2,x_3) = x_1+x_2+x_3$,
$h_2(x_1,x_2) = x_1^2+x_1x_2+x_2^2$, and 
$h_{i}\{x\} = 0$ for $i<0$. 

\subsection{Useful identities for $h_i\{x\}$} 
From (\ref{complete-symmetric-FW}), it is straightforward to 
show that
\begin{align}
h_i\{x\} = h_i\{\widehat{x}_m\} + x_m h_{i-1}\{x\}
\label{i1-FW}
\end{align}
Then from (\ref{i1-FW}
) one obtains
\begin{align}
(x_m-x_n)
h_{i-1}\{x\}
=
h_i\{\widehat{x}_n\}
-
h_i\{\widehat{x}_m\}
\label{i3-FW}
\end{align}
\begin{align}
(x_m-x_n)
h_i\{x\}
=
x_m
h_i\{\widehat{x}_n\}
-
x_n
h_i\{\widehat{x}_m\}
\label{i2-FW}
\end{align}

\subsection{Discrete derivatives}
The discrete derivative $\Delta_m h_i\{x\}$ of $h_i\{x\}$ with 
respect to any one variable $x_m \in \{x\}$ is defined using 
(\ref{i1-FW}) as  
\begin{align} 
\Delta_{m} h_i\{x\} = 
\frac{
h_i\{x\} - h_i\{ \widehat{x}_m \}
}
{x_m
} 
= h_{i-1}\{x\}
\label{discrete-derivative-FW}
\end{align}
Note that the effect of applying $\Delta_m$ to
$h_i\{x\}$ is a complete symmetric function $h_{i-1}\{x\}$ 
of degree $i-1$ in the same set of variables $\{x\}$.

\subsection{The discrete KP hierarchy}

Discrete KP is an infinite hierarchy of integrable 
partial {\it difference} equations in an infinite set of continuous 
Miwa variables $\{x\}$, where time evolution is obtained by changing 
the multiplicities $\{m\}$ of these variables. In this work, we are 
interested in the situation where the total number of continuous Miwa 
variables is finite, which corresponds to setting to zero all continuous 
Miwa variables apart from $\{x_1,\dots,x_N\}$. In this case, the discrete 
KP hierarchy can be written in bilinear form as the $n$ $\times$ $n$ 
determinant equations
\begin{align}
\det
\ll
\begin{array}{cccccc}
     1 & x_1    & \cdots & x_1^{n-2} & \, & x_1^{n-2} \tau_{+1}\{x\}
\tau_{-1}\{x\} \\
     1 & x_2    & \cdots & x_2^{n-2} & \, & x_2^{n-2} \tau_{+2}\{x\}
\tau_{-2}\{x\} \\
\vdots & \vdots & \vdots & \vdots    & \, & \vdots                         
        \\
     1 & x_n    & \cdots & x_n^{n-2} & \, & x_n^{n-2} \tau_{+n}\{x\}
\tau_{-n}\{x\} 
\end{array}
\rr
=0
\label{bilinear-difference-equation-FW}
\end{align}
where $3 \leq n \leq N$, and
\begin{align}
\tau_{+i}\{x\} 
&= 
\tau\{ x_1^{(m_1    )}, \dots, x_i^{(m_i + 1)}, \dots, x_N^{(m_N    )}\}
\\
\tau_{-i}\{x\} 
&= 
\tau\{ x_1^{(m_1 + 1)}, \dots, x_i^{(m_i    )}, \dots, x_N^{(m_N + 1)}\}
\nonumber
\end{align}
In other words, if $\tau\{x\}$ has $m_i$ copies of the variable $x_i$, 
then $\tau_{+i}\{x\}$ has $m_i + 1$ copies of $x_i$ and the multiplicities 
of all other variables remain the same, while $\tau_{-i}\{x\}$ has one 
more copy of each variable except $x_i$. Equivalently, one can use the 
simpler notation
\begin{align}
\tau_{+i}\{x\} 
&= 
\tau\{  m_1,      \dots, (m_i + 1), \dots,  m_N     \}
\label{notation-for-tau-FW}
\\
\tau_{-i}\{x\} 
&=
\tau\{ (m_1 + 1), \dots,  m_i     , \dots, (m_N + 1)\}
\nonumber
\end{align}
The simplest discrete KP bilinear difference equation, in the notation 
of (\ref{notation-for-tau-FW}), is 
\begin{multline}
\label{miwa-hirota-FW}
x_i(x_j-x_k) \tau\{ m_i+1, m_j,   m_k   \} \tau\{ m_i,   m_j+1, m_k+1 \} 
\\
+ 
x_j(x_k-x_i) \tau\{ m_i,   m_j+1, m_k   \} \tau\{ m_i+1, m_j,   m_k+1 \}
\\
+ 
x_k(x_i-x_j) \tau\{ m_i,   m_j,   m_k+1 \} \tau\{ m_i+1, m_j+1, m_k   \}   
= 0
\end{multline}
where $\{x_i, x_j, x_k\} \in \{x\}$ and 
$\{m_i, m_j, m_k\} \in \{m\}$ are
any two (corresponding) triples in the sets of continuous and discrete 
(integral valued) Miwa variables. Equation (\ref{miwa-hirota-FW}) 
is the discrete analogue of the KP equation in continuous time variables.

\subsection{Casoratian matrices and determinants}

A Casoratian matrix $\Omega$ of the type that appears in this paper
is such that its matrix elements $\omega_{ij}$ satisfy 
\begin{align}
\omega_{i,j+1}\{x\} = \Delta_m \omega_{ij}\{x\} 
\label{casoratian-condition-FW}
\end{align}
where the discrete derivative $\Delta_m$ is taken with 
respect to any one variable $x_m \in \{x\}$ (it is redundant to specify 
which variable, since $\omega_{ij}\{x\}$ is symmetric in $\{x\}$).
From the definition of the discrete derivative $\Delta_m$, it is clear 
that the entries of Casoratian matrices satisfy
\begin{multline}
\label{a1-FW}
\omega_{ij}\{x_1, \ldots, x_{m}^{(2)}, \ldots, x_{N}\}
=
\\
\omega_{i j  }\{x_1, \ldots,                      x_N \} 
+ 
x_m \omega_{i,j+1}\{x_1, \ldots, x^{(2)}_{m}, \ldots, x_N \}
\end{multline}
which, in turn, gives rise to the identity
\begin{multline}
(x_r - x_s) 
\  
      \omega_{ij} \{x_{1}, \ldots, x^{(2)}_{r},\dots, x^{(2)}_{s}, \ldots x_N \} = 
\\
x_{r} \ \omega_{ij} \{x_1,   \ldots, x_{r}^{(2)}, \ldots, x_{N} \}
-
x_{s} \ \omega_{ij} \{x_1,   \ldots, x_{s}^{(2)}, \ldots, x_{N} \}
\label{a2-FW}
\end{multline}
If $\Omega$ is a Casoratian matrix, then $\det \Omega$ is a Casoratian 
determinant. Casoratian determinants are discrete analogues 
of Wronskian determinants.

\subsection{Notation for column vectors with elements $\omega_{ij}$}

We need the column vector
\begin{align}
\vec\omega_{j} =
\ll
\begin{array}{c} 
\omega_{1j} \{ x^{(m_{1})}_1, \ldots, x^{(m_{N})}_{N} \} \\ 
\omega_{2j} \{ x^{(m_{1})}_1, \ldots, x^{(m_{N})}_{N} \} \\ 
\vdots                               \\ 
\omega_{Nj} \{ x^{(m_{1})}_1, \ldots, x^{(m_{N})}_{N} \} 
\end{array}
\rr
\end{align}
and write 
\begin{align}
\vec\omega_{j}^{[k_1,\ldots,k_n]} =
\ll
\begin{array}{c} 
\omega_{1j}\{ x^{(m_{1})}_1, \ldots, x^{(m_{k_1}+1)}_{k_1},\ldots,x^{(m_{k_n}+1)}_{k_n},\ldots, x^{(m_{N})}_{N} \} \\ 
\omega_{2j}\{ x^{(m_{1})}_1, \ldots, x^{(m_{k_1}+1)}_{k_1},\ldots,x^{(m_{k_n}+1)}_{k_n},\ldots, x^{(m_{N})}_{N} \} \\ 
\vdots \\ 
\omega_{Nj}\{ x^{(m_{1})}_1, \ldots, x^{(m_{k_1}+1)}_{k_1},\ldots,x^{(m_{k_n}+1)}_{k_n},\ldots, x^{(m_{N})}_{N} \}
\end{array}
\rr
\end{align}
for the corresponding column vector where the multiplicities of the variables 
$x_{k_1}, \dots,x_{k_n}$ are increased by 1. 

\subsection{Notation for determinants with elements $\omega_{ij}$}

We also need the determinant 
\begin{align}
\tau= 
\det 
\ll   \vec\omega_{1}\,\, 
      \vec\omega_{2}\,\,
      \cdots        \,\,
      \vec\omega_{N}
\rr
=
\big | \, 
      \vec\omega_{1}\,\, 
      \vec\omega_{2}\,\,
      \cdots        \,\,
      \vec\omega_{N}
\, \big |
\end{align}
and the notation 
\begin{align}
\tau^{[k_1, \ldots, k_n]} 
=
\big | \, 
\vec\omega_{1}^{[k_1, \ldots, k_n]} \, \, 
\vec\omega_{2}^{[k_1, \ldots, k_n]} \, \, 
\cdots                           \, \, 
\vec\omega_{N}^{[k_1, \ldots, k_n]}
\, \big |
\label{defn-FW}
\end{align}
for the determinant with shifted multiplicities. 

\subsection{Identities satisfied by Casoratian determinants}

Two identities, which are needed in the sequel, are
\begin{align}
x^{n-2}_{1} \ \tau^{[1]} =
\big | \,
\vec\omega_{1}   \,\,
\vec\omega_{2}   \,\,
\cdots           \,\, 
\vec\omega_{N-1} \,\,
\vec\omega_{N-n+2}^{[1]}
\, \big |
\label{A1-FW}
\end{align}
\begin{equation}
\prod_{1\leq r<s\leq n}(x_r-x_s)  \tau^{[1, \ldots, n]} =
%\\
\big | \,
\vec\omega_1            \,\,
\ldots                  \,\,
\vec\omega_{N-n}        \,\,
\vec\omega_{N-n+1}^{[n]}   \,\,
\vec\omega_{N-n+1}^{[n-1]} \,\,
\ldots                  \,\,
\vec\omega_{N-n+1}^{[1]}
\, \big | 
\label{A2-FW}
\end{equation}
These identities may be proved by using the (\ref{a1-FW}) 
and (\ref{a2-FW}) to perform column operations in the 
determinant expressions for $\tau^{[1]}$ and $\tau^{[1,\dots,n]}$. 
To keep the exposition concise we do not present these proofs, 
but full details can be found in \cite{FS-FW}. 

\subsection{Casoratians are discrete KP $\tau$-functions}

Following \cite{ohta-FW}, consider the $2N$ $\times$ $2N$ determinant
\begin{multline}
\det
\ll
\begin{array}{cccccccc}
\vec\omega_1          & 
\cdots                &
\vec\omega_{N-1}      & 
\vec\omega_{N-n+2}^{[1]} &    
0_1                   & 
\cdots                &    
0_{N-n+1}             &
\vec\omega_{N-n+2}^{[n]} 
\cdots 
\vec\omega_{N-n+2}^{[2]} 
\\
0_1                   & 
\cdots                &    
0_{N-1}               & 
\vec\omega_{N-n+2}^{[1]} & 
\vec\omega_1          & 
\cdots                & 
\vec\omega_{N-n+1}    &
\vec\omega_{N-n+2}^{[n]} 
\cdots 
\vec\omega_{N-n+2}^{[2]} 
\end{array}
\rr
\\
= 0
\label{2N-by-2N-FW}
\end{multline}
which is identically zero. For notational clarity, we have used 
subscripts to label the position of columns of zeros. Performing 
a Laplace expansion of the left hand side of (\ref{2N-by-2N-FW}) 
in $N$ $\times$ $N$ minors along the top $N$ $\times$ $2N$ block, 
we obtain 
\begin{multline}
\sum_{k = 1}^{n} 
(-)^{k - 1}
\big | \,
  \vec\omega_1 
  \cdots 
  \vec\omega_{N-1}  
  \vec\omega_{N-n+2}^{[k]} 
\, \big | 
\times \\
\big | \,
  \vec\omega_1 
  \cdots 
  \vec\omega_{N-n+1}         
  \vec\omega_{N-n+2}^{[n]} 
  \cdots 
  \vec\omega_{N-n+2}^{[k+1]} 
  \vec\omega_{N-n+2}^{[k-1]} 
  \cdots 
  \vec\omega_{N-n+2}^{[1]}
\, \big |
= 0
\label{Laplace-expansion-FW}
\end{multline}
By virtue of (\ref{A1-FW}
) and (\ref{A2-FW}
), (\ref{Laplace-expansion-FW}
) 
becomes
\begin{align}
\sum_{k = 1}^{n} 
(-)^{k - 1}
x^{n-2}_{k}\ \tau^{[k]}
\prod_{\substack{1\leq r<s\leq n \\ r,s\neq k}}(x_r-x_s)
\tau^{[1,\ldots\hat k \ldots,n]}
= 0
\label{57-FW}
\end{align}
Using the Vandermonde determinant identity
\begin{align}
\det
\ll
\begin{array}{cccccc}
     & 1 & x_1    & \cdots & x_1^{n-2} & \\
     & \vdots & \vdots & & \vdots & \\
\langle  & 1 & x_{k}  & \cdots & x_{k}^{n-2} & \rangle \\
	& \vdots & \vdots & & \vdots &  \\
      & 1 & x_n    & \cdots & x_n^{n-2}  &
\end{array}
\rr
=
\prod_{\substack{1\leq r<s\leq n \\ r,s\neq k}}(x_r-x_s)
\end{align}
with $\langle \begin{array}{cccc} 1 & x_k & \cdots & x_k^{n-2} \end{array} 
\rangle$ denoting the omission of the $k$-th row of the matrix, we recognize 
(\ref{57-FW}
) as the cofactor expansion of the determinant in 
(\ref{bilinear-difference-equation-FW}
) along its last column. Hence we 
conclude that Casoratian determinants satisfy the bilinear difference 
equations of discrete KP.

\subsection{Change of variables}
To interpret the Slavnov determinant (\ref{slavnov-FW}) as a $\tau$-function 
of discrete KP in the sense described above, it is necessary to adopt 
a change of variables as follows
\begin{align}
\label{set-FW}
\{
e^{-2 v_i},
e^{2 u_{\beta_i}},
e^{2 z_i},
e^{2 \eta}
\}
\rightarrow
\{
x_i, y_i, z_i, q
\}
\end{align}
In other words, our new variables (of which $\{x_1,\dots,x_N\}$ end up being 
the continuous Miwa variables of discrete KP) are expressed as exponentials 
of the original variables. Furthermore, we consider a new normalization of 
the scalar product, given by
\begin{equation}
\label{renorm1-FW}
\mathbb{S}[L,N,N]
=
%\\
e^{N^2 \eta}
\prod_{i=1}^{N}
e^{(L-1)(u_{\beta_i}-v_i)}
\prod_{i=1}^{L}
e^{2Nz_i}
\prod_{j=1}^{N} \prod_{k=1}^{L}
[v_j - z_k]
[u_{\beta_j} - z_k]
S[L,N,N]
\end{equation}
Applying this normalization to (\ref{slavnov-FW}), performing trivial 
rearrangements within the determinant and making the change of 
variables as prescribed by (\ref{set-FW}), we obtain 
\begin{multline}
\label{det1-FW}
\mathbb{S}[L,N,N]
=
\frac{\displaystyle{
(q-1)^N
\prod_{i=1}^{N}
\prod_{j=1}^{L}
(y_i-z_j)
}}
{\displaystyle{
\prod_{1\leq i<j \leq N}
(x_i-x_j)(y_i-y_j)
}}
\times
\\
\det\ll
\frac{
\displaystyle{
e^{- 2 i \theta}
q^{N-1}
\prod_{k\not= i}^{N}
\Big(1-x_j \frac{y_k}{q}\Big)
\prod_{k=1}^{L}
(1-x_j z_k)
}
}
{1-x_j y_i}
\right.
%\\
\left.
-
\frac{\displaystyle{
q^{\frac{L}{2}}
\prod_{k\not=i}^{N}
(1-q x_j y_k)
\prod_{k=1}^{L}
\Big(1 - x_j \frac{z_k}{q}\Big)
}
}
{1-x_j y_i}
\rr_{1\leq i,j \leq N}
\end{multline}
Our goal is to show that $\mathbb{S}[L,N,N]$ has the form of 
a Casoratian determinant, where the discrete derivative is 
taken with respect to the variables $\{x_1,\dots,x_N\}$.

\subsection{Removing the pole in the Slavnov scalar product}
For all $1\leq i \leq N$, define the function $\gamma_i$ as 
\begin{equation}
\gamma_i
=
e^{- 2 i \theta}
q^{N-1}
\prod_{j\not=i}^{N}
\ll
1 - \frac{y_j}{q y_i}
\rr
\prod_{j=1}^{L}
\ll
1-\frac{z_j}{y_i}
\rr
%\\
-
q^{\frac{L}{2}}
\prod_{j\not=i}^{N}
\ll
1 - \frac{q y_j}{y_i}
\rr
\prod_{j=1}^{L}
\ll
1-\frac{z_j}{q y_i}
\rr
\end{equation}
These functions provide a convenient way of expressing the Bethe 
equations (\ref{bethe-equations-FW}) under the change of variables 
(\ref{set-FW}
), namely 
\begin{align}
\gamma_i = 0, \quad \text{for all}\ 1\leq i \leq N.
\end{align}
Recalling that these equations are assumed to apply to the variables 
$\{y_1,$ $\dots,$ $y_N \}$, we see that the pole at $x_j = 1/y_i$ in 
the determinant of (\ref{det1-FW}) can be removed. We omit the details 
here as they are mechanical, and state only the result of this calculation, 
which reads
\begin{equation}
\label{det2-FW}
\mathbb{S}[L,N,N]
=
%\\
\frac{\displaystyle{
(q-1)^N
\prod_{i=1}^{N}
\prod_{j=1}^{L}
(y_i-z_j)
}}
{\displaystyle{
\prod_{1\leq i<j \leq N}
(x_i-x_j)(y_i-y_j)
}}
\det\ll
\sum_{k=0}^{L+N-2}
\Big[y_i^k \gamma_i\Big]_+
x_j^k
\rr_{1\leq i,j \leq N}
\end{equation}
where $[y_i^k \gamma_i]_{+}$ denotes all terms in the Laurent
expansion of $y_i^k \gamma_i$ which have non-negative degree in $y_i$. 

\subsection{The Slavnov scalar product is a discrete KP $\tau$-function}
Using identities (\ref{i1-FW}
) and (\ref{i3-FW}
) to perform elementary column 
operations in the determinant of (\ref{det2-FW}
), it is possible to remove 
the Vandermonde $\prod_{1\leq i<j \leq N} (x_i-x_j)$ from the denominator 
of this equation. This procedure is directly analogous to the proof of 
the Jacobi-Trudi identity for Schur functions \cite{macdonald-book-FW}. 
The result obtained is
\begin{equation}
\label{final-tau-FW}
\mathbb{S}[L,N,N]
=
%\\
\frac{\displaystyle{
(q-1)^N
\prod_{i=1}^{N}
\prod_{j=1}^{L}
(y_i-z_j)
}}
{\displaystyle{
\prod_{1\leq i<j \leq N}
(y_j-y_i)
}}
\det\ll
\sum_{k=0}^{L+N-2}
\left[y_i^k \gamma_i\right]_{+}
h_{k-j+1}\{x\}
\rr_{1\leq i,j \leq N}
\end{equation}
Up to an overall multiplicative factor which does not depend on the variables 
$\{x\}$, the normalized scalar product $\mathbb{S}[L,N,N]$ is a determinant 
of the form $\det \Omega$, where the matrix $\Omega$ has entries $\omega_{ij}$ 
which satisfy
\begin{align}
\omega_{i,j+1}
=
\Delta_m
\omega_{i,j},
\quad
\omega_{i,1}
=
\sum_{k=0}^{L+N-2}
\left[y_i^k \gamma_i\right]_{+}
h_{k}\{x\}
\end{align}
Hence $\mathbb{S}[L,N,N]$ has the form of a Casoratian determinant, making it 
a discrete KP $\tau$-function in the variables $\{x\} = \{x_1,\dots,x_N\}$.

\subsection{Restrictions of $\mathbb{S}[L,N_1,N_1]$}

Similarly to (\ref{renorm1-FW}), we define a new normalization of the restricted 
scalar product $S[L,N_1,N_2]$ as follows
\begin{multline}
\label{renorm2-FW}
\mathbb{S}[L,N_1,N_2]
=
e^{N_1^2 \eta}
\prod_{i=1}^{N_1}
e^{(L-1)u_{\beta_i}}
\prod_{i=1}^{N_2}
e^{-(L-1)v_i}
\prod_{i=1}^{L}
e^{(N_1+N_2)z_i}
\\
\times
\prod_{j=1}^{N_2} \prod_{k=1}^{L}
[v_j - z_k]
\prod_{j=1}^{N_1} \prod_{k=1}^{L}
[u_{\beta_j} - z_k]
S[L,N_1,N_2]
\end{multline}
Normalizing both sides of (\ref{restrict1-FW}) using (\ref{renorm1-FW}
) and (\ref{renorm2-FW}
), and working in terms of the variables introduced by (\ref{set-FW}
), we obtain the result 
\begin{equation}
\mathbb{S}[L,N_1,N_1]
\Big|_{\substack{x_{N_1}=1/z_1 \\ \vdots \\ x_{(N_2+1)}=1/z_{N_3}}}
=
%\\
(z_1 \dots z_{N_3})^{1/2}
\prod_{i=1}^{N_3} \prod_{j=1}^{L}
(q^{1/2}-q^{-1/2}z_j /z_i)
\mathbb{S}[L,N_1,N_2]
\end{equation}
Hence the function $\mathbb{S}[L,N_1,N_2]$ is (up to an overall 
multiplicative factor) a restriction of $\mathbb{S}[L,N_1,N_1]$, 
obtained by setting the variables $x_{N_1},\dots,x_{N_2+1}$ to 
the values $1/z_{1},\dots,1/z_{N_3}$. Since $\mathbb{S}[L,N_1,N_1]$ 
is a discrete KP $\tau$-function in the variables 
$\{x_1,$ $\dots,$ $x_{N_1}\}$, it is clear that 
$\mathbb{S}[L,N_1,N_2]$ is also a $\tau$-function in the unrestricted 
set of variables $\{x_1,\dots,x_{N_2}\}$. 

%SECTION07
\section{Summary and comments}
\label{section-summary-FW}

Following \cite{F-FW}, we obtained determinant expressions for 
two types of structure constants. 
{\bf 1.} structure constants of non-extremal 3-point functions 
of single-trace non-BPS operators in the scalar sector of SYM$_4$ 
and two close variations on it (an Abelian orbifolding of SYM$_4$ 
and a real-$\beta$-deformation of it. The operators involved map 
to states in closed XXX spin-$\frac{1}{2}$ chains, that are periodic 
in the case of SYM$_4$, and twisted in the other two cases.
{\bf 2.} structure constants of extremal 3-point functions 
of two non-BPS and one BPS single-trace operators in (not 
necessarily scalar, but spin-zero) sectors of pure gauge 
SYM$_2$, SYM$_1$ and QCD. 
The operators involved map to states in closed periodic XXZ 
spin-$\frac{1}{2}$ chains, with different values of the 
anisotropy parameter, 
as identified in \cite{DiV-FW, korchemsky-FW}. 
One of the operators must be BPS-like.

Our expressions are basically special cases of Slavnov's 
determinant for the scalar product of a Bethe eigenstate 
and a generic state in a (generally twisted) closed XXZ
spin chain. Finally, following \cite{FS-FW}, we showed that 
all these determinants are discrete KP $\tau$-functions, 
in the sense that they obey the Hirota-Miwa equations.

The study of 3-point functions is a continuing activity.
In \cite{plefka-FW}, a systematic study, using perturbation 
theory, of 3-point functions in planar SYM$_4$ at 1-loop level, 
involving scalar field operators up to length 5 is reported on. 
In \cite{GV-FW}, quantum corrections to 3-point functions of 
the very same type studied in this work planar SYM$_4$ are 
studied using integrability. At 1-loop level, new algebraic
structures are found that govern all 2-loop corrections to 
the mixing of the operators as well as automatically 
incorporate all 1-loop corrections to the tree-level 
computations. 

In \cite{AFN-FW}, operator product expansions of local single-trace
operators composed of self-dual components of the field
strength tensor in planar QCD are considered. Using methods 
that extend those used in this work to spin-1 chains, 
a determinant expression for certain tree-level structure
constants that appear in the operator product expansion is 
obtained. More recently, in \cite{Ivan-FW}, the classical limit 
of the determinant form of the structure constants that appear 
in this work, was obtained.

\section*{Acknowledgments}
\label{section-acknowledgements-FW}

OF thanks C Ahn, N Gromov, G Korchemsky, I Kostov, 
R Nepo\-me\-chie, D Serban, P Vieira and K Zarembo for discussions 
on the topic of this work and the Inst H Poincare for hospitality 
where it started.
Both authors thank the Australian Research Council for 
financial support, and the anonymous referee for remarks that 
helped us improve the text.

\end{document}